\documentclass[preprint,12pt]{aastex}
\usepackage{psfig}
\newcommand{\beq}{\begin{equation}}
\newcommand{\eeq}{\end{equation}}
\newcommand{\bea}{\begin{eqnarray}}
\newcommand{\eea}{\end{eqnarray}}
\newcommand{\bean}{\begin{eqnarray*}}
\newcommand{\eean}{\end{eqnarray*}}
\newcommand{\ba}{\begin{array}}
\newcommand{\ea}{\end{array}}
\newcommand{\bml}{\begin{mathletters}}
\newcommand{\eml}{\end{mathletters}}
\newcommand{\rem}[1]{{ }}
\newcommand{\dd}[2]{\frac{{\rm d} #1}{{\rm d} #2}}

\newcommand{\pp}[2]{\frac{\partial #1}{\partial #2}}
\newcommand{\rmmat}[1]{{\hbox{\rm{#1}}}}
\newcommand{\rmscr}[1]{{\hbox{\rm{\scriptsize #1}}}}
\bibpunct[,]{(}{)}{;}{a}{}{,}
\begin{document}
%\title{The Stability of Material Accreting onto a Neutron Star}
\title{Thermonuclear Stability of Material Accreting onto a Neutron Star}

\author{Ramesh Narayan\altaffilmark{1} and Jeremy S. Heyl\altaffilmark{1,2,3}}

\altaffiltext{1}{Harvard-Smithsonian Center for Astrophysics, Cambridge, 
MA 02138, U.S.A.}
\altaffiltext{2}{Chandra Fellow}
\altaffiltext{3}{Current Address: Department of Physics and Astronomy,
University of British Columbia, Vancouver, BC V6T 1Z1, Canada}
\email{rnarayan@cfa.harvard.edu; jheyl@cfa.harvard.edu}

\begin{abstract}
We present a global linear stability analysis of nuclear fuel
accumulating on the surface of an accreting neutron star and we
identify the conditions under which thermonuclear bursts are
triggered.  The analysis reproduces all the recognized regimes of
hydrogen and helium bursts, and in addition shows that at high
accretion rates, near the limit of stable burning, there is a regime
of ``delayed mixed bursts'' which is distinct from the more usual
``prompt mixed bursts.''  In delayed mixed bursts, a large fraction of
the fuel is burned stably before the burst is triggered.  Bursts thus
have longer recurrence times, but at the same time have somewhat
smaller fluences.  Therefore, the parameter $\alpha$, which measures
the ratio of the energy released via accretion to that generated
through nuclear reactions in the burst, is up to an order of magnitude
larger than for prompt bursts.  This increase in $\alpha$ near the
threshold of stable burning has been seen in observations.  We explore
a wide range of mass accretion rates, neutron star radii and core
temperatures, and calculate a variety of burst properties.  From a
preliminary comparison with data, we suggest that bursting neutron
stars may have hot cores, with $T_{\rm core}\gtrsim 10^{7.5}$ K,
consistent with interior cooling via the modified URCA or similar
low-efficiency process, rather than $T_{\rm core}\sim10^7$ K, as
expected for the direct URCA process.  There is also an indication
that neutron star radii are somewhat small $\lesssim 10$ km.  Both of
these conclusions need to be confirmed by comparing more careful
calculations with better data.
\end{abstract}

\section{Introduction}
\label{sec:intro}

When gas accretes onto a neutron star, nuclear reactions often occur
in an unstable fashion \citep{1975ApJ...195..735H}, leading to
thermonuclear explosions which are called Type I X-ray bursts.  These
bursts were first observed by \citet{1976ApJ...205L.127G}, and have
been studied intensively for many years \citep[see][ for summaries of
the observations]
{1988MNRAS.233..437V,1993SSRv...62..223L,Stro98,cor03}.  Recently,
there has been renewed excitement in the field, following the
discovery of high frequency oscillations in burst light curves
\citep[e.g.,][]{Stro96,Stro01a,vanstr01,mun01}.

The physics of Type I bursts has been widely studied, and the broad
features of the phenomenon are understood theoretically
\citep{1976Natur.263..101W,
1977Natur.270..310J,1978ApJ...224..210T,1983ApJ...264..282P,
1981ApJ...247..267F,1987ApJ...323L..55F,1996ApJ...459..271T,Bild98}.
Models show that there are three kinds of bursts, depending on whether
hydrogen or helium burning dominates.  At high mass accretion rates
$\dot M$, hydrogen is only partially burned before a burst is
triggered by unstable helium burning.  The result is a mixed burst in
which both hydrogen and helium burn explosively.  At somewhat lower
values of $\dot M$, all the hydrogen is consumed before the helium
instability is triggered.  This leads to a pure helium burst.  For yet
lower $\dot M$, hydrogen itself burns unstably, giving a hydrogen
burst.  The ranges of $\dot M$ corresponding to the different regimes
have been worked out approximately.

Two very different approaches have been pursued for theoretically
modeling the burst phenomenon.  In one approach, one simulates the
physics of the accreting gas with a fully time-dependent code that
includes a large network of nuclear reactions and sophisticated
thermodynamics
\citep[e.g.,][]{1978ApJ...225L.123J,1979ApJ...233..327T,1980ApJ...238..287J,
1982ApJ...258..761T,1987ApJ...319..902F,1993ApJ...413..324T}.  Such
studies are essential for understanding the details of the
thermonuclear explosion; indeed, the one-dimensional simulations of
the past have now been generalized to two-dimensional and even
three-dimensional simulations \citep[the FLASH effort at
Chicago,][]{zin01} which follow the physics of bursts in exquisite
detail.  Such simulations are, however, not very convenient for
parameter surveys or for detailed comparisons of theoretical
predictions with observational burst statistics.

An alternate approach, where one focuses on the thermonuclear
instability that triggers the burst rather than on the burst itself,
has been popular
\citep[e.g.,][]{1975ApJ...195..735H,1980A&A....84..123E,
1981ApJ...247..267F,1982ApJ...258..761T,1983ApJ...264..282P,
1987ApJ...323L..55F,1987ApJ...315..198F,2000ApJ...544..453C}.  Here
one treats the accumulating layer of gas on the surface of the neutron
star as a quasi-equilibrium system whose properties vary slowly with
time.  One first solves for the equilibrium structure and then
analyses the stability of the gas layer by considering the effect of
small perturbations on the underlying equilibrium.  If the
perturbations grow with time, one says that the system is unstable,
presumably resulting in a Type 1 X-ray burst.

The second approach, though not as general as the first, allows one to
explore a large range of parameter space and to study how the burst
phenomenon is affected by variations in control parameters such as the
mass accretion rate, the surface gravity of the star, etc.  However,
the analyses that have been published so far in the literature employ
fairly simple-minded criteria to decide exactly when an accreted layer
becomes unstable and are, therefore, not very accurate or complete.
The motivation of the present paper is to develop a more rigorous
stability analysis for accreting neutron stars.  Our hope is (i) to
verify the results of the earlier methods, (ii) to explore additional
phenomena that may have been missed previously, and (iii) to put the
theory on a rigorous footing to enable quantitative comparisons with
observations.  We reported some early results of this work in
\citet{Nara01typeibh}.

We begin the paper in \S\ref{sec:model} with a description of our
model of the accreted layer and the method we employ to calculate
equilibria and to study their stability.  We follow this in
\S\ref{sec:equil} with an exploration of sequences of equilibria, and
in \S\ref{sec:stability} with a discussion of the stability properties
of the equilibria.  Through this analysis, we reproduce the three
previously recognized regimes of burst activity, namely mixed bursts,
helium bursts, and hydrogen bursts.  However, we find that mixed
bursts themselves come in two kinds: prompt mixed bursts and delayed
mixed bursts.  The latter category has not been recognized previously.
In \S\ref{sec:results} we present results for neutron stars of various
radii and core temperatures, and explore a wide range of mass
accretion rates.  We calculate a number of burst observables such as
the recurrence time, the burst duration, and the dimensionless
parameter $\alpha$ (Eq.~\ref{alpha}) which is widely used in the burst
literature.  In \S\ref{sec:discussion}, we compare our theoretical
formalism to methods previously published in the literature.  We also
compare our predictions with selected observations.  From the latter,
we obtain preliminary results on the likely core temperatures and
radii of bursting neutron stars.  We conclude in \S\ref{sec:summary}
with a summary.

\section{The Model}
\label{sec:model}

\subsection{Governing Equations}
\label{sec:PDE}

We assume that gas accretes on the surface of a compact spherical star
of mass $M$ and radius $R$ at a rate $\dot \Sigma$ (mass per unit area
per unit time).  We consider all quantities to be functions of
$\Sigma$, the column density (mass per unit area) measured from the
top of the accreted layer.  We use partial derivatives
$\partial/\partial t$ and $\partial/\partial\Sigma$ to signify
Eulerian time and ``spatial'' derivatives at a fixed $\Sigma$, and
$d/dt$ to represent the ``Lagrangian'' time derivative following a
parcel of accreted gas \citep[see][]{1987ApJ...323L..55F}:
\begin{equation}
{d\over dt} \equiv {\partial \over\partial t} + \dot\Sigma
{\partial \over \partial\Sigma}. \label{eq:ddt}
\end{equation}
We consider only hydrogen and helium burning, and so we describe the
composition of the gas in terms of the hydrogen mass-fraction $X$, the
helium mass fraction $Y$, and the heavy element fraction $Z=1-X-Y$.
These quantities start off with values $X_\rmscr{out}$,
$Y_\rmscr{out}$, $Z_\rmscr{out}$ at $\Sigma=0$, corresponding to the
composition of the gas initially falling on the neutron star, and
evolve with increasing depth as a result of nuclear burning.  Because
$H$-burning is mostly done via the CNO-cycle, it is necessary to know
what fraction of $Z$ is in CNO elements (see Eq.~\ref{eq:epsilonCNO}).
For this, we assume that about 80\% of the initial $Z_\rmscr{out}$ is
in CNO, as appropriate for solar composition (Allen 2000), and that
all the $Z$ produced via helium burning is entirely in CNO; thus, we
take
\begin{equation}
Z_\rmscr{CNO}=0.8Z_\rmscr{out} +(Z-Z_\rmscr{out}). \label{eq:ZCNO}
\end{equation}

The time evolution of the accreting gas is described by a set of five
coupled partial differential equations:
\begin{eqnarray}
\pp{P}{\Sigma} &=& g, \label{eq:fulleqs1} \\
\pp{T}{\Sigma} &=&\frac{3\kappa F}{16 \sigma T^3}, \label{eq:fulleqs2} \\
\pp{F}{\Sigma} &=& - T \dd{s}{t} - (\epsilon_\rmscr{H}+\epsilon_\rmscr{He}), \label{eq:fulleqs3} \\
\dd{X}{t}&=&-{\epsilon_\rmscr{H}\over E_\rmscr{H}^*}, \label{eq:fulleqs4} \\
\dd{Y}{t} &=& \frac{\epsilon_\rmscr{H}}{E_\rmscr{H}^*} 
	- \frac{\epsilon_\rmscr{He}}{E_\rmscr{He}^*}. \label{eq:fulleqs5}
\end{eqnarray}
Table~\ref{tab:sym} gives the definitions of the various symbols.  We
assume that the accreted layer is thin relative to the stellar radius,
and so we take the gravitational acceleration $g$ to be independent
of $\Sigma$:
\begin{equation}
g = (1+z){GM\over R^2}, \qquad 1+z = \left(1-{2GM\over c^2R}
\right)^{-1/2}, 
\label{eq:grav}
\end{equation}
where $z$ is the gravitational redshift.  Note that $g$ and all other
quantities in equations (\ref{eq:fulleqs1}--\ref{eq:fulleqs5}) are
measured in the local frame of the gas.  In this spirit, $\dot \Sigma$
is the baryonic mass added per unit surface area per unit local time.

\begin{deluxetable}{l|l}
\tablecaption{Definition of symbols: \label{tab:sym}
}
\tablehead{ \colhead{Symbol} & \colhead{Meaning} }
\startdata
$R$, $M$, $R_S$ & stellar radius, stellar mass, 
Schwarzschild radius: $2 G M/c^2$ \\
$g$ $z$ & gravitational acceleration, redshift at the surface \\
${\dot \Sigma}$ & mass accretion rate per unit area \\
$L_{\rm acc}$, $L_{\rm Edd}$, $l_{\rm acc}$ & accretion luminosity, 
Eddington luminosity, $l_{\rm acc} = L_{\rm acc}/L_{\rm Edd}$ \\
$\Sigma$ & surface mass density measured from the surface; 
independent variable in eqs. \\ 
$\Sigma_{\rm layer}$ & $\Sigma$ of the accreted layer \\
$\Sigma_{\rm layer,crit}$ & critical $\Sigma_{\rm layer}$ at which a burst 
is triggered \\
$\Sigma_{\rm diff}$, $\Sigma_{\rm max}$ & Maximum integration depth for
the equilibrium, and for perturbations \\
$t$, $P$, $\rho$, $T$, $s$ & time, pressure, mass density, temperature,
entropy per unit mass \\ 
$F$ & outbound energy flux \\
$X$, $Y$, $Z$ & mass fractions of hydrogen, helium, metals \\
$Z_\rmscr{CNO}$ & mass fraction of CNO nuclei \\
$\rho_0$, $T_0$; $\rho_1$, $T_1$, & values in equilibrium; perturbations \\
$T_{\rm out}$, $X_\rmscr{out}$, etc. & values at the surface \\
$F_{\rm nuc}$ & expected surface flux if entire fuel is steadily burned \\
$F_\rmscr{out}$, $f_{\rm nuc}$ & escaping flux due to nuclear 
reactions and compression, $f_{\rm out} = F_{\rm out}/F_{\rm nuc}$  \\
$F_\rmscr{acc}$ & persistent flux from the surface due to accretion:
${\dot \Sigma} c^2 z/(1+z)$\\
$T_{\rm layer}$, $X_{\rm layer}$, etc. & values at the bottom of 
the accreted layer \\
$T_{\rm max}$, $T_{\rm core}$ & maximum $T$ in layer, core temperature \\
%$A_n$, $Z_n$ & mean atomic mass, mean atomic number, of nuclei \\
$\tau_{\rm diff}$ & thermal diffusion time \\
$\kappa$, $K$ & opacity, conductivity \\
$\epsilon_\rmscr{H,He}$ & energy-generation rate per unit mass
for H, He burning \\
$E_\rmscr{H,He}^*$ & total nuclear energy released per unit mass
of H, He burned \\
$\gamma$ & growth rate of mode, $\gamma=\Re(\gamma) + i \Im(\gamma)$ \\
$\gamma_\rmscr{acc}$, $t_{\rm acc}$ & $\gamma_{\rm acc} = {\dot \Sigma}/
\Sigma_\rmscr{layer}$, accretion rate; $t_{\rm acc} = \Sigma_{\rm layer}
/\dot\Sigma$, accretion time \\
$t_{\rm rec}$ & burst recurrence time: $(1+z)\Sigma_{\rm layer,crit}
/\dot\Sigma$ \\
$E_{\rm H}$, $E_{\rm He}$ & fluence in burst from burning hydrogen, helium \\
$t_{\rm H+He}$, $t_{\rm He}$ & effective burst duration: $(E_{\rm H}+
E_{\rm He})/L_{\rm Edd}$, $E_{\rm He}/L_{\rm Edd}$ \\
$\alpha$ & ratio of accretion energy to nuclear energy: 
$(t_{\rm rec}/t_{\rm H+He}) l_{\rm acc}$ \\
\enddata
\end{deluxetable}

Equation (\ref{eq:fulleqs1}) describes the condition of hydrostatic
equilibrium.  By making use of this equation rather than the full
momentum equation, we filter out sound waves and focus on variations
that occur on a much longer time scale than the sound-crossing time.
Equation (\ref{eq:fulleqs2}) describes energy transfer by radiation
and conduction.  Equation (\ref{eq:fulleqs3}) is the energy
conservation equation, and equations (\ref{eq:fulleqs4}) and
(\ref{eq:fulleqs5}) describe the evolution of the hydrogen and helium
mass-fractions as a result of nuclear burning.  Equations
(\ref{eq:fulleqs1})--(\ref{eq:fulleqs5}) require expressions for the
pressure $P$, the opacity $\kappa$, the entropy $s$, and the
energy-generation rates for hydrogen and helium-burning,
$\epsilon_\rmscr{H}$, $\epsilon_{\rm He}$.  These are discussed in
\S\ref{sec:quant}.

The calculations proceed as follows.  We start with a bare neutron
star and follow the properties of the accreted layer as gas piles up.
For a given column density $\Sigma_\rmscr{layer}$ of the accreted
layer, we calculate the following.  First, we solve for the
quasi-equilibrium state of the system.  To do this, we set the
Eulerian time derivative $\partial/\partial t$ to zero in equations
(\ref{eq:fulleqs1}--\ref{eq:fulleqs5}) and consider the following set of
ordinary differential equations \citep[see][]{1987ApJ...323L..55F}
\begin{eqnarray}
\dd{P}{\Sigma} &=& g, \label{eq:steadyeqs1} \\
\dd{T}{\Sigma} &=&\frac{3\kappa F}{16 \sigma T^3}, \label{eq:steadyeqs2} \\
\dd{F}{\Sigma} &=& - \dot\Sigma T \dd{s}{\Sigma} -
(\epsilon_\rmscr{H}+\epsilon_\rmscr{He}), \label{eq:steadyeqs3} \\
\dot\Sigma \dd{X}{\Sigma}&=&-{\epsilon_\rmscr{H}\over E_\rmscr{H}^*},
\label{eq:steadyeqs4} \\
\dot\Sigma \dd{Y}{\Sigma} &=& \frac{\epsilon_\rmscr{H}}{E_\rmscr{H}^*}
- \frac{\epsilon_\rmscr{He}}{E_\rmscr{He}^*}. \label{eq:steadyeqs5}
\end{eqnarray}
We solve these equations with boundary conditions (described in
\S\ref{sec:boundary-conditions}) to obtain the run of density
$\rho_0(\Sigma)$, temperature $T_0(\Sigma)$, etc., in quasi-steady
state.  Note that, because $\Sigma_{\rm layer}$ increases steadily
with time, the layer at any given time is not strictly in equilibrium,
and the above steady state equations are not precisely valid.
However, since $\Sigma_{\rm layer}$ increases only slowly with time
(on the accretion time scale defined in eq. \ref{tacc} below), and
since many of the physical processes in the layer have shorter
characteristic time scales, this is a reasonable approximation.

Having calculated the steady state equilibrium solution, we carry out
a linear perturbation analysis.  This is the principal contribution of
our work, and represents a significant advance over previous studies.
For the perturbation analysis, we assume that the various physical
quantities are functions of $\Sigma$ and $t$ in the form
\begin{eqnarray}
\rho(\Sigma,t) &=& \rho_0(\Sigma) + \exp (\gamma t) \rho_1(\Sigma), \\
\label{rhopert}
T(\Sigma,t) &=& T_0(\Sigma) + \exp (\gamma t) T_1(\Sigma), ~{\rm etc.},
\label{Tpert}
\end{eqnarray}
where $\rho_0$, $T_0$ represent the solutions obtained from solving
the steady state equations described above, and $\rho_1\ll\rho_0$,
$T_1 \ll\ T_0$, are small linear perturbations.  The frequency
$\gamma$ represents the growth rate of the perturbations.  We
substitute the perturbed solution into the original time-dependent
equations (\ref{eq:fulleqs1}--\ref{eq:fulleqs5}) and linearize in the
usual way to obtain a set of ordinary differential equations (in
$\Sigma$) for the first-order quantities $\rho_1$, $T_1$, etc.  These
equations are written down in Appendix A and some of their properties
are discussed there.  We solve the linearized perturbation equations
with appropriate boundary conditions and thereby obtain $\gamma$,
which plays the role of an eigenvalue.

In general, there are many solutions for $\gamma$ for a given steady
state solution; some values of $\gamma$ are real and some are complex.
If any solution for $\gamma$ has a real part sufficiently large
compared to the accretion rate --- see equation~(\ref{eq:mingrowth})
below for a precise statement of what the criterion is --- then we say
that the layer is unstable.  In this case, we identify $\Re(\gamma)$
with the growth rate of the instability in the system.  If no solution
for $\gamma$ satisfies equation (\ref{eq:mingrowth}), then we consider
the system to be stable.  Because $\gamma$ is in general complex, all
the quantities in the perturbation equations are complex and must be
handled with complex arithmetic (in contrast to the steady state
equations which involve purely real quantities).

We carry out the above two stages of calculations, namely steady state
and linear perturbation analysis, for each $\Sigma_\rmscr{layer}$ as
matter accumulates on the star.  If no instability is found for any
choice of $\Sigma_\rmscr{layer}$ up to a very large value, typically
$10^{13}-10^{14}$~g~cm$^{-3}$ --- see \citet{1978ApJ...224..210T} and
\citet{1998ApJ...496..915B} for a discussion of carbon flashes which
occur at yet larger column densities --- then we say that the system
is stable to bursts.  If, for some value of $\Sigma_\rmscr{layer}$, we
do obtain an instability, then we say that the system will undergo a
burst when it accumulates this much gas on its surface.

\subsection{Boundary Conditions}
\label{sec:boundary-conditions}

The solution of the five differential equations
(\ref{eq:steadyeqs1}--\ref{eq:steadyeqs5}) for the steady state
requires five boundary conditions.  Four are applied on the outside,
at the photosphere (where the optical depth is taken to be 2/3), and
one is applied below the accreted layer, deep inside the star.

The outer boundary conditions are very similar to the ones employed in
\citet{Nara01typeibh}.  The surface values of $X$ and $Y$ are set
equal to the corresponding values of the accreting gas, namely
$X=X_\rmscr{out}$, $Y=Y_\rmscr{out}$.  We have used a solar
composition, $X_\rmscr{out}=0.7$, $Y_\rmscr{out}=0.28$ (Allen 2000,
Table 3.1), in all the calculations reported in this paper.  Next, a
particular value is assumed for the escaping flux $F_\rmscr{out}$ from
the accreted layer.  This flux is the result of nuclear burning and
compression, and its precise value is determined only after applying
the inner boundary condition, as explained below.  The surface
temperature $T_\rmscr{out}$ is then obtained by the condition
\begin{equation}
\sigma T_\rmscr{out}^4 = F_\rmscr{out} + F_\rmscr{acc},
\qquad F_\rmscr{acc}= \dot \Sigma c^2 z(1+z), \label{eq:Tout}
\end{equation}
where $F_\rmscr{acc}$ is the gravitational energy released by the
accreting gas, and we have included the appropriate gravitational
redshift factor so that all quantities are calculated in the local
frame.  The above relation for $T_\rmscr{out}$ is approximate, but we
have confirmed that the results are very insensitive to the precise
choice of $T_\rmscr{out}$.  The fourth boundary condition is obtained
from the radiative transfer equation.  Given an assumed value for
$F_\rmscr{out}$ and the condition $P=0$ at $\Sigma=0$, this equation
directly gives the density profile $\rho(\Sigma)$.

As described above, the solution at the surface is completely
specified once a value of $F_\rmscr{out}$ is assumed.  The unknown value
of $F_\rmscr{out}$ is determined by requiring the solution to satisfy an
inner boundary condition.  For this, we assume that the accreting star
has a specified core temperature $T_\rmscr{core}$ and we require the
solution of the steady state equations to match this temperature.  The
key issue is where exactly to do the matching.  Our approach is as as
follows.  Associated with a given accreted column of depth
$\Sigma_\rmscr{layer}$, there is a characteristic accretion time
\begin{equation}
t_\rmscr{acc} = {\Sigma_\rmscr{layer} \over \dot\Sigma}. \label{tacc}
\end{equation}
We integrate the steady state equations from the surface down to the
bottom of the accreted layer at $\Sigma_\rmscr{layer}$, and then we
integrate further into the stellar substrate to a depth $\Sigma_{\rm
diff}$ such that the diffusion time from $\Sigma_\rmscr{layer}$ to
$\Sigma_\rmscr{diff}$ is equal to twice $t_\rmscr{acc}$ (the factor of 2
is arbitrary and was selected after some numerical experiments):
\begin{equation}
\tau_\rmscr{diff}(\Sigma_\rmscr{diff}) = 2t_\rmscr{acc}. \label{ibc1}
\end{equation} 
For applying this condition, we need an estimate of the thermal
diffusion time $\tau_\rmscr{diff}(\Sigma)$ for any choice of $\Sigma$
inside the star.  We obtain this by integrating a separate
differential equation:
\begin{equation}
\dd{\tau_\rmscr{diff}}{\Sigma}={9\kappa k_BT(\Sigma-\Sigma_\rmscr{layer})
\over 128 \sigma T^4 \mu m_u}. \label{taueq}
\end{equation}
The form of this equation and the numerical coefficient (which is
based on a simple toy model) are obtained by treating the energy
equation (\ref{eq:steadyeqs3}) as a diffusion problem.  The precise
details are unimportant for the final results.

For the calculations presented in this paper, we pick ``reasonable''
values for $T_\rmscr{core}$, trying a range that is likely to bracket
the true value.  In \S\ref{sec:constr-neutr-cool} we will estimate
the core temperature as a function of accretion rate for two kinds of
neutrino cooling in the core.  If we include both the inward directed
flux at the bottom of the layer that we calculate and the energy from
nuclear reactions in the deep crust \citep{1998ApJ...504L..95B}, we
find that the latter contribution dominates.  This yields a simple
relationship between the accretion luminosity and the core
temperature.

The above treatment of the inner boundary condition is superior to the
method used in Narayan \& Heyl (2002, and most other previous studies,
see \S6.1), where the temperature at the bottom of the accreted layer
was set equal to $T_\rmscr{core}$.  That is, in that calculation, the
substrate was assumed to be isothermal immediately below the accreted
layer.  By integrating down to a couple of diffusion lengths into the
substrate, we believe our present approach is physically better
motivated.  However, even with this approach, we are effectively
assuming that the core below $\Sigma_\rmscr{diff}$ is perfectly
isothermal --- hence the use of the term ``core temperature'' ---
which is again an approximation.  For a system that has been accreting
and bursting for a long time, there is expected to be a small
time-averaged flux flowing into the core even inside of
$\Sigma=\Sigma_\rmscr{diff}$.  In addition, there is a larger flux
from deep crustal reactions (Brown et al. 1998).  Both of these fluxes
will induce a temperature gradient, so that the temperature in the
crust, which is relevant for bursts, will not be equal to the
temperature $T_\rmscr{core}$ deep inside the neutron star (set for
instance by neutrino cooling, see Brown 2000 for a detailed analysis).
This effect may not be very serious, since the matter inside
$\Sigma_\rmscr{diff}$ is highly degenerate and very conductive.
Nevertheless, we mention the point here because the effect of the
approximation is presently not fully understood.  We believe that the
results we present in this paper for $T_\rmscr{core}\sim10^8$ K will
be hardly affected because the burning layer itself has a temperature
of this order.  However, for very cold cores, e.g.,
$T_\rmscr{core}\sim10^7$ K, the approximation may have a more serious
effect.

The above discussion pertains to the equilibrium solution.  The linear
perturbation equations have similar boundary conditions.  At the
surface, the first-order perturbations $X_1$ and $Y_1$ vanish, and we
assume an arbitrary value for the flux perturbation $F_1$; the latter
choice serves as the overall normalization of the perturbed
eigenfunction.  From the perturbed flux, the corresponding temperature
perturbation $T_1$ is readily obtained via equation (\ref{eq:Tout}),
and finally the density perturbation $\rho_1$ is calculated from the
radiative transfer equation.

At the bottom, we set the temperature perturbation $T_1(\Sigma_{\rm
max})$ at a prescribed depth $\Sigma_\rmscr{max}$ equal to zero.  In
analogy with what we did for the steady state solution, we determine
$\Sigma_\rmscr{max}$ by the condition that the diffusion time down to
this depth should be equal to twice the mode time scale 
$t_\rmscr{mode}$:
\begin{equation}
\tau_\rmscr{diff}(\Sigma_\rmscr{max}) = 2t_\rmscr{mode} = {2 \over
|\gamma|}, \label{ibc2}
\end{equation}
where $|\gamma|$ is the modulus of the complex eigenvalue $\gamma$.
We only consider modes that grow faster than the accretion time, see
below; therefore, $\Sigma_\rmscr{max}$ is always smaller than
$\Sigma_\rmscr{diff}$.  The condition $T_1(\Sigma_\rmscr{max})=0$
provides the final boundary condition that enables us to solve for the
eigenmode and the eigenvalue $\gamma$.

As already mentioned, there are many solutions for $\gamma$.  We
concentrate on modes that grow fast enough to be interesting,
specifically modes that satisfy
\begin{equation}
\Re(\gamma) \ge g_{\rm mode}\gamma_{\rm acc}, \qquad g_{\rm mode}=3,
\qquad \gamma_{\rm acc} \equiv {1\over t_\rmscr{acc}} = {\dot \Sigma
\over \Sigma_{\rm layer}}. \label{eq:mingrowth}
\label{eq:gammacrit}
\end{equation}
The factor 3 is arbitrary, but reasonable.  For an instability to have
any noticeable effect on the system, it needs to grow in a time
shorter than the lifetime $t_\rmscr{acc}$ of the system.  Also, our
approach of treating the accreted layer as a quasi-steady system, and
analysing the perturbations as if they occur in a time-steady system,
is valid only if the time scale of the perturbations is sufficiently
small compared to $t_\rmscr{acc}$.  For both reasons, the choice,
$g_{\rm mode}=3$, in equation (\ref{eq:mingrowth}) seems appropriate.

\subsection{Auxiliary Prescriptions}
\label{sec:quant}

The solution of the set of differential equations described in 
\S\ref{sec:PDE}
requires knowledge of the thermodynamic and other physical properties
of the gas.  We describe here the particular prescriptions we have
used for the calculations presented in this paper.

\subsubsection{Equation of State}
\label{sec:EOS}

The pressure is assumed to be supplied by photons, nuclei and
electrons.  For the photons we use the blackbody formula and for the
nuclei we assume an ideal non-degenerate gas.  In the case of the
electrons, we write the pressure as the quadrature sum of two
terms, one equal to the pressure of an ideal non-degenerate gas and
the other equal to the pressure of a zero-temperature degenerate
electron gas.  Thus we take \citep[see][]{1983ApJ...267..315P}
\begin{equation}
P = {1\over3}aT^4 + P_{\rm nuc,nd} +\left[ P_{\rm e,nd}^2
+ P_{\rm e,d}^2 \right]^{1/2}, \label{pressure}
\end{equation}
\begin{equation}
P_{\rm nuc,nd} = {\rho k_BT\over \mu_{\rm nuc} m_u}, \qquad
P_{\rm e,nd} = {\rho k_BT\over \mu_{\rm e} m_u},
\label{pressure2}
\end{equation}
where the molecular weights $\mu_{\rm nuc}$ and $\mu_{\rm e}$ are
determined in the standard way \citep[e.g.][]{Clay83}.  In determining
$\mu_{\rm nuc}$, $\mu_{\rm e}$, we assume that a fraction $0.2Z_{\rm
out}$ of the heavy elements in the accreted layer consists of
$^{56}$Fe and the rest (what we have called $Z_{\rm CNO}$, see eq. 2)
consists of $^{14}$N (as a surrogate for CNO elements).  The latter
choice is an approximation.  The initial CNO elements in the accreted
gas consist mostly of $^{12}$C and $^{16}$O.  During H-burning via the
CNO cycle, the composition is mostly $^{14}$O and $^{15}$O, while the
composition reverts back to $^{12}$C and $^{16}$O after the CNO cycle
is done.  Considering the other simplifications we have employed,
approximating the composition as $^{14}$N for the purposes of
calculating the molecular weights seems reasonable.  We take the
stellar substrate below the accreted layer to be made of pure
$^{56}$Fe.

For the zero-temperature degenerate electron pressure $P_\rmscr{deg}$,
we use an exact expression that is valid for all Fermi energies
\citep{Shap83}.  We do not include Coulomb corrections on the
pressure.  Separating the electron pressure into a non-degenerate part
and a zero-temperature degenerate part and taking the quadrature sum
is, again, an approximation.  The separation works well in various
asymptotic limits and is reasonably accurate even in the transition
regime between non-degeneracy and degeneracy
\citep[]{1983ApJ...267..315P}.  We believe that the approximation is
adequate for our purposes since the transition zone is usually fairly
narrow in $\Sigma$.

The entropy term in equation (\ref{eq:fulleqs3}) is important since it
is the origin of the compressional flux.  We write the entropy as the
sum of contributions from each nuclear species and the electrons.  The
entropy per unit mass for species $i$ takes the form
\begin{equation}
s_i = N_ik \left[ -\ln(\rho N_i) +{3\over2}\ln T +{5\over2}
+{3\over2}\ln\left({2\pi m_ik\over h^2}\right) \right],
\label{entropy}
\end{equation}
where $N_i$ is the number of particles of the particular species per
unit mass of the gas, $m_i$ is the mass of each particle, and $h$ is
Planck's constant.  This expression corresponds to an ideal gas.  In
the case of the electrons, we use the above entropy so long as the
quantity is positive, and replace it with zero when the expression
becomes negative.  This allows us to handle both the non-degenerate
and degenerate limits adequately.

\subsubsection{Opacity}
\label{sec:opacity}

We include radiative and conductive energy transfer, and model the
opacity $\kappa$ as
\begin{equation}
{1\over\kappa} = {1\over\kappa_\rmscr{rad}} +
{1\over\kappa_\rmscr{cond}}. \label{opacity}
\end{equation}
For the radiative opacity $\kappa_\rmscr{rad}$, we use the formulae
given in Appendix A of \citet{1999ApJ...524.1014S}, using the analytic
formulae of \citet{1993ApJS...84..101A} to calculate the electron
chemical potential.

To calculate the conductive opacity we have used the software of
\citet{1999A&A...351..787P}.  Up to a density of $10^9$~g~cm$^{-3}$,
we include the effects of impurities among the nuclei.  Above this
density we assume that the material is pure, and we interpolate
between the two regimes.  The density of the fuel layer is nearly
always less than $10^9$~g~cm$^{-3}$, except for some rare helium
bursts.

\subsubsection{Nuclear Energy Generation Rates}
\label{sec:nucl-energy-gener}

For $\epsilon_\rmscr{H}$, we include the pp chain and the CNO cycle.
Because the temperature of the burning material is on the order of
$10^8$~K, the CNO cycle usually dominates.  We include fast-CNO
burning, saturated CNO burning and electron capture reactions, as
described in \citet{1984ApJ...287..969M} and
\citet{1998ApJ...506..842B}, and write the energy-generation rate as
\begin{equation}
\epsilon_\rmscr{H,CNO} = 4 E_\rmscr{H}^* r_\rmscr{CNO} {Z_\rmscr{CNO}
\over 14}, \label{eq:epsilonCNO}
\end{equation}
where $r_\rmscr{CNO}$ is the rate of reactions per CNO nucleus, and we
have assumed that in equilibrium the majority of the CNO nuclei are
$^{14}$N \citep{Clay83}.  For the reaction rate, we assume
\begin{equation}
r_\rmscr{CNO} = \left ( \frac{1}{\tau_{13}} +
 \frac{1}{862.0~\rmmat{s}} \right )^{-1} \left [
 \left(\frac{1}{\tau_{13}}\right) \left(\frac{1}{\tau_{13} + \tau_{14}
 + 278.2~\rmmat{s}}\right) + \left(\frac{1}{862.0~\rmmat{s}}\right)
 \left(\frac{1}{\tau_{14} + 1038.0~\rmmat{s}}\right) \right ].
\label{eq:CNO}
\end{equation}
Here, $\tau_{13}$ is the lifetime of $^{13}$N against the reaction
$^{13}\rmmat{N}(p,\gamma)^{14}\rmmat{O}$ \citep{1984ApJ...287..969M},
\begin{eqnarray}
\tau_{13} &=& \left(X\rho/{\rm g\,cm^{-3}}\right )^{-1} \biggr [ 3.35
\times 10^{7}\ T_9^{-2/3} \exp \left ( -15.202\ T_9^{-1/3} - 0.8702\
T_9^2 \right ) \nonumber \\ & & ~~~~~ \times \left ( 1 + 0.027\
T_9^{1/3} + 0.9\ T_9^{2/3} + 0.173\ T_9 + 4.61\ T_9^{4/3} + 2.26\
T_9^{5/3} \right ) \\ & & ~~~~~ + 3.03 \times 10^{5}\ T_9^{-3/2} \exp
\left ( -6.348\ T_9^{-1} \right ) \biggr ]^{-1} \rmmat{s}, \nonumber
\label{tau13}
\end{eqnarray}
where $T_9=T/(10^9 \rmmat{K})$, and $\tau_{14}$ is the lifetime of
$^{14}$N against the reaction $^{14}\rmmat{N}(p,\gamma)^{15}\rmmat{O}$
\citep[e.g.][]{1983ApJ...264..282P},
\begin{equation}
\tau_{14} = 3.1\times 10^{10} \left(X\rho/{\rm g cm^{-3}}\right)^{-1}
T_6^{2/3} \exp \left ( 152.313\ T_6^{-1/3} \right ) ~{\rm s},
\label{tau14}
\end{equation}
where $T_6=T/(10^6 \rmmat{K})$.  The time scale $862.0$~s refers to
the beta-decay timescale of $^{13}$N, $278.2$~s is the sum of the
beta-decay timescales of $^{14}$O and $^{15}$O, and $1038.0$~s is the
sum of the beta-decay timescales of $^{13}$N and $^{15}$O.  In
deriving the above rates we have assumed that the various species have
reached their equilibrium abundances.

\begin{figure}
\plotone{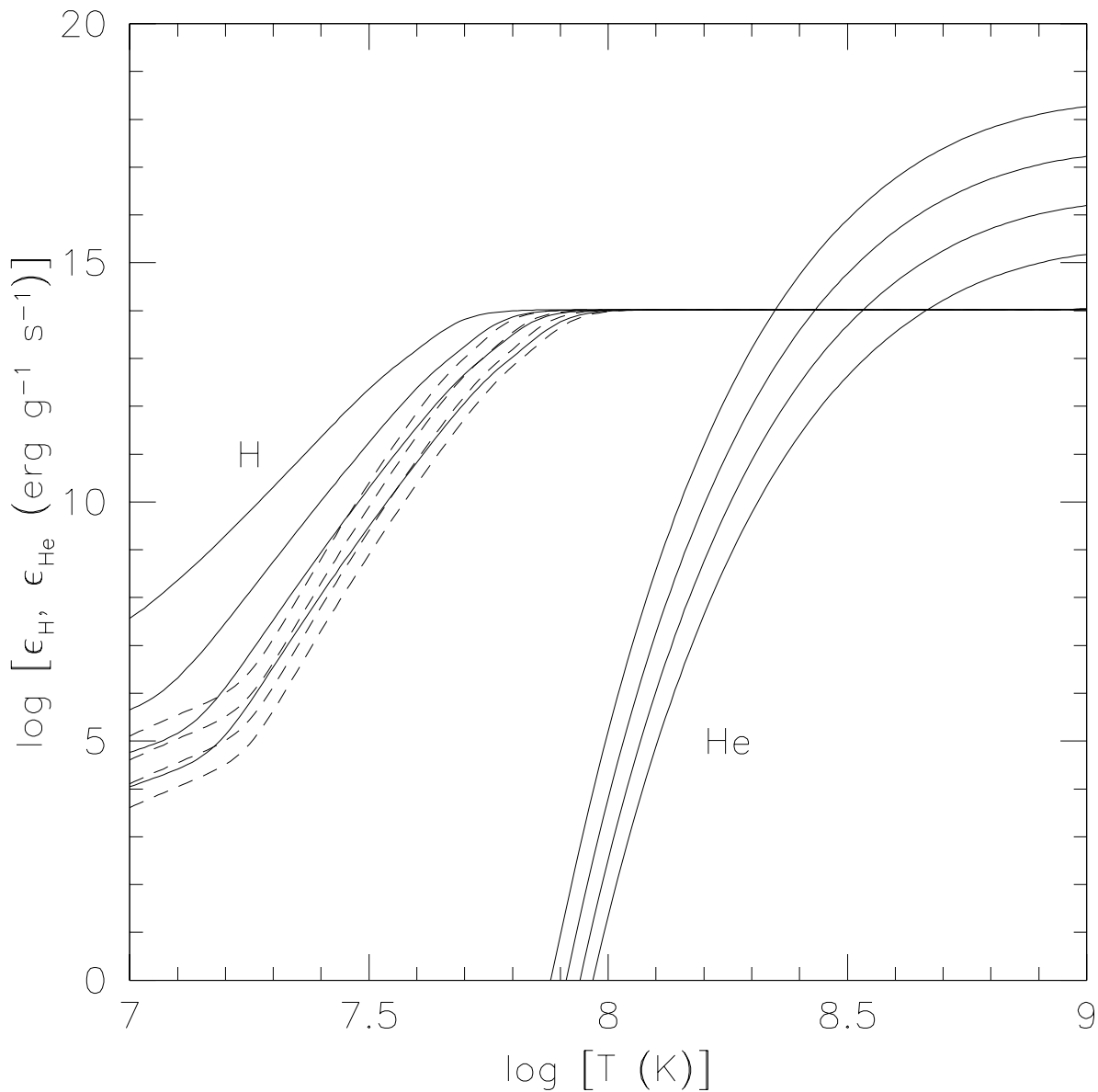}
\caption{Nuclear reaction rates for hydrogen and helium, plotted as a
function of temperature.  From bottom to top, the four curves
correspond to densities of $10^5, ~10^{5.5}, ~10^6, ~10^{6.5} ~{\rm
g\,cm^{-3}}$.  The hydrogen-burning curve is dominated by the pp chain
at low temperatures, the CNO cycle at intermediate temperatures, and
saturates above about $10^{7.8}-10^{7.9}$ K.  The dashed lines trace
the results if one ignores screening.  The helium rates are taken from
Fushiki \& Lamb (1987b).}
\label{fig:nuclear}
\end{figure}

Figure~\ref{fig:nuclear} shows the variation of $\epsilon_\rmscr{H}$
and $\epsilon_\rmscr{He}$ with temperature for some typical densities.
For low and moderate temperatures, the hydrogen-burning rate is a
steeply increasing function of temperature, and it is this steep
dependence that drives a thermonuclear instability.  For a temperature
greater than about $10^{7.8}-10^{7.9}$ K, however, hydrogen-burning
switches rather abruptly to the saturated burning regime.  Beyond this
temperature, hydrogen-burning is stable.  As we discuss in
\S\ref{sec:equil}, this change has a noticeable effect on the
sequence of equilibria.  We have corrected the proton capture rates
for screening using the formulae of \citet{1973ApJ...181..439D} for
the non-resonant reactions and the formulae of
\citet{2003ApJ...586.1436I} for the resonant reaction
($^{13}$N$(p,\gamma)^{14}$O).  Screening increases the reaction rates at
temperatures where the CNO cycle is not saturated.

For $\epsilon_\rmscr{He}$, we use the fitting formula given in
equations (4.7), (4.8a) and (4.8b) of \citet{1987ApJ...317..368F}
which include screening.  We introduce a smooth transition between the
various regimes defined by these authors in order to more faithfully
reproduce the numerical results they have depicted in their Fig.~3.
Our fitting results are shown in Fig.~\ref{fig:nuclear}.

\section{Equilibrium Solutions}
\label{sec:equil}

\citet{1983ApJ...264..282P} has presented a very helpful analysis of
the stability of nuclear burning on the surface of a compact star.
His model involves numerous simplifications: he uses a one-zone
approximation, he considers only helium burning, and he assumes an
inner boundary condition on the flux rather than on the temperature.
Nevertheless, many of the insights he has obtained via his simple
analysis carry over to our more detailed work.  In particular,
following his work, we have found it very helpful to consider
equilibria in the $F_\rmscr{out}$-$\Sigma_\rmscr{layer}$ plane.
Appendix A discusses why the insights from Paczy\'nski's analysis
apply to our more complicated model.

\begin{figure}
\plotone{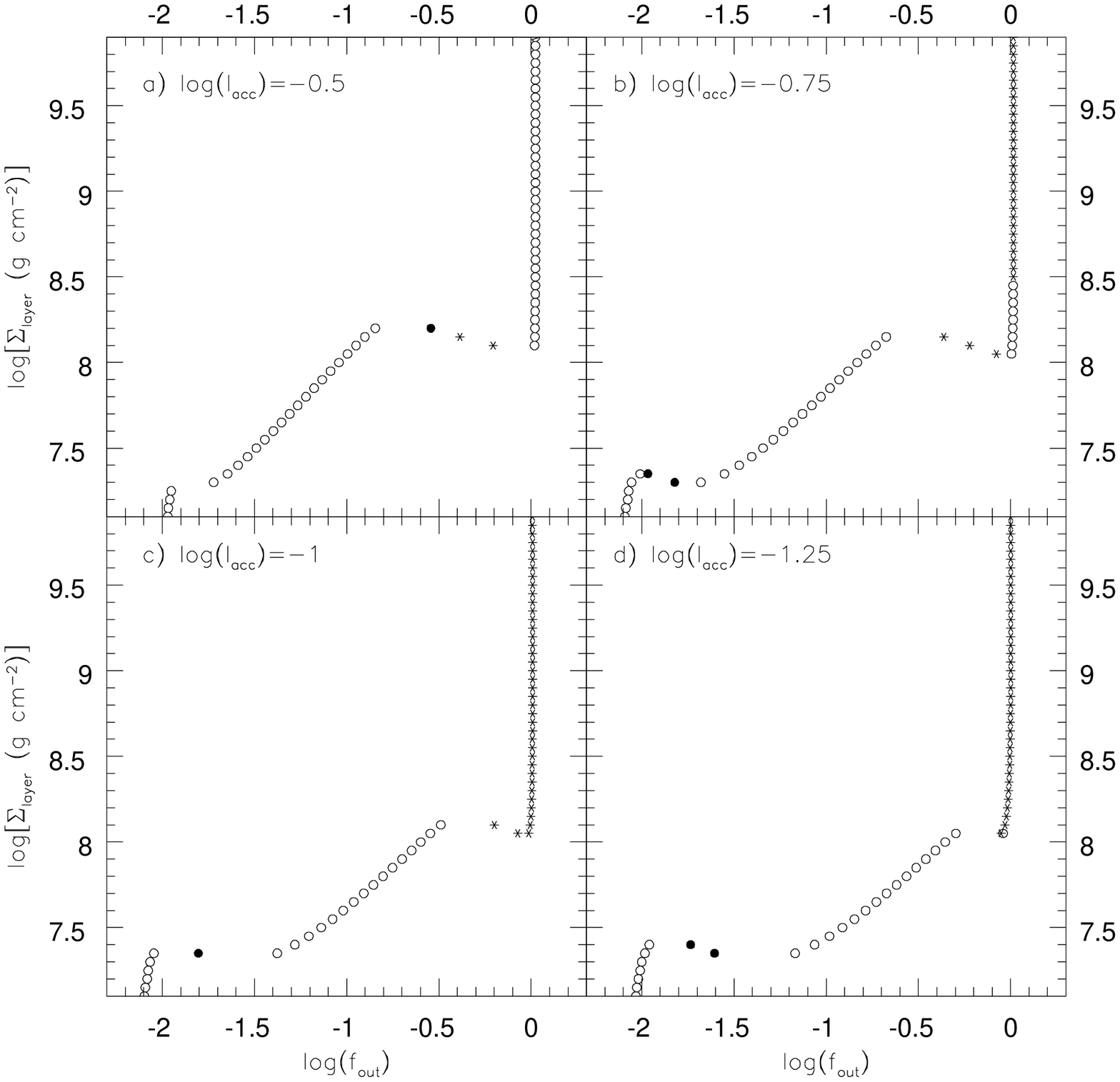}
\caption{The points trace the locus of values of the surface density
of the accreted layer $\Sigma_\rmscr{layer}$ and the normalized
escaping flux $f_{\rm out} = F_\rmscr{out}/F_{\rm nuc}$ which satisfy
the outer and inner boundary conditions.  The calculations assume a
neutron star of mass $M=1.4$~M$_\odot$, radius $R = 10^{0.4} R_s =
10.4$ km, and core temperature $T_{\rm core}=10^8$ K.  The four panels
correspond to four relatively large accretion luminosities.  An open
circle indicates that the corresponding equilibrium is stable.  A
filled circle indicates that the equilibrium is unstable with a real
mode frequency $\gamma$, while a star indicates an unstable
equilibrium with complex $\gamma$, i.e., an overstability.}
\label{fig:scurve1}
\end{figure}

\begin{figure}
\plotone{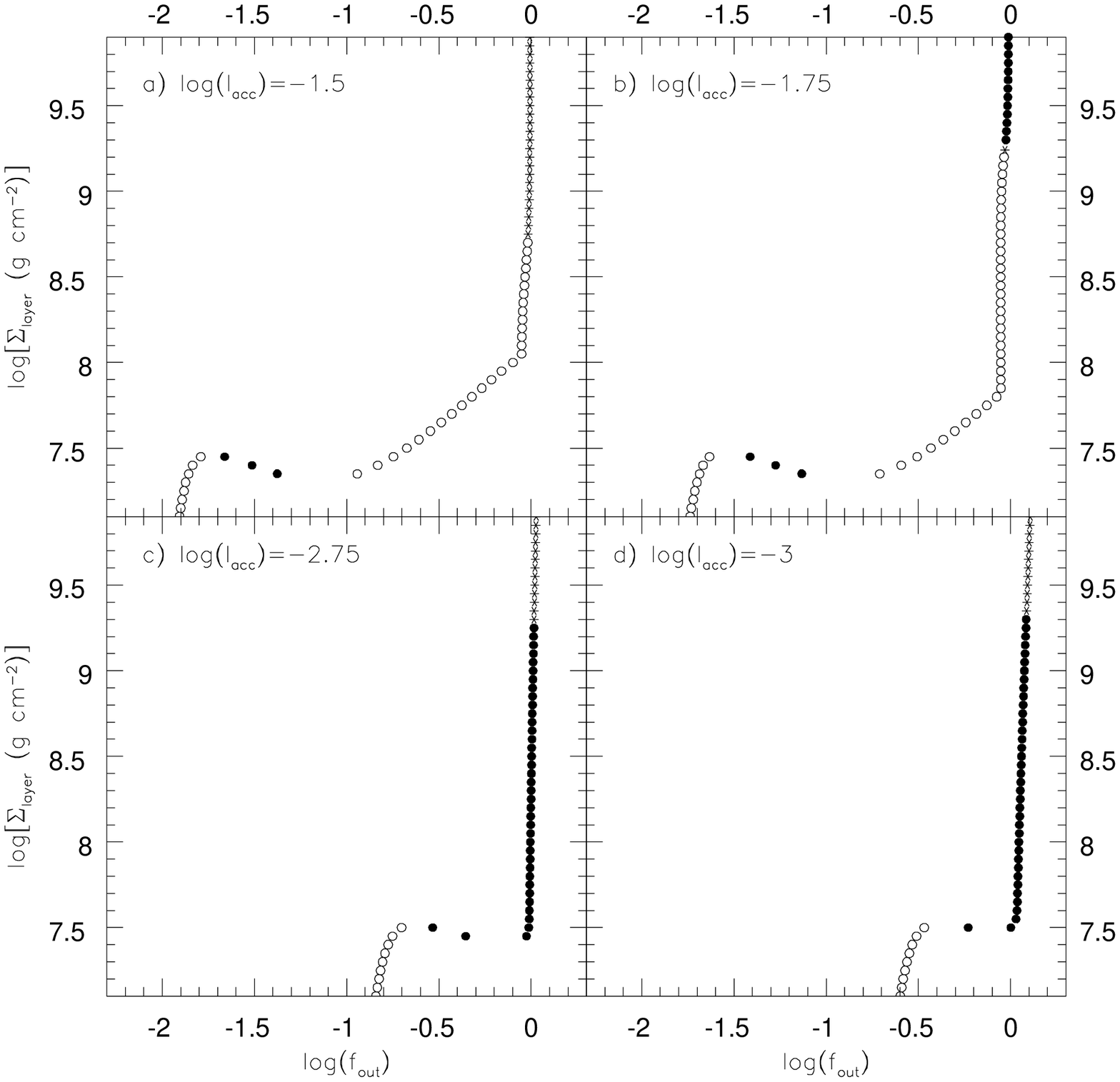}
\caption{Similar to Fig. \ref{fig:scurve1}, but for four lower
accretion luminosities.}
\label{fig:scurve2}
\end{figure}

Figures~\ref{fig:scurve1} and~\ref{fig:scurve2} show sequences of
equilibria of the accreted layer for a $1.4M_\odot$ neutron star with
a radius of 10.4 km and a core temperature of $10^8$K.  The different
panels correspond to different mass accretion rates $\dot M$,
parameterized by the accretion luminosity $L_\rmscr{acc}$ measured at
infinity,
\begin{equation}
L_\rmscr{acc} = \dot M c^2 {z\over(1+z)}
\equiv l_{\rm acc} L_{\rm Edd}, \label{Lacc}
\end{equation}
where $L_\rmscr{Edd}=4\pi GM c/\kappa_\rmscr{es}$, with
$\kappa_{\rm es} =0.4 ~{\rm cm^2\,g^{-1}}$, is the Eddington
luminosity measured at infinity.  The local surface mass accretion
rate $\dot\Sigma$ is related to $\dot M$ by
\begin{equation}
\dot\Sigma = {\dot M\over 4\pi R^2}(1+z). \label{Sigmadot}
\end{equation}  
The mass accretion rate $\dot M$, or equivalently $\dot \Sigma$, is a
key parameter that determines the nature of bursts.  Since it is
however not directly measured, we prefer to give all our results in
terms of  the dimensionless luminosity $l_{\rm acc} = L_{\rm
acc}/L_{\rm Edd}$.  In doing this, we assume that the accretion is
radiatively efficient and satisfies the relation given in equation
(\ref{Lacc}).

The eight panels in Figs. \ref{fig:scurve1}, \ref{fig:scurve2}
correspond to accretion luminosities $\log(l_{\rm acc}) = -0.5$,
$-0.75$, $-1$, $-1.25$, $-1.5$, $-1.75$, $-2.75$ and $-3$ in Eddington
units.  In each panel, the horizontal axis shows the escaping flux
$f_\rmscr{out}$ from the accreted layer, normalized by the maximum
nuclear burning energy available in the accreting gas:
\begin{equation}
f_{\rm out} \equiv {F_{\rm out}\over F_{\rm nuc}}, \qquad
F_\rmscr{nuc} = {\dot \Sigma} \left [ X_\rmscr{out} E^*_\rmscr{H} +
(X_\rmscr{out} + Y_\rmscr{out}) E^*_\rmscr{He} \right ]. \label{Fnuc}
\end{equation}
The vertical axis shows the column density of the layer, $\Sigma_{\rm
layer}$.

The most obvious feature of the various panels is that the equilibria
do not form monotonic sequences in the $F_\rmscr{out}$-$\Sigma_{\rm
layer}$ plane.  At a given $\Sigma_\rmscr{layer}$, there can be one,
three, or even five, distinct solutions for $F_\rmscr{out}$.  (There
are no cases of five solutions in the sequences shown here, but it is
fairly common when the core temperature is lower, e.g., $10^{7.5}$ K).
Because the equations are nonlinear and include many different
physical effects, it is not surprising to have multiple solutions.

\begin{figure}
\plottwo{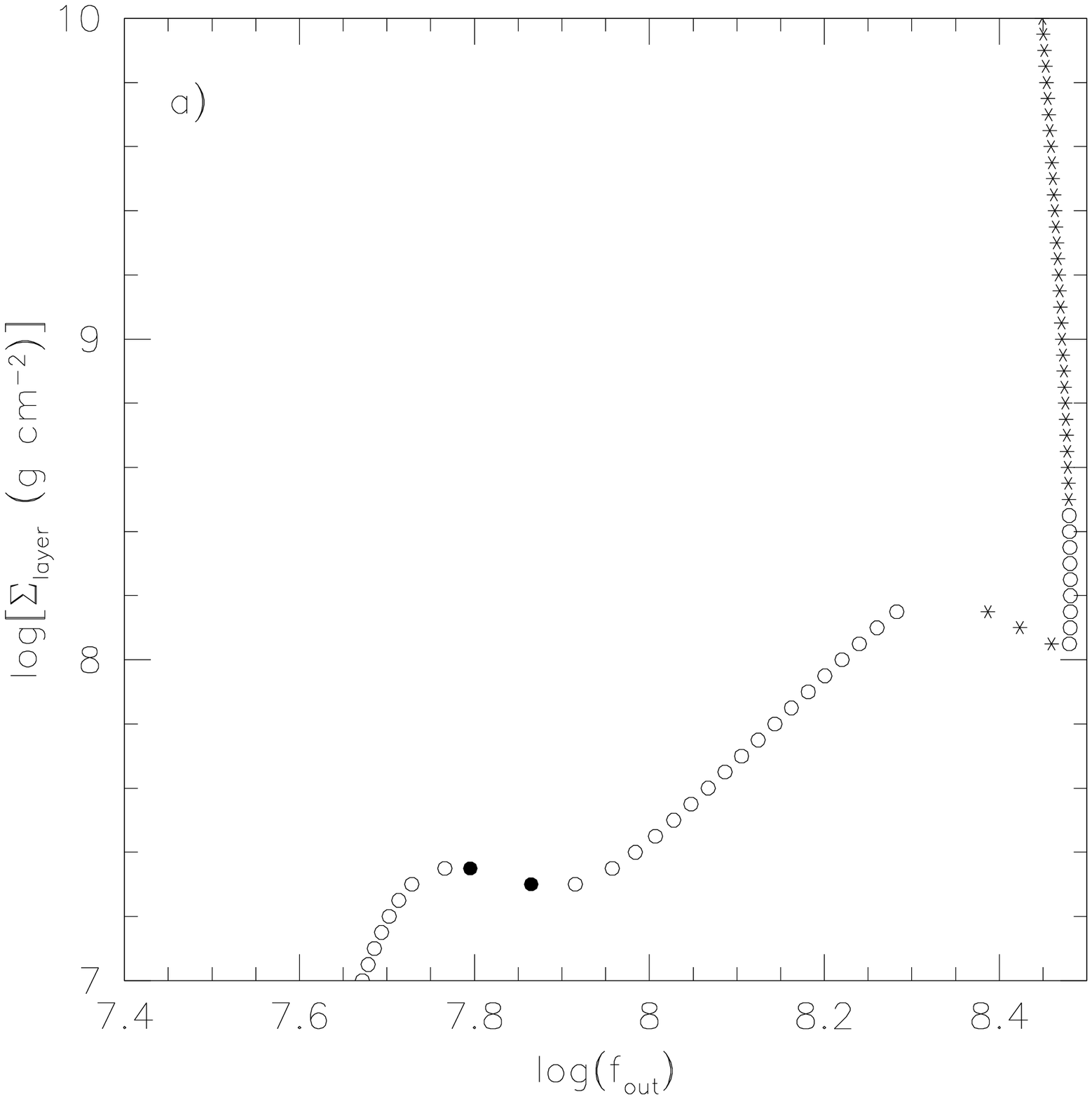}{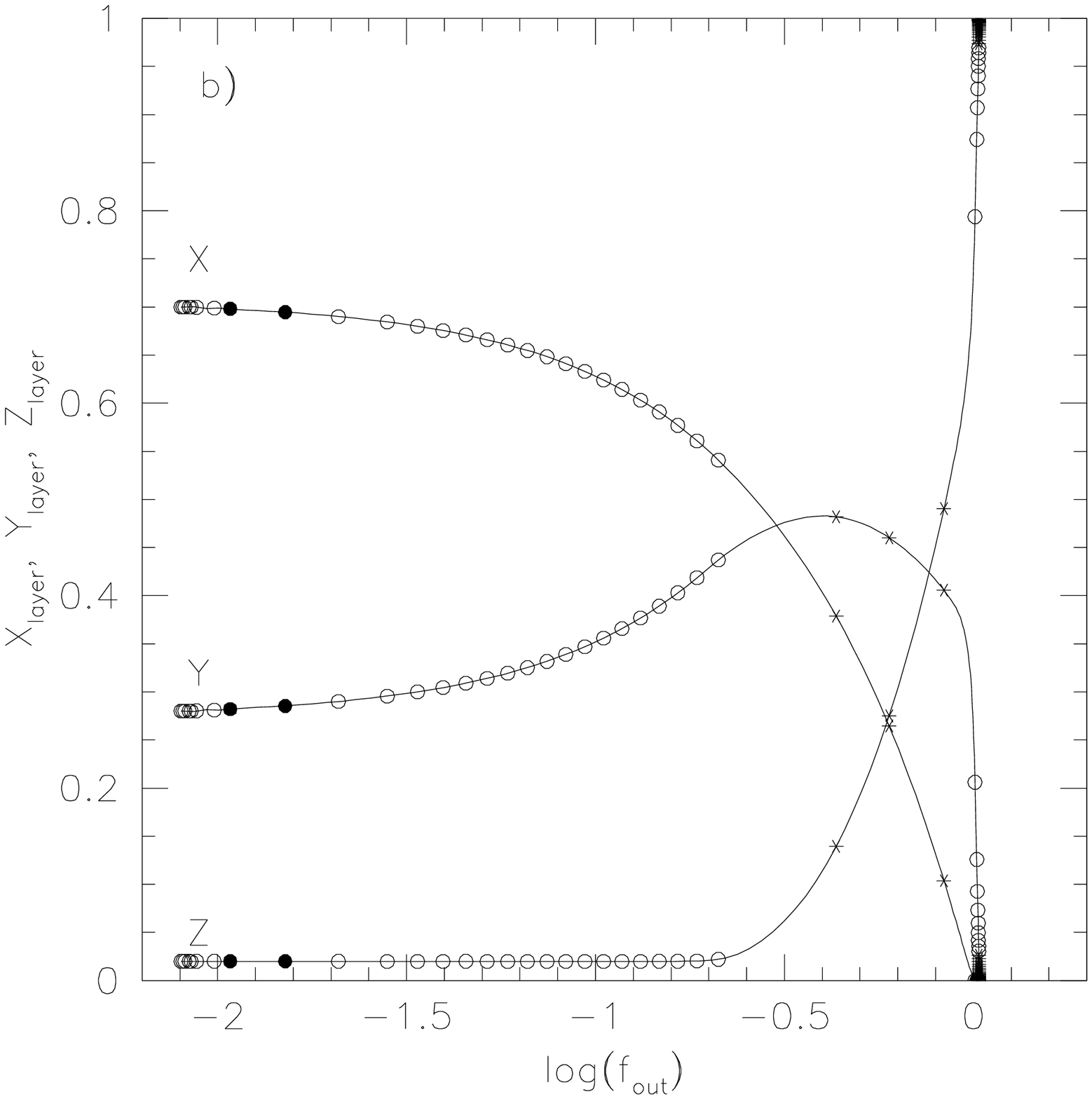}
\caption{Additional details of the equilibrium sequence shown in
Fig. \ref{fig:scurve1}b.  (a) Shows the column density $\Sigma_{\rm
layer}$ of the accreted layer as a function of the temperature $T_{\rm
layer}$ at the bottom of the layer.  The open circles, the filled
circles and the stars have the same meaning as in
Fig. \ref{fig:scurve1}.  (b) Shows the mass fractions of hydrogen,
helium and metals at the bottom of the layer as a function of the
normalized outgoing flux.}
\label{fig:onetenth_detail}
\end{figure}

Consider as an example Fig.~\ref{fig:scurve1}(b), which corresponds to
an accretion luminosity $\log(l_\rmscr{acc})=-0.75$.  The sequence of
equilibria shows two peaks, one at $\log(f_{\rm out}) \sim -2$ and one
at $\sim -0.5$.  In addition, at the very right, there is a steep
vertical segment which we refer to as the ``wall.''
Fig.~\ref{fig:onetenth_detail} shows other details of the equilibria
for this choice of $l_\rmscr{acc}$; panel (a) is a plot of
$\Sigma_\rmscr{layer}$ vs the temperature $T_\rmscr{layer}$ at the
base of the layer, while panel (b) shows the variation of the
H-fraction $X_\rmscr{layer}$, He-fraction $Y_\rmscr{layer}$ and heavy
element fraction $Z_\rmscr{layer}$ at the bottom of the layer.

When the column depth of the accreted layer $\Sigma_\rmscr{layer}$ is
very small, the gas is neither hot nor dense and does not undergo any
nuclear burning.  Thus, $X_\rmscr{layer}=X_\rmscr{out}$, $Y_{\rm
layer}=Y_\rmscr{out}$, $Z_\rmscr{layer}=Z_\rmscr{out}$, and the flux
$F_{\rm out}$ that emerges is mostly what is released as a result of
compressing the accreting gas, plus any flux that escapes from the
$10^8$K core.  This stage corresponds to the nearly vertical segment
at the left of Fig. 2(b).  As $\Sigma$ increases, the gas at the
bottom of the layer becomes hotter.  Ultimately, when
$\Sigma\sim10^{7.3} ~{\rm g\,cm^{-2}}$ and the temperature is about
$10^{7.7}$ K, hydrogen-burning begins.  Soon after this,
hydrogen-burning takes over as the dominant term in the energy
equation, and at this point, the
$\Sigma_\rmscr{layer}$-$F_\rmscr{out}$ curve in Fig. 2(b), as well as
$\Sigma_\rmscr{layer}$-$T_\rmscr{layer}$ curve in Fig. 4(a), reverse
direction, producing the peak on the left in the two plots.  We
identify this peak as the ``hydrogen peak.''  The onset of
hydrogen-burning is also evident in the run of $X_\rmscr{layer}$ in
Fig. 4(b) (see also $Y_\rmscr{layer}$ whose variations simply reflect
the amount of helium produced by hydrogen-burning).

As we descend from the top of the hydrogen peak in the direction of
increasing $F_\rmscr{out}$ and $T_\rmscr{layer}$, a point is reached
when hydrogen-burning becomes saturated (because the reactions are
beta-limited).  This happens around $T_\rmscr{layer}\sim10^{7.9}$ K.
Beyond this point, the sequence of equilibria start rising again in
the $\Sigma_\rmscr{layer}$-$F_\rmscr{out}$ and $\Sigma_{\rm
layer}$-$T_\rmscr{layer}$ planes.  The rise continues for a while,
with more and more hydrogen being burnt into helium until at
$\Sigma_{\rm layer}\sim10^{8.1} ~{\rm g\,cm^{-2}}$ and
$T_\rmscr{layer}\sim10^{8.3}$ K, helium-burning is initiated.  At this
point there is a second peak in the sequence of equilibria, the
``helium peak.''

As the sequence of equilibria fall off from the helium peak, the
helium is burned rapidly, and the CNO elements that this produces
cause the hydrogen-burning also to pick up since the rate of hydrogen
burning is proportional to $Z_\rmscr{CNO}$ (eq.~\ref{eq:epsilonCNO}).
When the hydrogen and helium are almost all exhausted, the sequence of
equilibria turn round again and march up the wall.  In this segment of
the curve, very little changes as a function of increasing $\Sigma$.
The hydrogen and helium are burned close to the surface within a
column of order $10^8 ~{\rm g\,cm^{-2}}$, and the rest of the accreted
layer consists just of burned CNO material which is progressively
compressed by the weight of the overlying gas.  Note that, since
helium burning by the triple-$\alpha$ reaction is not beta-limited,
there is no regime of saturated helium burning.  Therefore, the
transition from the declining slope of the helium peak to the rapidly
rising wall is generally quite abrupt.  Figure 4(a) shows that, as
$\Sigma_{\rm layer}$ increases on the wall, the temperature at the
base of the layer falls.  This is because there is a net flux flowing
from the layer into the star.  The nuclear burning occurs at a fixed
$\Sigma$ in this sequence of wall models.  With increasing
$\Sigma_{\rm layer}$, there is a larger dead column between the
burning layer and the base of the layer, and the ingoing flux causes
the temperature at the base to drop.

We should note at this point that not all equlibria shown in
Figs.~\ref{fig:scurve1}--\ref{fig:onetenth_detail} are accessible to a
real system. Imagine starting with a bare neutron star and adding gas
at the specified rate $\dot\Sigma$.  As $\Sigma_\rmscr{layer}$
increases, the system will ride up the left slope of the hydrogen
peak.  When $\Sigma_\rmscr{layer}$ is equal to the maximum $\Sigma
\sim10^{7.3} ~{\rm g\,cm^{-2}}$ of the hydrogen peak, the system no
longer has any equilibria available in the vicinity of the peak.
Therefore, it will relax towards the nearest available equilibrium,
which is a solution with the same value of $\Sigma_\rmscr{layer} \sim
10^{7.3} ~{\rm g\,cm^{-2}}$ on the left slope of the helium peak.
Thus, all the equilibria on the right slope of the hydrogen peak and
the lower left slope of the helium peak will be by-passed, since they
correspond to lower values of $\Sigma_{\rm layer}$ than the current
layer thickness.  With increasing $\Sigma_\rmscr{layer}$, the system
will ride up the helium peak until it reaches the maximum $\Sigma\sim
10^{8.1} ~{\rm g\,cm^{-2}}$ of the helium peak.  At this point, the
system will once again move across to the nearest available solution,
which is located on the wall, bypassing the equilibria in the valley
to the right of the helium peak.  Having reached the wall, the system
will continue rising up the wall with increasing
$\Sigma_\rmscr{layer}$.

Let us turn now to the other sequences of equilibria shown in
Figs.~\ref{fig:scurve1} and~\ref{fig:scurve2}.
Fig.~\ref{fig:scurve1}(a) corresponds to $\log(l_{\rm acc})=-0.5$.  In
this case, because of the rapid accretion rate, the accreted layer is
quite hot, and so hydrogen burning is already in the saturated limit
when it turns on.  There is, therefore, no hydrogen peak (though there
is the semblance of a plateau).  Fig.~\ref{fig:scurve1}(b) was
discussed above, and has a modest hydrogen peak.  Note that the helium
peak in Fig. 2(b) is pushed to the right relative to the peak in
Fig. 2(a).  This is because helium requires a temperature of at least
about $2\times10^8$ K to burn (Fig. \ref{fig:nuclear}), and such high
temperatures are obtained at the lower accretion rate only at larger
values of $F_\rmscr{out}$.  With decreasing accretion rate, the gas
temperature continues to fall and the hydrogen peak becomes more
pronounced (Figs. \ref{fig:scurve1}c,d).  The helium peak also gets
pushed progressively farther to the right, until it finally hits the
wall and disappears.  For accretion luminosities below about
$10^{-1.25}L_\rmscr{Edd}$, there is no helium peak.  Helium now starts
burning on the wall, at progressively larger values of $\Sigma_{\rm
layer}$ with decreasing $l_\rmscr{acc}$.

The above results are for a core temperature of $T_\rmscr{core}=10^8$ K.
The pattern of peaks can be different for other core temperatures.  If
the core temperature is a few times $10^8$ K, the hot core keeps the
accreted layer hot under all circumstances and hydrogen burning is
always in the saturated regime.  In this case, there is no separate
hydrogen peak.  On the other hand, for lower values of $T_\rmscr{core}$,
e.g. a few times $10^7$ K, the hydrogen peak becomes quite pronounced
and also has a higher peak value of $\Sigma_\rmscr{layer}$.  In this
case, it is often possible to have situations in which the helium peak
is lower than the hydrogen peak.  If such is the case, when the system
reaches the top of the hydrogen peak it will move directly across to
the wall without stopping at the helium peak.

\section{Stability}
\label{sec:stability}

The previous section described sequences of equilibria for various
mass accretion rates.  As explained in \S\ref{sec:model}, for every
equilibrium we carry out a linear stability analysis to check whether
the system is stable or unstable (the latter is determined by the
criterion given in eq. \ref{eq:mingrowth}).  In
Figs.~\ref{fig:scurve1} and~\ref{fig:scurve2}, open circles indicate
stable equilibria, filled circles indicate unstable equilibria in
which the most unstable mode has a real $\gamma \ge 3\gamma_{\rm
acc}$, and stars indicate unstable equilibria in which the
fastest-growing mode has a complex $\gamma$, with $\Re(\gamma) \ge
3\gamma_{\rm acc}$.

From the various panels in Figs.~\ref{fig:scurve1}
and~\ref{fig:scurve2}, we quickly discern the following patterns.  The
left slopes of the hydrogen and helium peaks always correspond to
stable equilibria, while the right slopes of both are always unstable;
this is similar to the pattern described by \citet[][ who only
considered He-burning]{1983ApJ...264..282P} and is easily understood
from his one-zone analysis.  As explained in Appendix A, the same
pattern is expected in our problem as well.  The stability of
equilibria on the wall is variable.  Sometimes the equilibria on the
wall are fully stable, sometimes they are fully unstable, and
sometimes some equilibria are stable and some are unstable.  As we
discuss now, it is the stability of the wall that determines whether
or not a particular accretion rate leads to thermonuclear bursts.

Consider as an example the case shown in Fig.~\ref{fig:scurve1}(b),
corresponding to $\log(l_\rmscr{acc})=-1$.  Starting from the bare
neutron star, as gas piles up, the system moves up the hydrogen peak
along a sequence of equilibria that are all stable (open circles).
When $\Sigma_{\rm layer}$ hits the top of the hydrogen peak, the
system moves across to a point on the left slope of the helium peak
with the same value of $\Sigma_\rmscr{layer}$.  There is some
adjustment of the accreted layer in order to switch to the new
solution, but because the equilibrium solution here is stable the
system is able to make the adjustment.  With increasing $\Sigma_{\rm
layer}$, the system climbs up the helium peak along a sequence of
stable equilibria.  As before, once it reaches the top of the peak, it
moves across to the right, this time to the wall, and the solution
here is again stable.  However, as the system climbs up the wall, it
reaches a point at which the equilibrium is unstable.  At this point,
there is no stable equilibria available for any value of
$f_\rmscr{out}$.  The system therefore undergoes a thermonuclear
burst.

We may now quickly identify which of the other cases shown in
Fig.~\ref{fig:scurve1} and~\ref{fig:scurve2} are stable and which
unstable.  The high accretion rate system in Fig.~\ref{fig:scurve1}(a)
is stable because the states on the wall are all stable.  The system
will move across to the wall after it hits the helium peak, and it
will then climb up the wall along a sequence of stable equilibria to
arbitrarily large values of $\Sigma_\rmscr{layer}$.  The cases shown
in Figs.~\ref{fig:scurve1}(b), \ref{fig:scurve2}(a),(b) are all
similar.  These systems reach the wall stably, climb part way up the
wall and then go unstable; we call all these cases ``delayed bursts.''
In contrast, the cases shown in Figs.  \ref{fig:scurve1}(c),(d),
\ref{fig:scurve2}(c),(d) become unstable the moment they hit the top
of the last peak since there are no stable equilibria available on the
wall; we call these ``prompt bursts.''

\subsection{Unstable vs Overstable Modes}
\label{sec:unstable-overstable}

We pause to discuss the distinction between modes with real $\gamma$
and those with complex $\gamma$.  The former correspond to a simple
instability, where the mode amplitude grows exponentially with time,
whereas the latter correspond to an overstability, where the mode
oscillates at a frequency equal to the imaginary part of $\gamma$ even
as the amplitude grows.  As Figs.~\ref{fig:scurve1}
and~\ref{fig:scurve2} show, we see both kinds of behavior for our
unstable modes.  We should note that complex eigenvalues $\gamma$
always appear in complex conjugate pairs, which is obvious from the
structure of the governing equations.

As explained in Appendix A, from an inspection of the steady state and
linear perturbation equations, it can be seen that any system that is
at the top of a peak or the bottom of a valley, i.e., where
$d\Sigma_\rmscr{layer}/ dF_{\rm out}=0$, has a zero-frequency mode,
$\gamma=0$.  Once we recognize this important rule, we can understand
why the right slopes of the hydrogen and helium peaks are unstable.
The argument goes as follows: the left slopes are stable, with all
modes having $\Re(\gamma)<0$; the peak has a mode with $\gamma=0$; by
continuity, this particular mode should have $\gamma>0$ on the right
slope and should be real.

The above rule has a corollary: if the system makes a transition from
stability to instability at a point where
$d\Sigma_\rmscr{layer}/dF_{\rm out} \ne 0$, then the unstable mode at
this point must have a complex $\gamma$.  This rule applies, for
instance, to all the systems that have delayed bursts, i.e., become
unstable half-way up the wall, e.g., Figs.~\ref{fig:scurve1}(b),
\ref{fig:scurve2}(a),(b).  In these cases, when the system moves into
the unstable zone, it always first hits an overstable region.  The
overstability may in some cases then become a pure instability as a
pair of modes with complex $\gamma$ merge to spin off a pair of real
$\gamma$.  Also, it is not possible to tell ``how complex'' the mode
is, i.e., the relative magnitudes of $\Re(\gamma)$ and $\Im(\gamma)$.

Finally, in cases where the wall is entirely unstable, we have not
been able to find any helpful rule.  Whether the instability
corresponds to real or complex $\gamma$ can be determined only with a
full calculation.

\subsection{Burst Energetics and Trigger Mechanism}
\label{sec:trigger-energy}

Because there are two burning fuels, hydrogen and helium, we would like
to know which fuel determines the burst properties.  There are two
aspects to this question.

\begin{figure}
\plotone{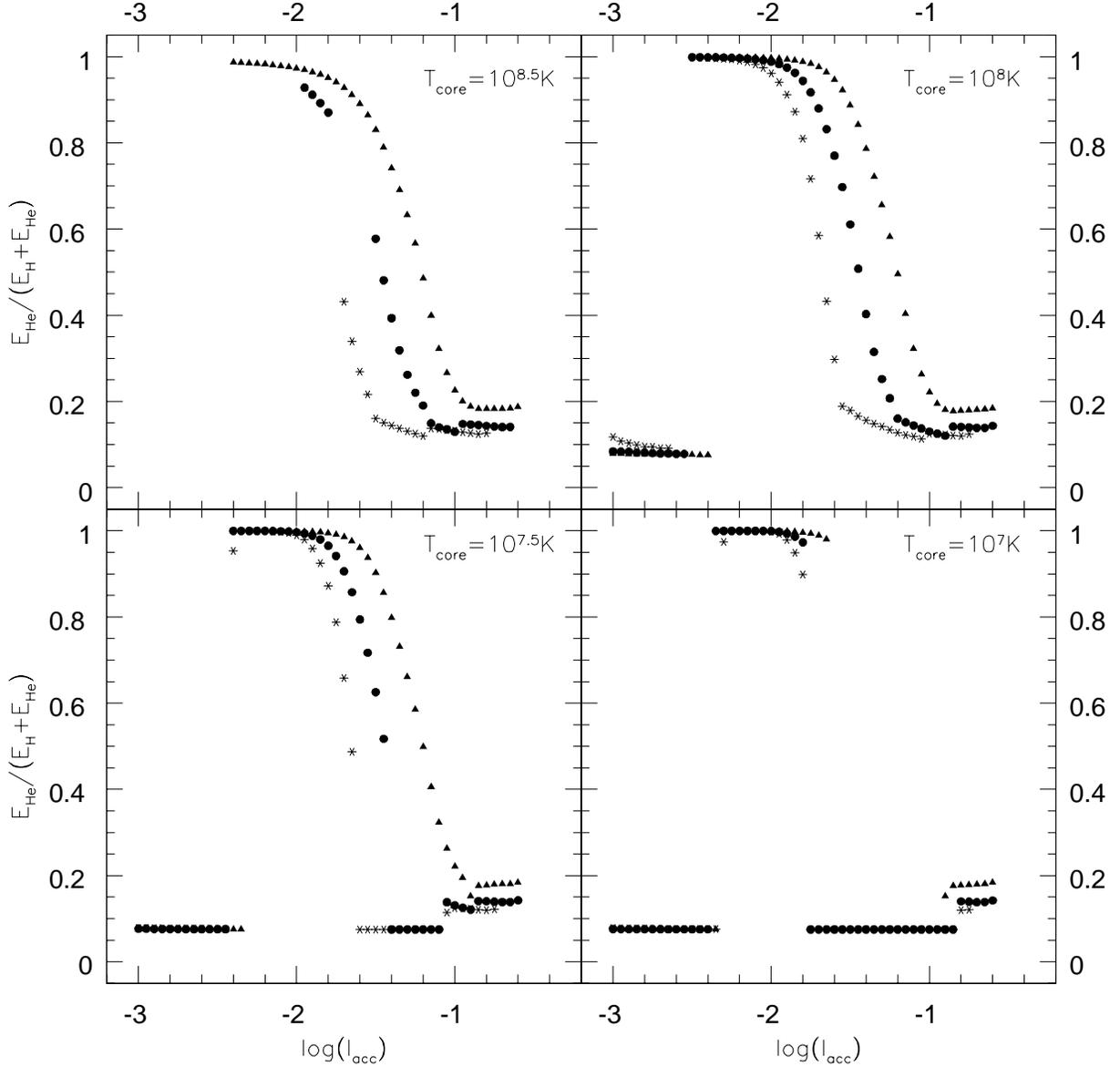}
\caption{Ratio of the fluence in the burst due to helium burning to
the fluence from hydrogen burning, assuming that each fuel is burned
to $^{56}$Fe.  The four panels correspond to core temperatures $T_{\rm
core} = 10^{8.5}, ~10^8, 10^{7.5}$ and~$10^7$K, respectively.  In each
panel, the triangles, circles and stars trace the results for a
neutron star of mass $1.4M_\odot$ and radii $R=10^{0.6}R_S=16.4$ km,
$10^{0.4}R_S=10.4$ km, and $10^{0.2}R_S=6.5$ km, respectively.  Note
that hydrogen and helium burning both contribute at high and low
luminosities (with hydrogen being about five times more important
energetically per unit mass), while helium burning dominates at
intermediate luminosities $\log(l_{\rm acc})\sim -1.5$ to $-2.5$. }
\label{fig:fluenceratio}
\end{figure}

First, once an instability has been triggered, we would like to know
how much energy is available for the burst in unburnt helium compared
to the energy in unburnt hydrogen.  Figure~\ref{fig:fluenceratio}
shows this quantity as a function of accretion rate for three choices
of the neutron star radius (assuming a mass of $1.4M_\odot$) and four
core temperatures.  We have assumed that all the energy in the
unburnt fuel is emitted as radiation during the burst.  In actuality,
some energy is lost into the core and some comes out as neutrinos; we
neglect this complication.

In interpreting the results, we should note that hydrogen-burning to
iron provides nearly 5 times more energy per gram than helium-burning
to iron.  Keeping this in mind, we see that for high mass accretion
rates roughly in the range $\log(l_{\rm acc}) \sim -0.6$ to $-1.5$
(the precise values depend on the neutron star radius and core
temperature), the bursts are of a mixed kind with energy being
contributed by both hydrogen and helium.  For intermediate accretion
rates, $\log(l_{\rm acc}) \sim -1.5$ to $-2.5$, we have pure
helium-burning bursts.  Finally, for $\log(l_{\rm acc}) \lesssim
-2.5$, we again have mixed bursts. Actually, this last range
corresponds to pure hydrogen bursts and it is not clear if helium
burning will be triggered during the burst.  In making our estimates
we assumed that all the unburnt fuel is consumed.  The ranges of
$\log(l_{\rm acc})$ given here correspond to a core temperature of
$10^8$ K and a neutron star radius of 10.4 km.

\begin{figure}
\plotone{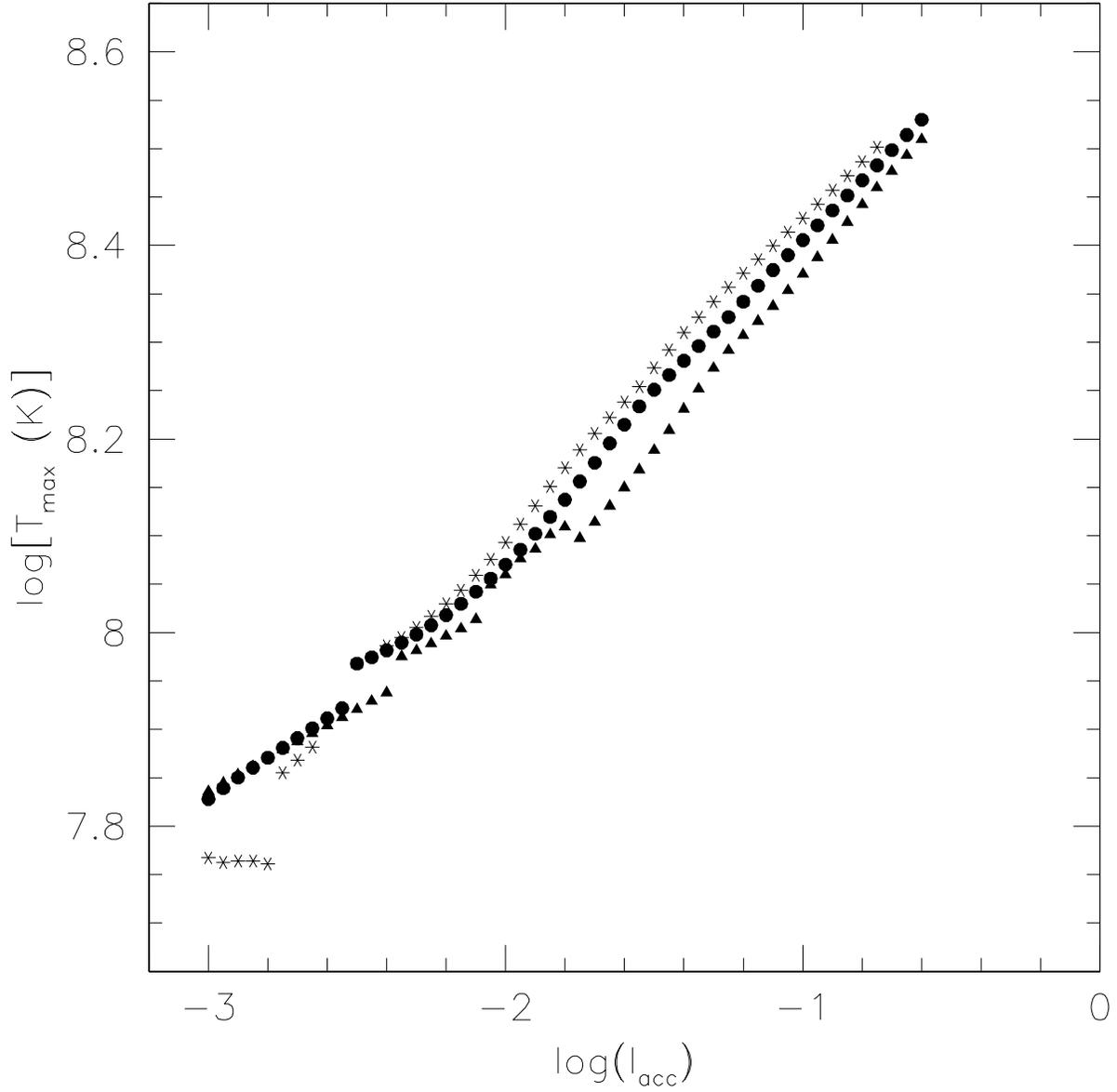}
\caption{The maximum temperature $T_{\rm max}$ achieved in the
accreted layer when the layer first becomes unstable, plotted as a
function of the accretion luminosity.  The core temperature is assumed
to be $T_\rmscr{core}=10^8$~K, and the calculations correspond to the
equilibrium at the ``wall.''  The symbols have the same meanings as in
Fig.~\ref{fig:fluenceratio}.}
\label{fig:tmax}
\end{figure}

Another interesting question is to understand what precisely triggers
the burst instability.  One way to do this is to look at the maximum
temperature in the fuel layer (Fig.~\ref{fig:tmax}) and to thereby
identify which fuel might be important.  As we discussed in
\S\ref{sec:quant}, hydrogen burning becomes saturated, and hence
stable, for temperatures above about $10^{7.9}$ K (see
Fig.~\ref{fig:nuclear}).  Therefore, we expect hydrogen-triggered
bursts at lower temperatures and helium-triggered bursts at higher
temperatures.  From Fig.~\ref{fig:tmax} we see that the former occur
at lower accretion rates and the latter at higher accretion rates.
However, there is no clear way to distinguish the kinds of bursts from
this plot since the variation of $T_{\rm max}$ with accretion
luminosity is smooth.  In our models, we do not find unstable modes
for layers with maximum temperatures above $3.5\times10^8$ K.  This
temperature limit is reached at $\log(l_{\rm acc})\sim -0.6$.

Gaining a deeper understanding of what triggers the bursts requires an
examination of the eigenfunctions of the unstable modes.
Specifically, we compute the contribution of helium burning integrated
over the eigenfunction and take the ratio of this to the hydrogen
burning integrated over the eigenfunction (here we do not consider the
additional burning to iron as we did with the burst fluence ratio).
The ratio should be large for a pure helium-triggered burst and small
for a pure hydrogen-triggered burst, and on the order of $0.1-0.2$ for
mixed-triggered burst (the ratio is not 0.5 in this case since
hydrogen-burning releases nearly an order of magnitude more energy per
gram than helium-burning).
\begin{figure}
\plotone{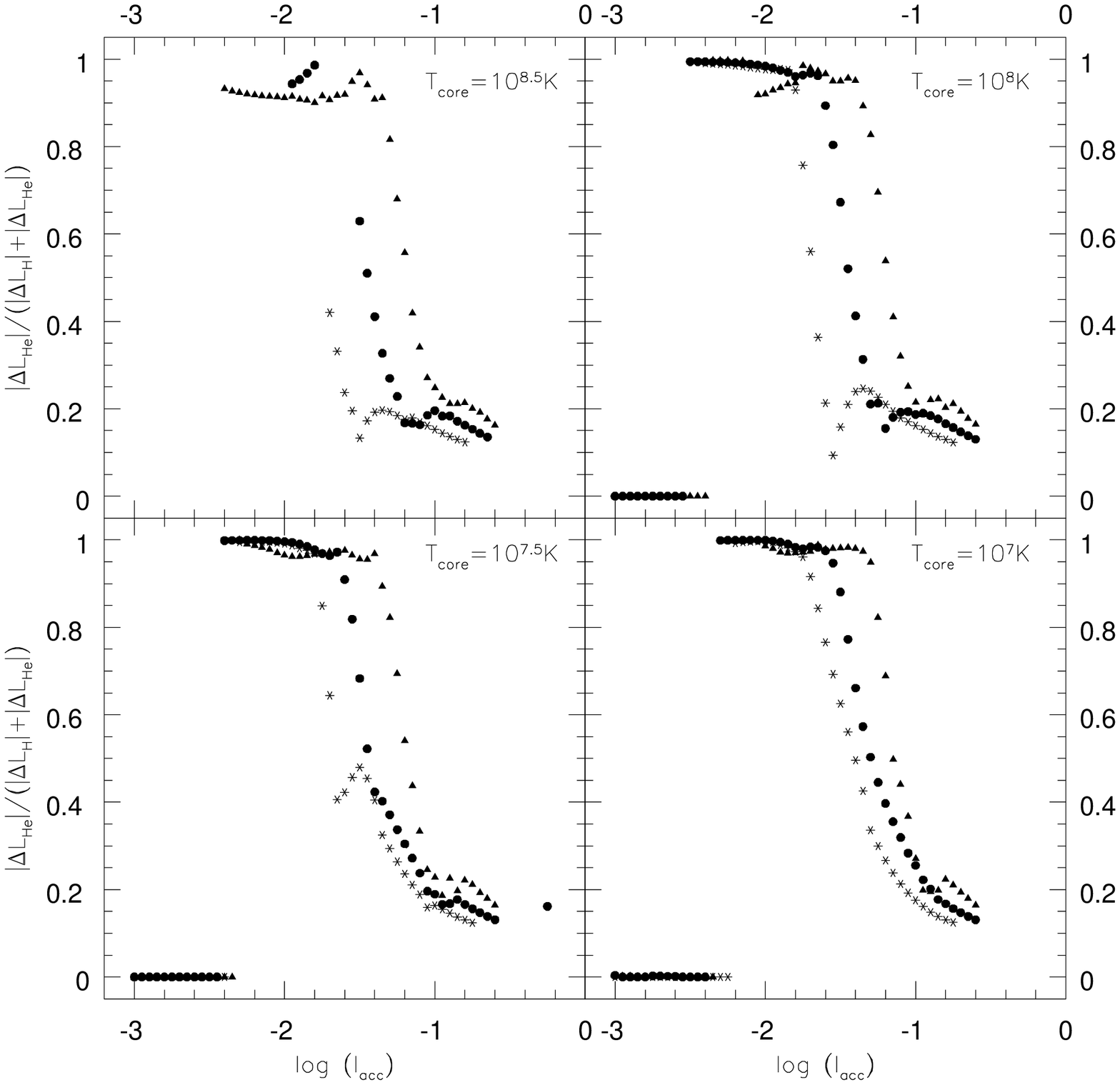}
\caption{Ratio of the energy produced by helium burning in the linear
eigenmode to the energy produced by hydrogen burning.  The symbols and
the layout are the same as in Fig.~\ref{fig:fluenceratio}.  Note that
at high luminosities there is mixed burning with significant
contributions from both hydrogen and helium, at intermediate
luminosities helium burning dominates, while at low luminosities
hydrogen burning dominates.  }
\label{fig:perratio}
\end{figure}

Figure~\ref{fig:perratio} shows the above ratio as a function of
$\log(l_\rmscr{acc})$ for the same three choices of the neutron star
radius and four choices of the core temperature.  We see that, at high
mass accretion rates $\log(l_{\rm acc}) \sim -0.6$ to $-1.5$, the
bursts are triggered by both hydrogen and helium.  Within this range,
the higher accretion rates give delayed bursts, while the lower rates
give prompt bursts (as seen from Fig. 2).  For intermediate accretion
rates $\log(l_{\rm acc}) \sim -1.5$ to $-2.5$, we have pure
helium-triggered bursts.  These all involve overstable modes.
Finally, for $\log(l_{\rm acc}) < -2.5$, we have hydrogen-triggered
bursts.  These are pure unstable modes (real $\gamma$).
\begin{figure}
\plotone{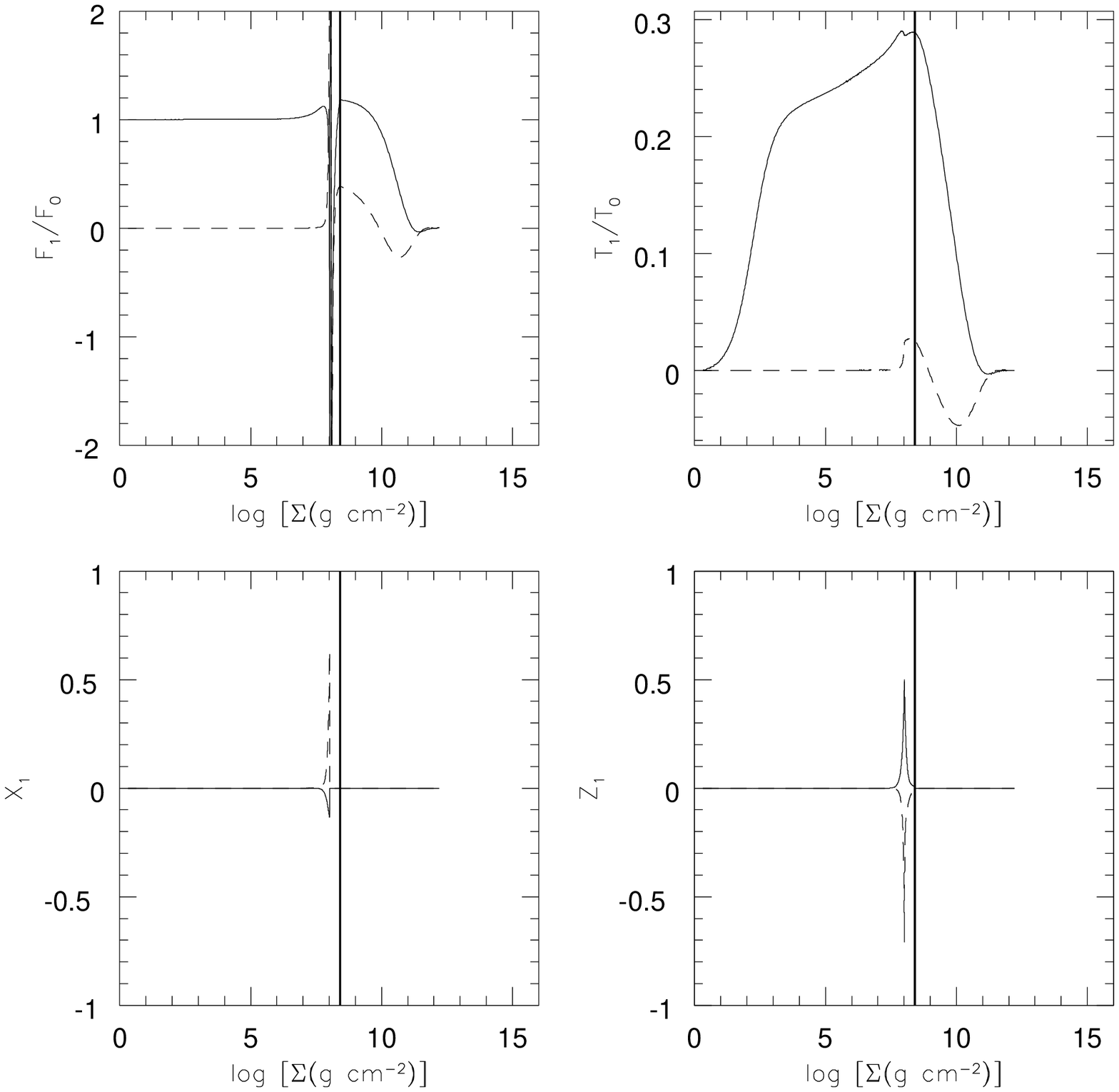}
\caption{Overstable eigenmode for $\log(l_\rmscr{acc})=-0.9$,
$M=1.4M_\odot$, $R=10.4$ km and $T_\rmscr{core}=10^8$~K.  The solid
and dashed lines trace the real and imaginary parts of the mode.  The
bold vertical line shows the bottom of the accreted layer at $\Sigma =
\Sigma_\rmscr{layer}$.  The eigenmode continues into the substrate to
a depth $\Sigma_{\rm max}$, at which point the amplitude of the mode
is substantially decreased.  The boundary condition at $\Sigma_{\rm
max}$ is that the temperature perturbation should vanish. However, it
is seen that the perturbations in essentially all variables vanish
there.}
\label{fig:per-1}
\end{figure}

\begin{figure}
\plotone{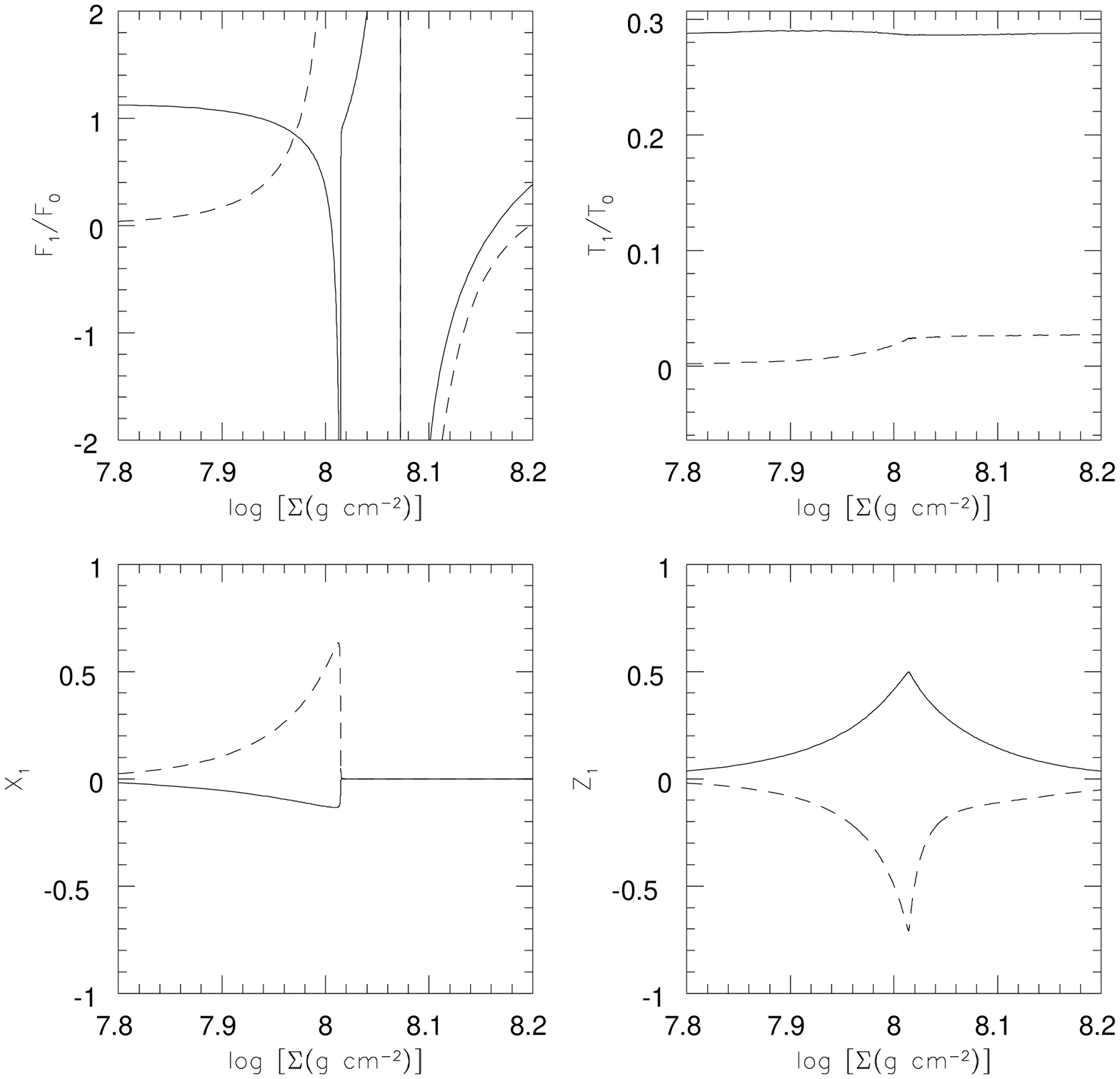}
\caption{Expanded version of the eigenmode in Fig. \ref{fig:per-1},
showing the burning layer in more detail.}
\label{fig:per-1z}
\end{figure}

Figures~\ref{fig:per-1}--\ref{fig:per-1z} show a typical eigenmode
corresponding to $\log(l_{\rm acc})=-0.9$.  Near the surface
($\Sigma\to 0$), all perturbations vanish according to our boundary
conditions, except the flux, which is given a unit perturbation (in
arbitrary units --- it merely sets the normalization).  The accreted
layer extends up to $\Sigma_{\rm layer}$, which is indicated by the
thick vertical line.  The eigenfunction then continues on into the
substrate to a depth such that the diffusion time to that point is
equal to twice the mode time scale (see
\S\ref{sec:boundary-conditions}).  Note that, even though the inner
boundary condition corresponds to a vanishing temperature
perturbation, in fact the flux perturbation also vanishes at this
point.  This shows that our choice of the inner boundary is physically
well-motivated.  We have tried integrating the eigenmode to larger or
smaller depth in the substrate and generally obtained the same
results.

Within the accreted layer, the finite flux perturbation at the surface
causes a temperature perturbation over the entire layer.  However, the
perturbations in $X$ and $Z$ are limited to a very narrow layer.  This
is the burning layer, which is shown in an expanded view in
Fig.~\ref{fig:per-1z}.  The fluctuation in $X$ indicates hydrogen
burning and the fluctuation in $Z$ corresponds to helium burning.  For
this mode, we see that both fuels burn at roughly the same depth,
indicating that it is an example of a mixed burst.  Also, since the
eigenmode has nontrivial real and imaginary parts, we have a complex
eigenvalue $\gamma$.  In this particular case, the real part of
$\gamma$ is equal to $8.9\times10^{-5} ~{\rm s^{-1}}$ and the
imaginary part is equal to $6.6\times10^{-4} ~{\rm s^{-1}}$.

At lower accretion rates, the eigenmodes tend to be simpler.  If we
consider the critical values of $\Sigma_{\rm layer}$ at which the
systems shown in Figs. \ref{fig:scurve2}(a),(b) become unstable, the
burning is dominated by helium and there is only a small contribution
from hydrogen.  Moreover, the two fuels tend to burn at different
values of $\Sigma$.  The eigenmode is still complex, though the
imaginary part is not very large.  If we go to yet lower accretion
rates as in Figs. \ref{fig:scurve2}(c),(d), then there is no helium
burning at all and the burning layer is dominated by hydrogen. In this
case, the eigenmode is fully real.

\subsection{Putting Things Together}
\label{sec:together}

Pulling together the various ideas discussed above, we identify five
regimes of bursting behavior as a function of decreasing accretion
luminosity.  In the following, the numerical values correspond to the
particular calculations presented earlier, in which $M=1.4M_\odot$,
$R=10^{0.4}R_S=10.4$ km, $T_{\rm core}=10^8$ K.  Tables
\ref{tab:burst-regimes1} and \ref{tab:burst-regimes2} give results for
other selected choices of $R$ and $T_{\rm core}$.

\noindent
I. For very high accretion rates, with $\log(l_\rmscr{acc}) > -0.6$,
there are no thermonuclear bursts.  This is the regime of stable
accretion, where the accreting gas is able to burn the nuclear fuel
without instability.

\noindent
II. For somewhat lower accretion rates $\log(l_\rmscr{acc}) \sim -0.6$
to $-0.85$, we have mixed bursts triggered by both hydrogen and
helium.  These bursts burn substantial amounts of both fuels.  In this
particular accretion range, the instability is delayed and is
triggered only after the system climbs part way up the wall.  We call
these ``delayed mixed bursts.''  As the system climbs up the wall, a
large fraction of the nuclear fuel is burned stably in the accreted
layer before the burst is triggered.\footnote{Note that, on the wall,
$f_{\rm out}$ is nearly equal to unity, which means that the escaping
flux in the equilibrium solution is almost equal to the nuclear energy
content of the new fuel being added to the surface.  Nevertheless, the
accreted layer does possess energy in unburnt fuel.  Loosely
speaking, it is the unburnt fuel that accumulated in the layer prior
to reaching the wall.  The energy from burning this fueld is released
during the burst.}  As we shall see, this has dramatic consequences
for both the recurrence times of the bursts and the ratio of the burst
fluence to the total accretion fluence between bursts.  When the
system becomes unstable, the mode has a complex $\gamma$, which
corresponds to an overstability.  Thus, we expect the system to show
an oscillatory behavior as it approaches the instability.

\noindent
III. For yet lower accretion rates $\log(l_\rmscr{acc}) \sim -0.9$ to
$-1.25$, we continue to have mixed triggered bursts burning mixed fuel.
However, here the entire wall is unstable and so the instability is
triggered when the system reaches the top of the helium peak.  Because
the burst happens as soon as the system hits the wall, we refer to
these as ``prompt mixed bursts,'' to distinguish them from the
previous class of delayed mixed bursts.  When the instability first
begins at the peak, $\gamma$ is real, but when the system reaches the
wall, $\gamma$ may possibly be complex (though not always).  We cannot
tell from our present analysis whether or not the system will show
oscillatory behavior prior to the burst.

\noindent
IV. For still lower accretion rates $\log(l_\rmscr{acc}) \sim -1.3$ to
-2.5, bursts are again triggered only after the system climbs part way
up the wall, i.e., we have delayed bursts.  Here the instability is
helium-triggered and the burst is dominated by helium burning.  These
correspond to the classic ``helium bursts'' that have been studied by
previous authors.  According to our calculations, $\gamma$ is complex
when these systems go unstable, and this means that there should be
some kind of oscillatory behavior prior to the burst.  In practice,
the imaginary parts of $\gamma$ tend to be small for these modes.

\noindent
V. Finally, for the lowest accretion rates $\log(l_\rmscr{acc}) <
-2.5$, the wall is again unstable, and now the burst happens as soon
as the system hits the top of the hydrogen peak.  The instability is
triggered by hydrogen burning, and we call these ``hydrogen bursts.''
They do not have any oscillatory behavior.  It is not clear if these
hydrogen bursts will trigger helium burning.  If they do not, then the
helium will accumulate as the ashes of hydrogen burning and the system
is likely to have enormous but very rare helium bursts.

\begin{deluxetable}{lllll}
\tablecaption{Burst Regimes for $T_{\rm core}=10^{7.5}$ K
\label{tab:burst-regimes1}
}
\tablehead{ 
\colhead{} & 
\multicolumn{2}{c}{Range in $\log(l_\rmscr{acc})$} &
\colhead{} \\
\colhead{} & 
\colhead{$R=6.5$ km} &  
\colhead{$R=10.4$ km} & 
\colhead{Type of Burst} &
\colhead{Symbol in Fig.~\ref{fig:regime}}}
\startdata
I.   &  $\ge (-0.7)$      & $\ge (-0.55)$       & No Bursts & None \\
II.  & $(-0.75)$ to ($-1.0)$  & $(-0.6)$ to $(-0.85)$   & Delayed
Mixed & Naked Star \\ 
III. & $(-1.05)$ to $(-1.6)$ & $(-0.9)$ to $(-1.4)$   & Prompt
Mixed & Circled Symbol \\
IV.  & $(-1.65)$ to $(-2.3)$ & $(-1.45)$ to $(-2.4)$ & (Delayed)
Helium & Naked Star \\ 
V.   & $\le (-2.35)$ & $\le (-2.45)$   & (Prompt)
Hydrogen &
Circled Symbol \\
\enddata
\end{deluxetable}

\begin{deluxetable}{lllll}
\tablecaption{Burst Regimes for $T_{\rm core}=10^8$ K
\label{tab:burst-regimes2}
}
\tablehead{ 
\colhead{} & 
\multicolumn{2}{c}{Range in $\log(l_\rmscr{acc})$} &
\colhead{} \\
\colhead{} & 
\colhead{$R=6.5$ km} &  
\colhead{$R=10.4$ km} & 
\colhead{Type of Burst} &
\colhead{Symbol in Fig.~\ref{fig:regime}}}
\startdata
I.   &  $\ge (-0.7)$      & $\ge (-0.55)$       & No Bursts & None \\
II.  & $(-0.75)$ to ($-1.0)$  & $(-0.6)$ to $(-0.85)$   & Delayed
Mixed & Naked Star \\ 
III. & $(-1.05)$ to $(-1.4)$ & $(-0.9)$ to $(-1.25)$   & Prompt
Mixed & Circled Symbol \\
IV.  & $(-1.45)$ to $(-2.4)$ & $(-1.3)$ to $(-2.5)$ & (Delayed)
Helium & Naked Star \\ 
V.   & $\le (-2.65)$ & $\le (-2.55)$   & (Prompt)
Hydrogen &
Circled Symbol \\
\enddata
\end{deluxetable}

The various regimes described above are summarized in Tables
\ref{tab:burst-regimes1}, \ref{tab:burst-regimes2}.  Much of our
discussion in the rest of the paper is in terms of the burst regimes
discussed in this section.  Many of the regimes have been recognized
by previous authors
\citep{1981ApJ...247..267F,1987ApJ...323L..55F,2000arxt.confE..65B},
but one of the new ideas to come out of our analysis is the
distinction between delayed mixed bursts and prompt mixed bursts.
This distinction has not been made earlier.  The other new aspect is
our ability to distinguish between simple exponential instability and
oscillatory overstability.  It is unique to our work since this is the
first study to carry out a full linear stability analysis to calculate
eigenmodes and (complex) eigenfrequencies.

The other comment we should make is that the pattern of five regimes
described above is valid only for intermediate core temperatures
$\sim10^{7.5}-10^8$ K.  We have focused on this case since it
represents the likely core temperature for standard neutrino cooling
models of accreting neutron stars.  For higher core temperatures,
e.g., $10^{8.5}$ K, the hydrogen-burning layer is hotter than
$10^{7.8}$ K so that the systems are in the saturated hydrogen-burning
limit.  They therefore lack the hydrogen-triggered bursts at low
accretion rates.  On the other hand, if the core temperature is lower,
e.g., $10^{7}$ K, the prompt mixed bursts are no longer triggered by
the helium peak.  Instead, the bursts are triggered when the accreted
layer reaches the top of the hydrogen peak, which is now higher than
the helium peak.  The bursts are still of the mixed variety though.
In addition, there are quantitative differences as a function of
neutron star radius and core temperature.  These are discussed in the
next section.

\section{Results}
\label{sec:results}

\subsection{Different Regimes of Bursting}
\label{sec:results-regimes}

\begin{figure}
\plotone{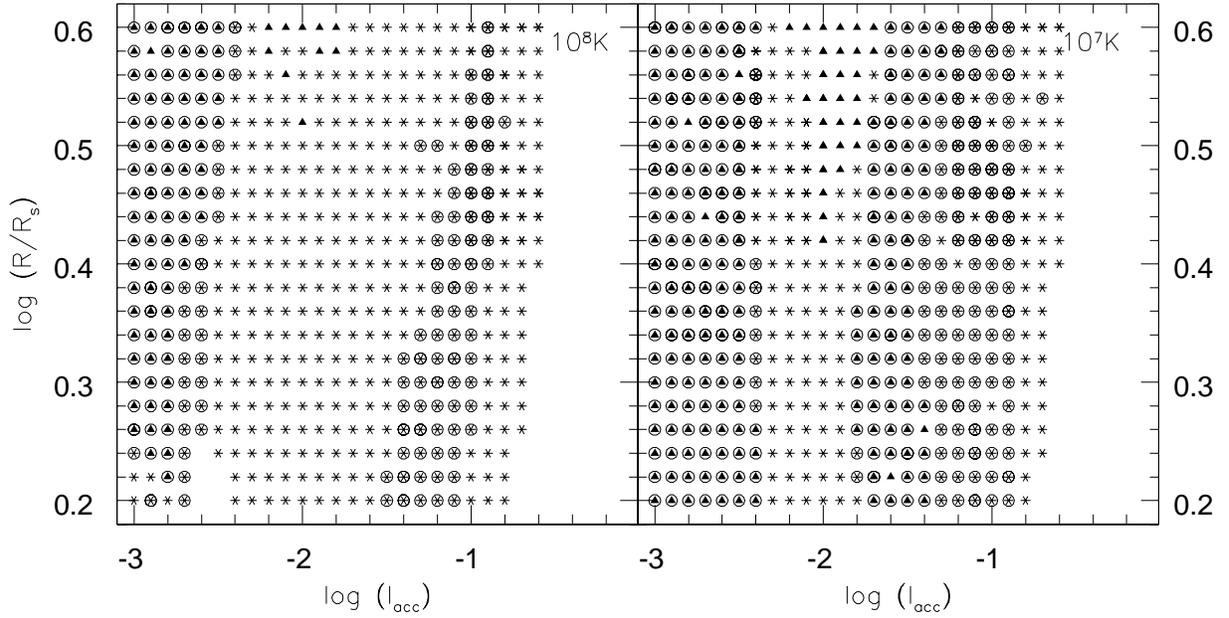}
\caption{Regimes of burning on the surface of a 1.4$M_\odot$ neutron
star for $T_\rmscr{core}=10^8$~K and $10^7$~K.  The empty areas
correspond to stable burning, the filled triangles denote regions with
unstable modes (real $\gamma$), and the stars to regions with
overstable/oscillatory modes (complex $\gamma$).  Prompt bursts (see
the text for an explanation) are shown by symbols circumscribed with a
circle, and delayed bursts are shown by symbols without circles.
Starting from the right, with decreasing luminosity, one goes through
the following sequence: stable, delayed mixed bursts, prompt mixed
bursts, helium bursts, hydrogen bursts (compare with Table
\ref{tab:burst-regimes2}).}
\label{fig:regime}
\end{figure}

Figure~\ref{fig:regime} depicts the various bursting regimes as a
function of the radius of the neutron star and the accretion rate.
The results shown are for core temperatures of $T_\rmscr{core}=10^8$~K
and $10^7$ K.  The blank areas correspond to stable nuclear burning,
the symbols enclosed within circles correspond to prompt bursts and
the naked symbols correspond to delayed bursts.  The five regimes of
bursting behavior described in \S\ref{sec:together} and summarized in
Tables \ref{tab:burst-regimes1}, \ref{tab:burst-regimes2} are clearly
delineated.

Moving from right to left in the panel corresponding to $10^8$ K, we
see that at the highest mass accretion rates there is no burst
activity since these systems consume hydrogen and helium stably
(region I in \S\ref{sec:together}).  Immediately to the left is a
region of delayed mixed bursts (region II).  At lower accretion rates,
there is a zone of prompt mixed bursts (symbols with circles, region
III).  Then there is a relatively broad zone of delayed helium bursts
(region IV), and finally, at the lowest accretion rates, a region of
prompt hydrogen bursts (symbols with circles, region V).

Some of the boundaries between the different zones are remarkably
insensitive to the neutron star radius, though the zone of prompt
mixed bursts does vary with radius, at least for $T_{\rm core}=10^8$
K.  For very compact neutron stars, e.g., $\log (R/R_S)=0.2$, this
zone is reasonably wide, whereas for $\log(R/R_S)=0.6$, the zone
almost disappears.  The zone of prompt mixed bursts is much broader
for $T_{\rm core}=10^7$ K.  These variations with radius and $T_{\rm
core}$ lead to noticeable signatures in many of the diagnostics that
we discuss later.

In Fig.~\ref{fig:regime}, the points represented with stars correspond
to overstable modes with complex $\gamma$ and the triangles correspond
to unstable modes with real $\gamma$.  In accordance with the
discussion in \S\ref{sec:together}, we see that delayed bursts are
nearly always overstable, while prompt bursts are either overstable or
unstable.

The map of the different burst regimes is roughly the same at other
core temperatures, though the widths and locations of the various
zones tend to vary.  Generally, the regime of delayed mixed bursts
becomes somewhat narrower with decreasing $T_{\rm core}$.  Also, for
$T_\rmscr{core}=10^{8.5}$~K, there are no hydrogen bursts, i.e., the
leftmost region of filled triangles within circles is absent.  These
patterns may be noticed in several of the plots discussed later.

\subsection{Burst Recurrence Time}
\label{sec:results-recurrence}

We begin our overview of the observable properties of thermonuclear
bursts on neutron stars with the recurrence time $t_{\rm rec}$, i.e.,
the average time between bursts.  Because we only treat the physics of
the accreted layer until a burst is initiated, but not the burst
itself, we need to make some assumption as to how much of the
available fuel is consumed in a burst.  In our calculations we assume
that all the fuel is burned, which means that our estimates of $t_{\rm
rec}$ are really upper limits to the observed recurrence time.  With
this approximation, if $\Sigma_{\rm layer,crit}$ is the critical
column depth of the accreted layer at which a burst is initiated, then
the recurrence time $t_{\rm rec}$ measured by a distant observer is
given by
\begin{equation}
t_{\rm rec} = (1+z){\Sigma_{\rm layer,crit}\over\dot\Sigma}.
\label{eq:trec}
\end{equation}

\begin{figure}
\plotone{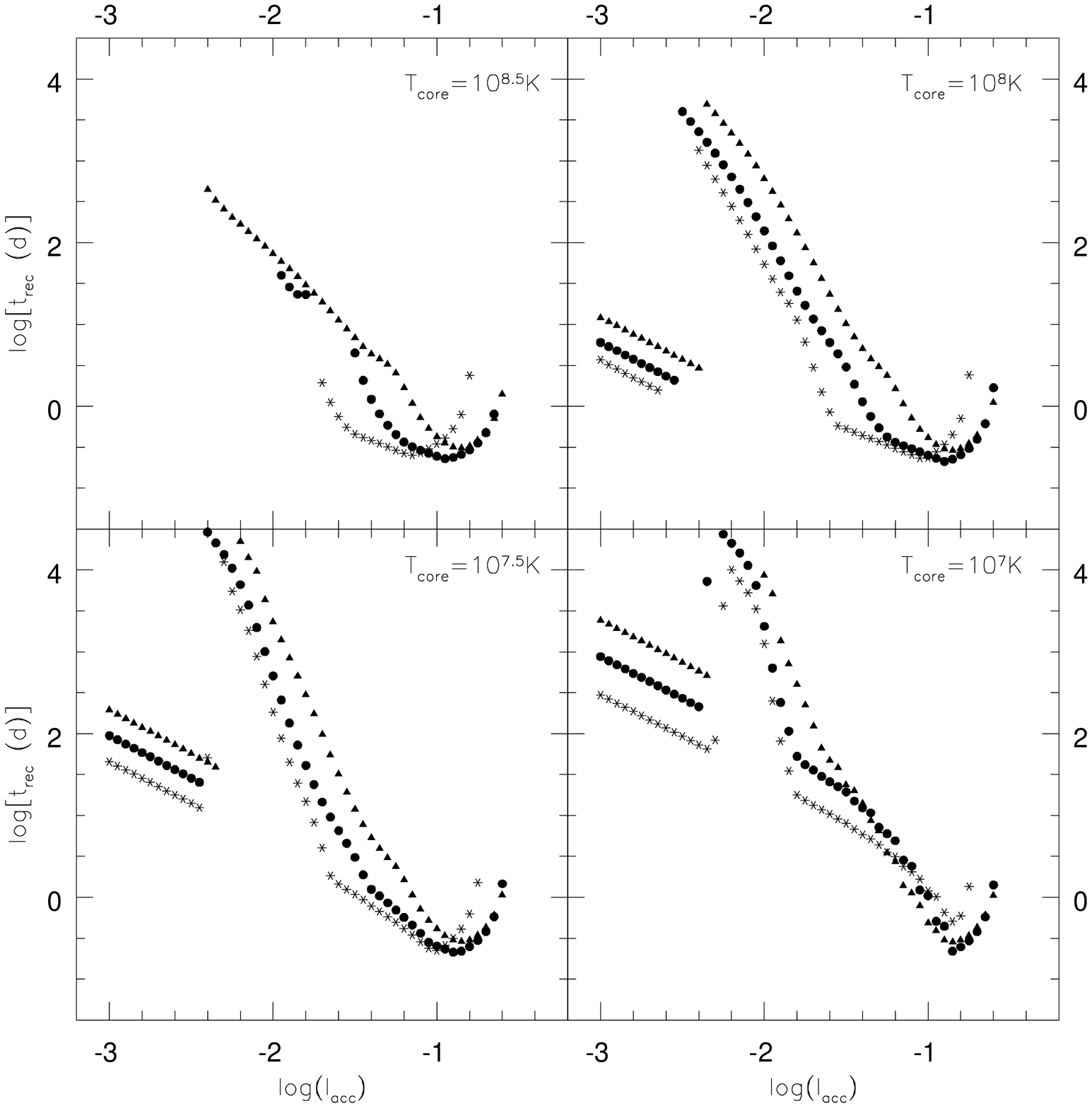}
\caption{ The burst recurrence time $t_{\rm rec}$ for a distant
observer, assuming that all the fuel is consumed.  The four panels
correspond to four core temperatures, and the triangles, circles and
stars correspond to neutron star radii of 16.4 km, 10.4 km and 6.5 km,
respectively.  The four regimes of bursting described in
\S\ref{sec:together} and summarized in Tables
\ref{tab:burst-regimes1}, \ref{tab:burst-regimes2} are clearly seen in
many of the sequences.}
\label{fig:trec}
\end{figure}

\begin{figure}
\plotone{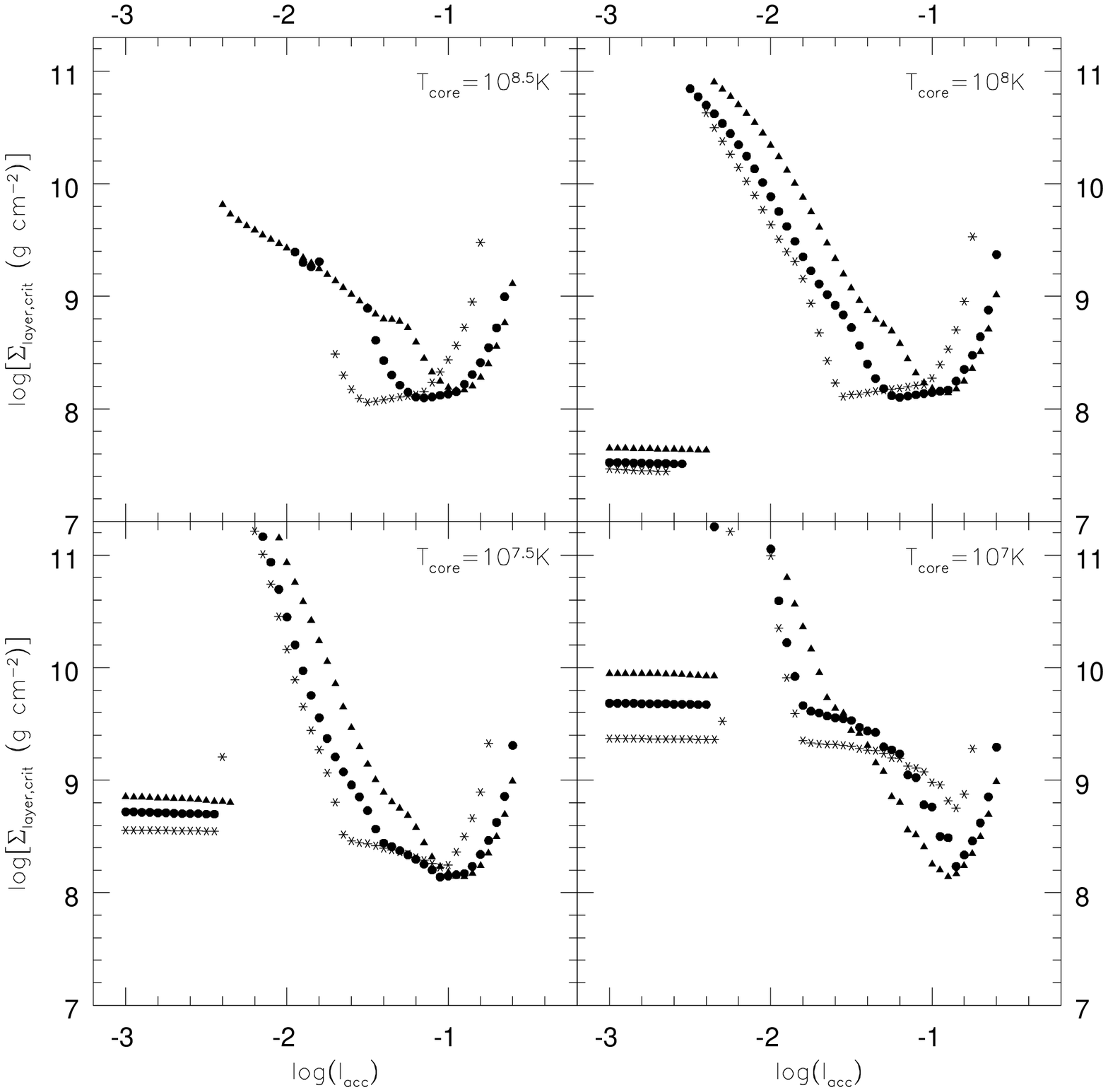}
\caption{Critical column density of accreted material $\Sigma_{\rm
layer,crit}$ at the onset of the burst.  The symbols and layout are
the same as in Fig.~\ref{fig:trec}.  }
\label{fig:sigma}
\end{figure}

Figure~\ref{fig:trec} shows $t_{\rm rec}$ as a function of the
accretion luminosity for three choices of the neutron star radius,
$\log(R/R_S)=0.2, ~0.4, ~0.6$ (i.e., $R=16.4$, 10.4, 6.5 km), and four
core temperatures, $T_{\rm core}=10^{8.5}, ~10^8, ~10^{7.5}, 10^7$ K.
The $10^8$ K panel, and to some extent the $10^{7.5}$ K panel, clearly
show the different regimes of bursting behavior discussed in the
previous subsection, especially for smaller neutron star radii.  The
recurrence time tends to become larger for delayed bursts compared to
prompt bursts.  The straight band of points between $\log(l_{\rm acc})
\sim -1$ and $-1.6$ for the two smaller radii corresponds to prompt
mixed bursts.  This band is missing for the largest radius because, as
seen in Fig.~\ref{fig:regime}, the region of prompt mixed bursts
disappears for that radius.

One obvious pattern in Fig. \ref{fig:trec} is that there is a general
overall increase of $t_{\rm rec}$ with decreasing $L_{\rm acc}$.  This
is because $\dot \Sigma$ becomes lower with decreasing accretion rate,
so it takes longer to build up a given accretion column.  To
illustrate this point, Fig.~\ref{fig:sigma} shows the dependence of
$\Sigma_{\rm layer,crit}$ for the various cases.  We see that,
for the prompt mixed bursts, $\Sigma_{\rm layer,crit}$ is more or less
constant.

The results for the other core temperatures are generally reasonable.
For $T_{\rm core}=10^{8.5}$ K, the recurrence times are fairly similar
to the $10^8$ K case, except that there are no hydrogen bursts at low
mass accretion rates.  The case of $T_{\rm core}=10^{7.5}$ K is very
similar to $10^8$ K.  However, for $T_{\rm core} = 10^7$ K, the column
needed to produce bursts tends to be higher.  This leads to
significantly longer recurrence times.

The shortest recurrence time we find in the model is about 5 hours,
which occurs at $\log(l_{\rm acc})\sim-1$ for intermediate core
temperatures $T_{\rm core} \sim 10^{7.5}-10^8$ K.  The corresponding
maximum burst rate is about 5 bursts/day.

\subsection{Burst Fluence}
\label{sec:results-fluence}

\begin{figure}
\plotone{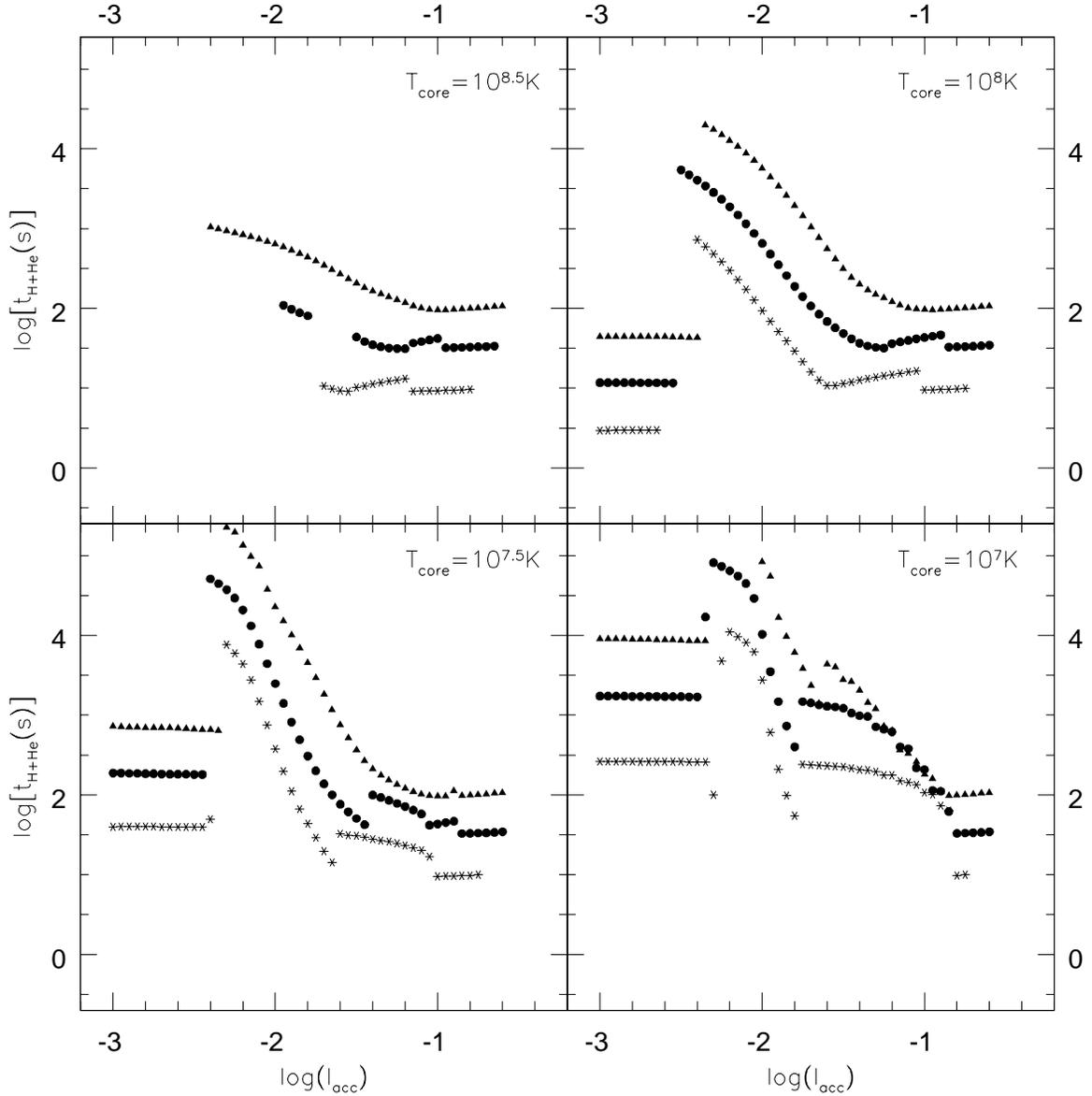}
\caption{Effective burst duration, calculated by dividing the total
burst fluence by the Eddington luminosity.  Triangles, circles and
stars correspond to neutron star radii of 16.4~km, 10.4 km and 6.5 km
respectively.  Notice that, at the extreme right of the various
panels, where we have the regime of delayed mixed bursts, the burst
duration is either level or falls relative to the prompt mixed bursts,
even though the recurrence time $t_{\rm rec}$ (Fig. \ref{fig:trec})
and the layer thickness $\Sigma_{\rm layer,crit}$
(Fig. \ref{fig:sigma}) increase substantially.  This indicates that
there is considerable steady burning of nuclear fuel in this regime.}
\label{fig:bfluences}
\end{figure}

Probably the most easily observed property of bursts is the total
fluence, $E_{\rm H}+E_{\rm He}$, resulting from burning the fuel in
the accreted layer.  We assume that all of the unburnt hydrogen and
helium in the accreted layer is fully burned to iron in the burst and
radiated during the burst (neglecting any flux going into the star or
any energy loss through neutrino emission).  We then divide the burst
fluence by the Eddington luminosity to calculate an effective burst
duration:
\begin{equation}
t_{\rm H+He} = {E_{\rm H}+E_{\rm He} \over L_{\rm Edd}}.
\label{tburst}
\end{equation}
This quantity is plotted in Fig.~\ref{fig:bfluences} for our standard
three neutron star radii and four core temperatures.  As in the case
of the recurrence times discussed above, the calculated values are
upper limits because of the assumption of complete burning of the
fuel.

\begin{figure}
\plotone{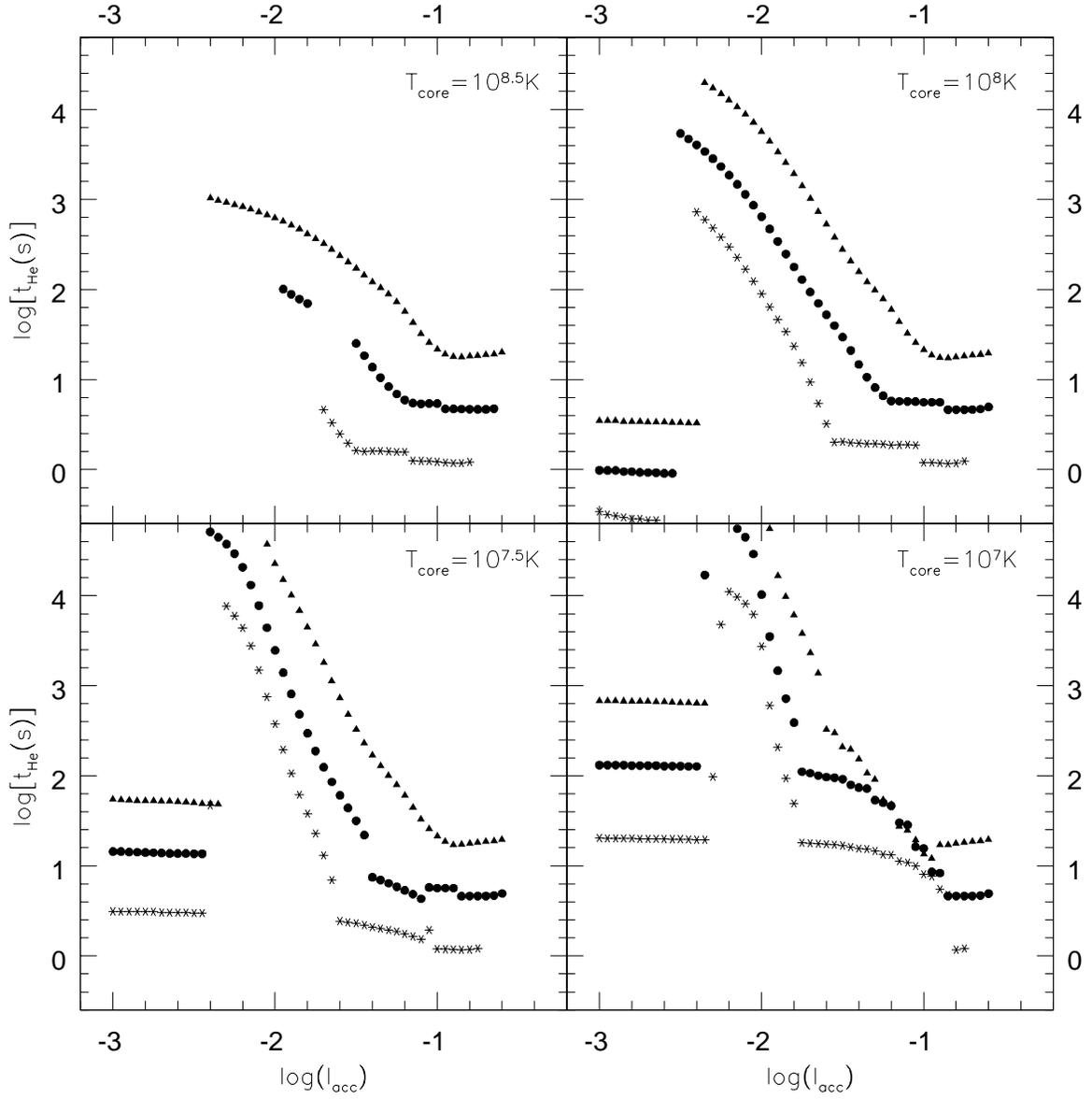}
\caption{Similar to Fig. \ref{fig:bfluences}, but for only helium
burning.}
\label{fig:hefluences}
\end{figure}

Since hydrogen burning is typically beta-limited during the burst,
the hydrogen fluence may in some cases emerge as a long plateau
following the initial spike of helium burning (which is not
beta-limited).  Therefore, another possibly useful measure of burst
duration is the effective duration $t_{\rm He}$ of helium burning
alone, calculated again as fluence divided by the Eddington
luminosity.  This quantity is plotted in Fig.~\ref{fig:hefluences}.

Consider first the regime of helium bursts, which corresponds to the
segments of $t_{\rm H+He}$ and $t_{\rm He}$ with steep negative slopes
in the range $\log(l_{\rm acc})\sim-1.5$ to $-2.5$.  These are delayed
bursts in which the hydrogen burns stably at a shallow depth, while
the helium accumulates over a fairly large column before the system
goes unstable.  The bursts are very much dominated by helium burning
(see Fig. 5), and so $t_{\rm H+He}$ and $t_{\rm He}$ are nearly equal.
Also, the fluence is directly proportional to the accretion column
$\Sigma_{\rm layer}$, as can be seen by comparison with
Fig.~\ref{fig:sigma}.

There are two segments of prompt bursts, one at higher luminosities
and the other at lower luminosities, on either side of the helium
bursts.  Both involve significant unburnt hydrogen and helium, and
therefore for both $t_{\rm H+He}$ is significantly greater than
$t_{\rm He}$ (since hydrogen burning releases more energy per gram, by
a factor of about 5, compared to helium burning).  Both burst
durations are roughly proportional to $\Sigma_{\rm layer}$ as can be
seen by comparison with Fig.~\ref{fig:sigma}.

Finally, the delayed mixed bursts, which are present for $\log(l_{\rm
acc}) > -1$, show a different behavior.  These are systems in which
considerable stable burning of nuclear fuel occurs in the accreted
layer before the layer actually bursts.  Therefore, the amount of
nuclear fuel available to power the burst is much less than one might
expect for the given column depth.  The burst durations are at most
comparable to, and are often less than, the durations of adjacent
prompt bursts, even though these delayed bursts have larger values of
$\Sigma_{\rm layer}$ and correspondingly much longer recurrence times
(compare Figs. \ref{fig:bfluences}, \ref{fig:hefluences} with
Fig. \ref{fig:sigma}).

From Figs.~\ref{fig:bfluences}, \ref{fig:hefluences} we see that the
typical burst durations are predicted to be a few seconds to few tens
of seconds at high mass accretion rates.  However, the duration can be
quite large, an hour or longer, for helium bursts with $\log(l_{\rm
acc}) \lesssim -2$, and as much as a day for cooler core temperatures.

\subsection{Burst $\alpha$}
\label{sec:results-burst-alpha}

The recurrence times and burst durations discussed in the previous
subsections were calculated assuming that the entire fuel reservoir is
consumed in a burst.  However, this assumption may not be valid
\citep[e.g.][]{1982ApJ...258..761T,1987ApJ...319..902F}, especially
for cold neutron stars \citep{1993ApJ...413..324T}.  It is, therefore,
useful to consider the ratio $\alpha$ of the nuclear-burning energy
that is emitted during the burst to the total energy released between
bursts by accretion.  In terms of quantities we have introduced
earlier, $\alpha$ is given by
\begin{equation}
\alpha \equiv {t_{\rm rec}L_{\rm acc}\over E_{\rm H}+E_{\rm He}} =
{t_{\rm rec}\over t_{\rm H+He}} l_{\rm acc}.
\label{alpha}
\end{equation}
The quantity $\alpha$ should be independent of what fraction of the
fuel is burned in bursts; if only a fraction of the fuel is burned,
then both $t_{\rm rec}$ and $t_{\rm H+He}$ will decrease by the same
fraction and $\alpha$ ought to remain the same.  This makes it a
particularly useful parameter.

\begin{figure}
\plotone{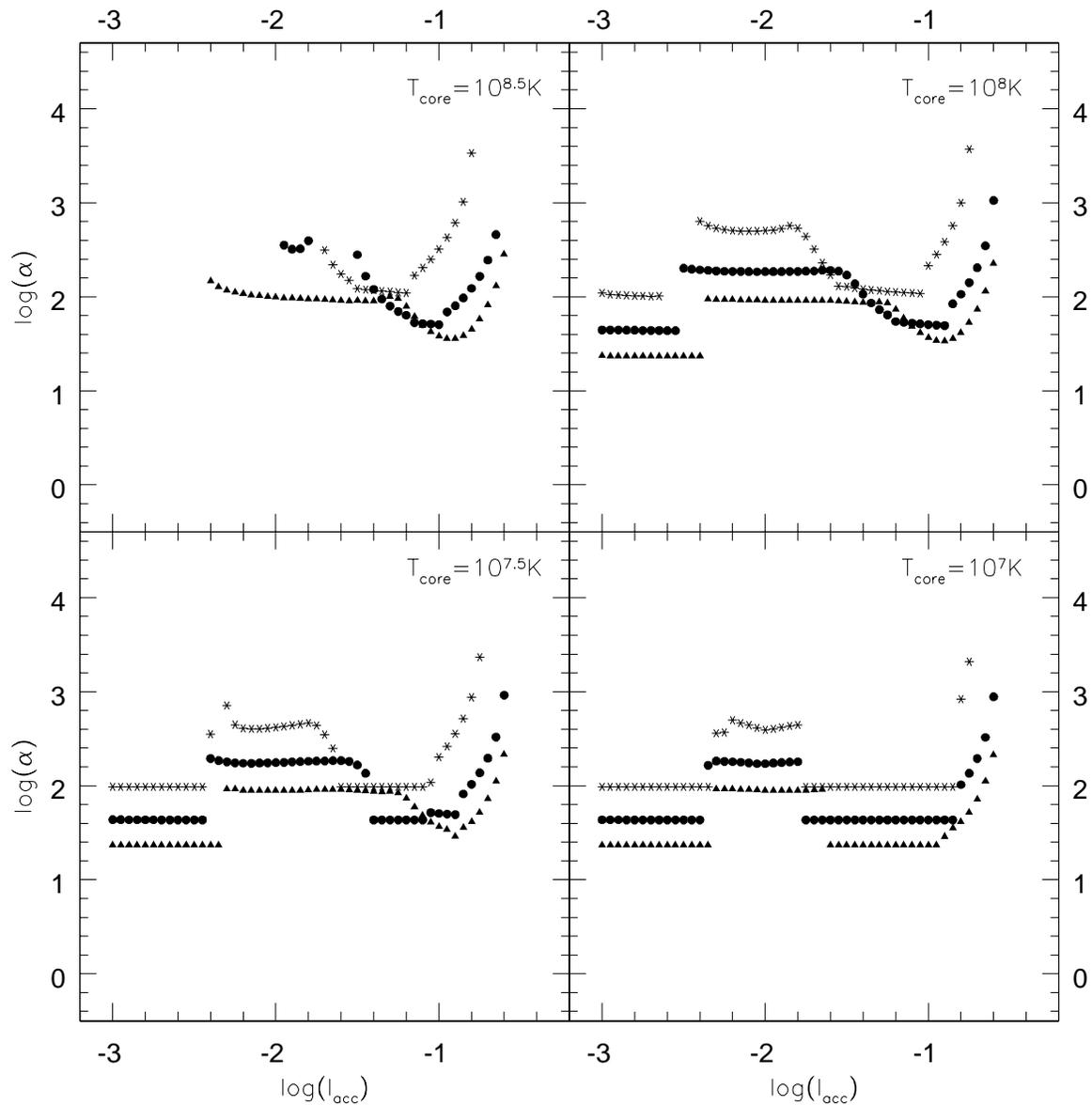}
\caption{Dependence of the parameter $\alpha$, the ratio of the energy
generated by accretion between bursts to the nuclear energy emitted
during bursts.  Triangles, circles and stars correspond to neutron
star radii of 16.4 km, 10.4 km and 6.5 km, respectively.  Notice the
steep rise of $\alpha$ at high accretion luminosities, near the stable
burning limit.  The rise is characteristic of delayed mixed bursts.}
\label{fig:alpha}
\end{figure}

Fig.~\ref{fig:alpha} shows $\alpha$ from our models for several
neutron star radii and core temperatures, and a range of accretion
rates.  Two trends are obvious.  First, more compact neutron stars
have larger values of $\alpha$ compared to less compact stars; the
former release more gravitational energy per unit accreted mass
compared to the latter, whereas both release roughly the same amount
of nuclear energy in bursts.  Second, prompt bursts tend to have
smaller values of $\alpha$ compared to delayed bursts.  Generally,
prompt bursts have a lot of unburnt hydrogen available.  This gives a
larger burst fluence per unit mass and thus a smaller $\alpha$.  The
delayed helium bursts have very little hydrogen.  Since helium burning
releases a factor of about 5 less energy per gram, the $\alpha$ of
these bursts is larger by about this factor.  The increase of $\alpha$
for the delayed mixed bursts at $\log(l_{\rm acc}) > -1$ is for a
different reason.  Here, the accreting layer has both hydrogen and
helium, but both fuels are burned substantially before the burst is
initiated.  This is the reason for the upturn of $\alpha$ at the
highest accretion rates in all the panels in Fig.~\ref{fig:alpha}.
This trend has not been noted in previous work.

\subsection{Constraints on Neutrino Cooling}
\label{sec:constr-neutr-cool}

If we assume that the neutron star is in thermal equilibrium as it
accretes we may estimate the temperature of its core as a function of
its mean accretion rate.  There are two contributions to energy
flowing into the core.  The first is the inward flux at the inner
boundary of the accreted layer and the substrate; this may be either
positive or negative (see Brown 2000).  The second is the energy of
released by nuclear reactions deep in the crust
\citep{1998ApJ...504L..95B}, typically $\sim 1$~MeV/baryon.  Our
calculations show that the former contribution is much smaller than
the latter; this is consistent with the discussion in Cumming \&
Bildsten (2000) where the authors estimate that, out of the 1 MeV per
baryon released by deep crustal nuclear reactions, only about 150 keV
per baryon escapes to the surface.  Therefore, for an approximate
estimate of the core temperature, it is reasonable to assume that the
flux flowing into the core is equal to 1 MeV/baryon (see Brown 2000
for a more detailed discussion including the effect of different
assumptions on the thermal conductivity).  If we further assume the
simple estimates of the neutrino emissivity given in \citet{Shap83},
we may estimate the equilibrium temperature of the core for a given
accretion rate.  For the modified URCA process we find
\begin{equation}
\left ( \frac{T_\rmscr{core}}{10^8~\rmmat{K}} \right )^8 = 2500
\frac{E_\rmscr{DH}}{1 \rmmat{MeV/baryon}} \frac{(1+z)^2}{z}
\frac{{\dot M}}{{\dot M}_\rmscr{Edd}} \left (
\frac{\rho}{\rho_\rmscr{nuc}} \right )^{1/3},
\label{eq:mUrca}
\end{equation}
which gives a typical temperature in the range $(1-3)\times10^8$ K.
If the neutron star cools via the direct URCA process from (say) a
pion core, then the core temperature is
\begin{equation}
\left ( \frac{T_\rmscr{core}}{10^7~\rmmat{K}} \right )^6 = 9
\frac{E_\rmscr{DH}}{1 \rmmat{MeV/baryon}} \frac{(1+z)^2}{z}
\frac{{\dot M}}{{\dot M}_\rmscr{Edd}} \frac{\rho}{\rho_\rmscr{nuc}}
\theta^{-2},
\label{eq:pUrca}
\end{equation}
where $\theta \sim 0.3$.  In this case, the core is significantly
cooler, $T_{\rm core}\sim10^7$ K.  In both cases, the core temperature
depends only weakly on the mean accretion rate (as ${\dot M}^{1/8}$
and ${\dot M}^{1/6}$, respectively) and on the exact flux flowing into
the core (whether it is 1 MeV per baryon or a little more or less.)

An important caveat is that our results for cool neutron stars with
$T_\rmscr{core} \sim 10^7$~K depend somewhat sensitively on how we
treat the inner boundary condition (see
\S\ref{sec:boundary-conditions}), so the bursting behavior of stars
with direct URCA cooling may not be precisely as depicted by our
results for $10^7$~K; however, it should be noticeably different from
that of stars with modified URCA cooling.

\section{Discussion}
\label{sec:discussion}

We begin in \S\ref{sec:sensitivity} with a brief summary of how the
results depend on the input physics.  We follow that in
\S\ref{sec:thermalmodes} with a discussion of different kinds of modes
present in our model and use this to explain the occurrence of delayed
mixed bursts.  In \S\ref{sec:comp-earl-theor}, we discuss how our
methods compare with the many previous theoretical analyses of Type I
bursts in neutron stars.  We then present in
\S\ref{sec:comp-with-observ} a preliminary comparison of our
theoretical predictions with observations, making extensive use of the
review of EXOSAT data published by \citet{1988MNRAS.233..437V}.  We
defer a detailed comparision with more recent observations with the
Rossi X-ray Timing Explorer and BeppoSAX to a subsequent paper, but
briefly touch on the work of \citet{cor03}.  (The latter paper was
posted after our paper was submitted to the journal.  Since their
results are very relevant for this work, we have included some
discussion.)  Finally, in \S\ref{sec:future-prospects}, we discuss
some improvements to this work that may be worth pursuing.

\subsection{Sensitivity to Input Physics}
\label{sec:sensitivity}

In the course of doing this work, we changed several prescriptions for
the input physics, usually starting with a simple approximation and
graduating later to more realistic prescriptions.  By monitoring how
the results changed we have developed a sense of which aspects of the
input physics are most important if one is interested in accurate
results.

In the equation of state, the only complication is the electron
pressure.  Before settling on the quadrature prescription given in
equation (\ref{pressure}), which is taken from
\citet{1983ApJ...267..315P}, we tried a simpler prescription in which
we wrote the electron pressure as the straight sum of $P_{\rm e,nd}$
and $P_{\rm e,d}$.  The differences in the results are insignificant.
Similarly, we tried modeling the nuclear reactions both with and
without screening.  There was little difference, except perhaps at the
lowest accretion rates, $\log(l_{\rm acc})\sim-3$, where the hydrogen
bursts showed some small variations.

The effect of changing the opacity prescriptions is more serious.  For
instance, the opacities of \citet{Hern84b}, which we tried first, are
generally lower than those of \citet{1999A&A...351..787P}.  Therefore,
with the former, a thicker layer must accumulate before the layer
becomes unstable.  This effect is strongest for the hydrogen bursts,
and yields longer recurrence times and larger fluences.  Also, because
the burning layer is more poorly insulated from the core, the results
tend to be more sensitive to the core temperature than when we use the
modern conductive opacities of \citet{1999A&A...351..787P}.  As a
result, with the Hernquist \& Applegate (1984) opacities we find that
models with $T_{\rm core}=10^{7.5}$ K are more similar to models with
$T_{\rm core} =10^7$ K, whereas with the Potekhin (1999) opacities,
they are more similar to $10^8$ K.

The results are even more sensitive to the radiative opacity
prescription.  Prior to settling on the \citet{1999ApJ...524.1014S}
prescription, we employed the simple fitting function of
\citet{1975ApJ...196..525I}.  The results were qualitatively the same
as the ones presented here, except that the boundaries between the
different regimes moved towards lower values of $\log(l_{\rm acc})$ by
about 0.2.  That is, the patterns in Figs. 5, 10--15 were shifted to
the left by this amount, but not altered very much in shape.  In view
of this, it may be worthwhile to incorporate an even better
approximation for the radiative opacity in future calcualtions.

The frequencies and structure of the unstable modes are not changed
significantly by the opacity with the exception that the modes
penetrate deeper into the substrate for lower opacities.

\begin{figure}
\plottwo{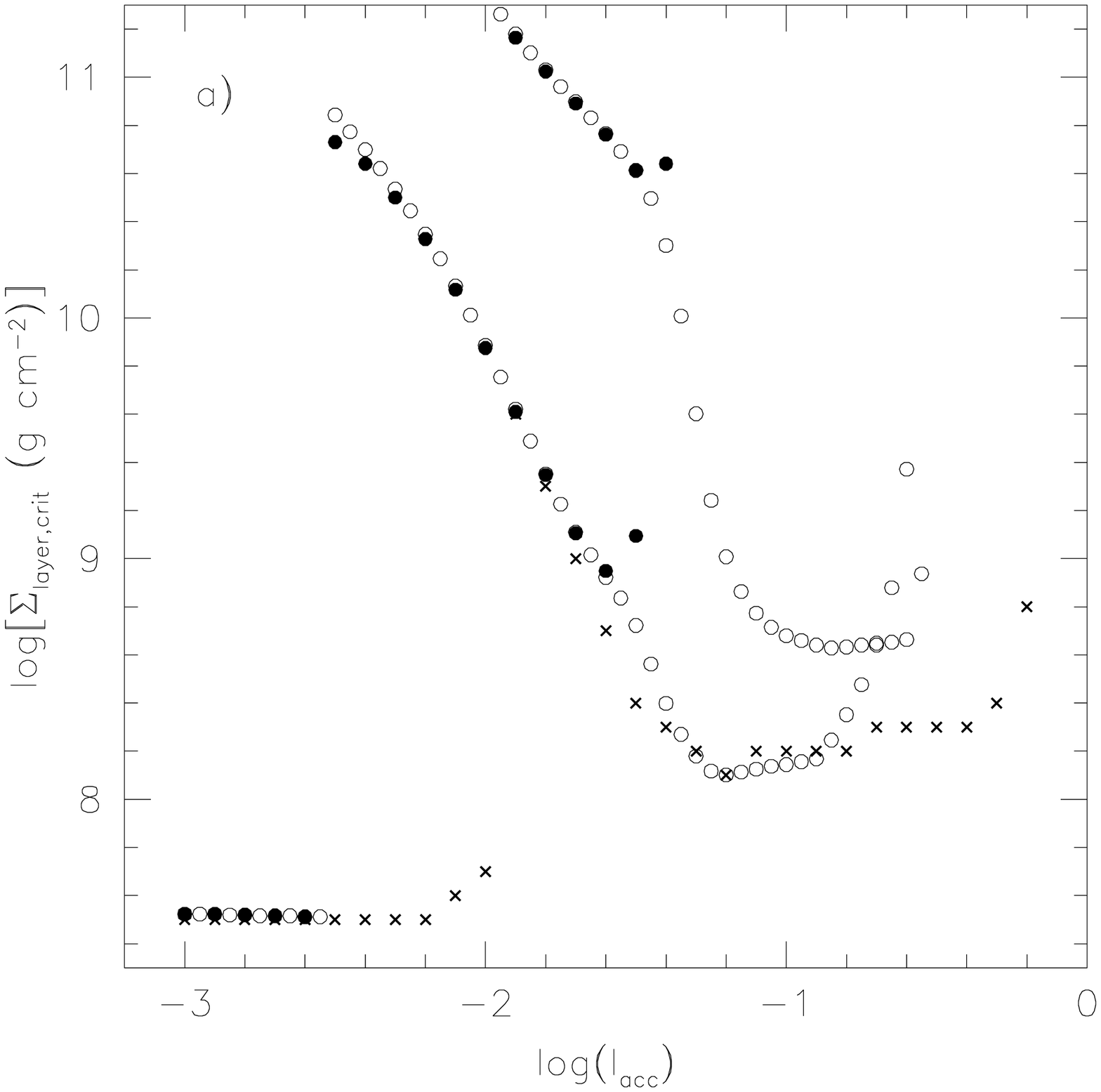}{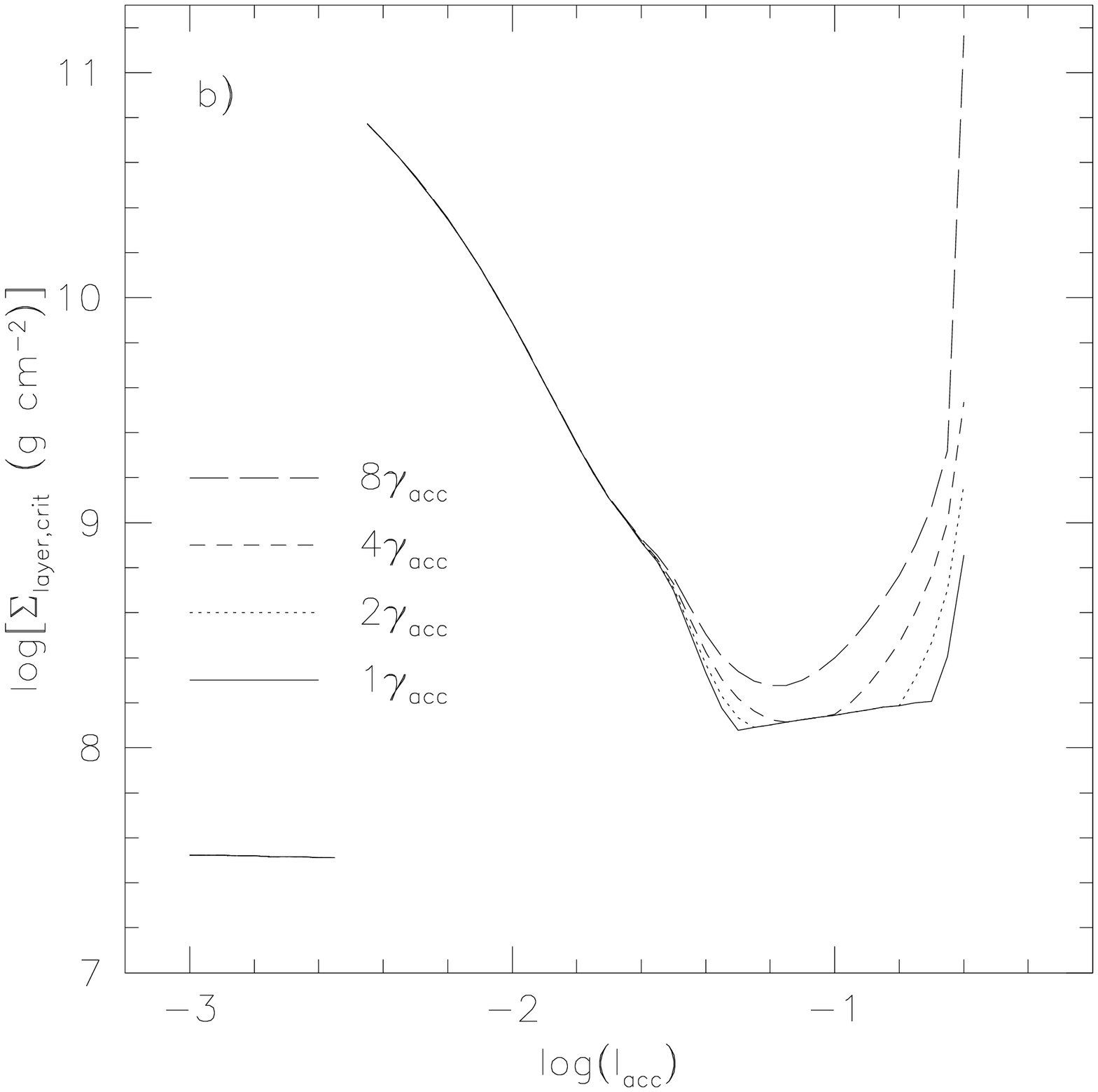}
\caption{Results for $T_{\rm core}=10^8$~K, $R=10.4$ km,
$M=1.4$M$_\odot$.  (a) Shows how the restriction to thermal modes
(filled circles) affects the results.  Compared to the full
instability calculation (open circles), only a fraction of the range
of $l_{\rm acc}$ is found to be unstable for purely thermal
perturbations.  The results depicted by crosses use the instability
criterion of \citet{1987ApJ...323L..55F}.  The lower set of circles
corresponds to our standard boundary condition in which the
temperature boundary condition is applied inside the star at a
diffusion depth.  The upper set of circles depicts the results if the
temperature boundary condition is applied at the bottom of the fuel
layer [as in \citet{Nara01typeibh} and \citet{1987ApJ...323L..55F}].
(b) Shows how the choice of $g_\rmscr{mode}$ affects the onset of
instability.  The larger the value of $g_{\rm mode}$, the more the
nuclear modes near the right end of the plot are delayed.  The thermal
modes on the left are hardly affected.}
\label{fig:gammacrit}
\end{figure}

Finally, we have found that the inner boundary condition plays an
important role.  Figure~\ref{fig:gammacrit}(a) shows two sets of
calculations, both of which set the temperature equal to $T_{\rm
core}$ at the inner boundary.  In one case, the matching is done
inside the substrate at a depth $\Sigma_{\rm diff}$ determined by a
diffusion criterion (see \S\ref{sec:boundary-conditions} and
eq. \ref{ibc1} for details), while in the other the boundary condition
is set at $\Sigma_{\rm layer}$, the bottom of the accreted layer.
There is an enormous difference in the results.  In the former case,
which is more physical, the temperature of the accreted layer tends to
remain high and the temperature profile relaxes to $T_{\rm core}$ only
well inside the star.  In the latter case, however, the boundary
condition forces the gas layer to be cool near the bottom and this
causes the rest of the layer to be cooler than in the previous case.
The lower temperature delays the onset of nuclear burning, causing the
critical surface density for instability to go up significantly.

We feel that we have captured a good fraction of the important physics
with the boundary condition that we use.  Nevertheless, there is
probably room for further improvement.

\subsection{Thermal Modes, Nuclear Modes, and Delayed Mixed Bursts}
\label{sec:thermalmodes}

Because the eigenmodes and growth rates are calculated by perturbing
five coupled differential equations, the physics of the modes is not
always apparent.  We make a beginning here by sorting the modes into
two major classes.

To understand the division, we note that the perturbation equations
written in Appendix A involve three time derivatives, one each in the
energy equation (\ref{app10}), the H-evolution equation (\ref{app6})
and the He-evolution equation (\ref{app12}).  The characteristic time
scale of the energy equation is something like the diffusion time,
which is usually quite short, whereas the characteristic time scale of
the other two equations is the accretion time, since this is the time
on which the composition of a parcel of gas changes.  We thus expect
fast-growing modes to mostly involve the the energy equation (plus the
equations of hydrostatic equilibrium and energy transfer which do not
involve time derivatives), while slower modes should involve
significant perturbations in all variables, including $X$ and $Y$.  We
refer to the former as ``thermal modes'' and the latter as ``nuclear
modes.''

It is straightforward to filter out the nuclear modes in the
calculations --- we just switch off perturbations in $X$ and $Y$ by
setting $X(\Sigma)=X_0(\Sigma)$, $Y(\Sigma)=Y_0(\Sigma)$, or
equivalently, $X_1(\Sigma) =Y_1(\Sigma)=0$.
Figure~\ref{fig:gammacrit}(a) shows the results of such a calculation.
Compared to our standard calculation in which all variables are
allowed to vary (open circles), we see that thermal modes (filled
circles) are present only for lower accretion rates, $\log(l_{\rm
acc})<-1.5$, i.e., in the regime of helium bursts and hydrogen bursts,
but not at higher accretion rates where mixed bursts occur.  An
inspection of the mode growth rates confirms that these modes grow on
relatively short time scales; typically, the growth time for thermal
modes is a factor of tens to hundreds shorter than the accretion time
scale.  The modes above $\log(l_{\rm acc})\sim-1.5$, which are present
only when all variables are allowed to vary, are the nuclear modes in
our classification.  These modes have slow time scales, and their
eigenfunctions involve important fluctuations in composition, which is
not the case with the thermal modes.  The eigenfunction shown in
Figs. 8, 9 is a nuclear mode.

The identification of the slow nuclear modes provides a natural
explanation for the category of delayed mixed bursts that we have
newly identified in this paper.  Recall that we do not consider a
system to be unstable unless the growth rate $\Re(\gamma)$ is greater
than $g_{\rm mode}\gamma_{\rm acc}$ (see eq.~\ref{eq:gammacrit}).  For
our nominal choice of $g_{\rm mode}=3$, the thermal modes are not in
the least affected by the value of $g_{\rm mode}$, since their growth
rates are typically much greater than the limit.  However, the slower
nuclear modes are strongly influenced by the criterion, and the effect
is particularly severe at higher accretion rates.  The right panel of
Figure~\ref{fig:gammacrit} shows the critical column for bursts
$\Sigma_{\rm layer,crit}$ as a function of $\log(l_{\rm acc})$ for
different choices of $g_{\rm mode}$.  The delayed burst regime becomes
more and more prominent as $g_{\rm mode}$ increases.  The calculations
show that, for high accretion rates, the accreted layer is marginally
unstable already at small values of $\Sigma_{\rm layer}$.  However,
the growth rate is low, and the instability takes a while to grow.  By
the time the mode has grown enough to produce a burst, $\Sigma_{\rm
layer}$ is significantly larger; this is the reason for the delay in
the burst.  In addition, because of the delay, much of the fuel is
burned stably prior to the burst, and the amount of unburnt fuel
available for the burst is reduced.  This causes a dramatic increase
of $\alpha$ in this regime.  The extent of these effects depends on
the choice of $g_{\rm mode}$.  The value we have used for the results
presented in \S5, $g_{\rm mode}=3$, are in our opinion reasonable.

\subsection{Comparison with Earlier Theoretical Work}
\label{sec:comp-earl-theor}

Previous investigations have identified the various burning regimes
that we have found here
\citep{1981ApJ...247..267F,1987ApJ...323L..55F,2000arxt.confE..65B},
with the exception of the regime of delayed mixed bursts (see
\S\ref{sec:together}, Tables \ref{tab:burst-regimes1},
\ref{tab:burst-regimes2}, Fig. \ref{fig:regime}).  The ranges of
$l_{\rm acc}$ of the different regimes obtained by these workers
generally agree with our results, though we have found several new
features, as discussed in previous sections.  Our method, being more
rigorous, promises to provide better quantitative predictions for
comparison with observations.  Also, we calculate mode frequencies,
which enables us to identify overstable modes and to estimate the
oscillation periods.  It is unclear if this will have clear
observational signatures, but it is a topic worthy of further
investigation.

As described in \S1, our method is similar in spirit to previous
investigations that evaluated steady-state configurations and then
studied the stability of the equilibria to small perturbations
\citep[for example][]{1975ApJ...195..735H,
1980A&A....84..123E,1981ApJ...247..267F,1987ApJ...323L..55F,
2000ApJ...544..453C}.  The computation of the equilibria in the
various studies probably do not differ a great deal since the physics
is basically well understood (\S\ref{sec:model}).  The approximations
that we and others use (\S\ref{sec:quant}) appear to be harmless,
except perhaps the opacity which does make a difference
(\S\ref{sec:sensitivity}).  Once one has calculated a sequence of
equilibria such as those shown in Figs. \ref{fig:scurve1},
\ref{fig:scurve2}, it is necessary to determine at what $\Sigma_{\rm
layer,crit}$, if any, the accreted layer becomes unstable.  It is at
this stage that the major differences in methods appears.

The technique that we have used in this paper is new and avoids any
prejudices as to which processes are important and which are not for
triggering a burst.  We carry out a full linear stability analysis of
the equilibrium solution by solving for the eigenmodes and their
complex eigenfrequencies.  In particular, the regime of delayed mixed
bursts that we have found (\S\ref{sec:together}, Tables
\ref{tab:burst-regimes1}, \ref{tab:burst-regimes2}), and the
identification of nuclear modes (\S\ref{sec:thermalmodes}), are
entirely the result of the more rigorous formalism we employ.  As we
shall see below, the delayed mixed burst regime may help to explain
some puzzling observations.  Also, none of the previous studies was
able to identify whether the growing mode is a simple instability or
an overstability.  This too may have some observational implications.

Earlier works have generally used various local or approximately
global criteria, calculated at or near the bottom of the layer, to
determine the stability of the accreted layer.  With the exception of
\citet{1987ApJ...323L..55F}, no one to our knowledge has discussed
global perturbations of the steady-state configuration.  Even the
Fushiki \& Lamb (1987a) study was relatively crude since it assumed a
constant temperature perturbation as a function of depth and did not
treat perturbations in the substrate below the layer.  In contrast,
our approach involves a computation of the entire eigenfunction from
first principles, without any preconceptions as to the shape of the
mode, and allows for perturbations in the substrate.  It should be
emphasized that we consider perturbations in all variables, including
the composition parameters $X$, $Y$ and $Z$ ($=1-X-Y$).  This enables
us to find both thermal modes and nuclear modes
(\S\ref{sec:thermalmodes}).  In contrast, all previous studies
restricted themselves to thermal perturbations, that too
approximately, and thus were sensitive only to thermal modes.

To get some sense of how good the approximate criteria of the past
are, we present here some quantitative comparisons with two methods
described in the literature, those of \citet{1987ApJ...323L..55F} and
\citet{2000ApJ...544..453C}.  These papers use two different criteria
for the onset of instability.  Both express the criterion as
\begin{equation}
{d\epsilon_\rmscr{heat}\over d T} > {d \epsilon_\rmscr{cool}
\over d T},
\label{eq:criterion}
\end{equation}
but differ in their definitions of the quantities.
\citet{1987ApJ...323L..55F} define $\epsilon_\rmscr{heat}$ to be the
total nuclear energy generation rate while \citet{2000ApJ...544..453C}
define it to be 1.9 times the energy generation rate from the
triple-alpha reacton.  In order to include the hydrogen bursts in the
latter calculation we generalized it by adding also the hydrogen
energy generation rate.  The differences in the cooling rates in the
two approaches are more striking.  \citet{1987ApJ...323L..55F} asume
that $dT_1/d\Sigma=0$ and use a formula that is equivalent to (this is
our best guess since their paper does not explain in sufficient
detail)
\begin{equation}
\frac{d \epsilon_\rmscr{cool}}{d T} 
={d\over dT} \left({dF\over d\Sigma}\right)
= \frac{d \ln K}{d T} \epsilon -
	\frac{d^2 \ln K}{d T d\Sigma} F,
\label{eq:flcrit}
\end{equation}
where $\epsilon$ is the total energy generation rate and $K=16\sigma
T^3 /3\kappa$ is the thermal conductivity (see eq. \ref{eq:fulleqs2}).
\citet{2000ApJ...544..453C}, on the other hand, start from the work of
Fujimoto et al. (1981) and write approximately
\begin{equation}
{d\epsilon_\rmscr{cool}\over dT} =
{d\over dT} \left( \frac{a c T^4}{3 \kappa \Sigma^2} \right),
\label{eq:cbcrit}
\end{equation}
where the quantity on the right is roughly the thermal energy density
divided by the diffusion time to the surface.  Clearly both criteria
are approximate.  Both are also local since they focus on values at a
particular depth.  The Fushiki \& Lamb (1987a) method could be
generalized to an approximate global criterion (it is possible that
the authors did make such a generalization, but we have not been able
to understand exactly how they might have done it).  We limit
ourselves to the local version given above.  The Cumming \& Bildsten
(2000) criterion is of a mixed kind, since it compares a local
quantity for the heating rate to a pseudo-global quantity for the
cooling rate.

\begin{figure}
\plotone{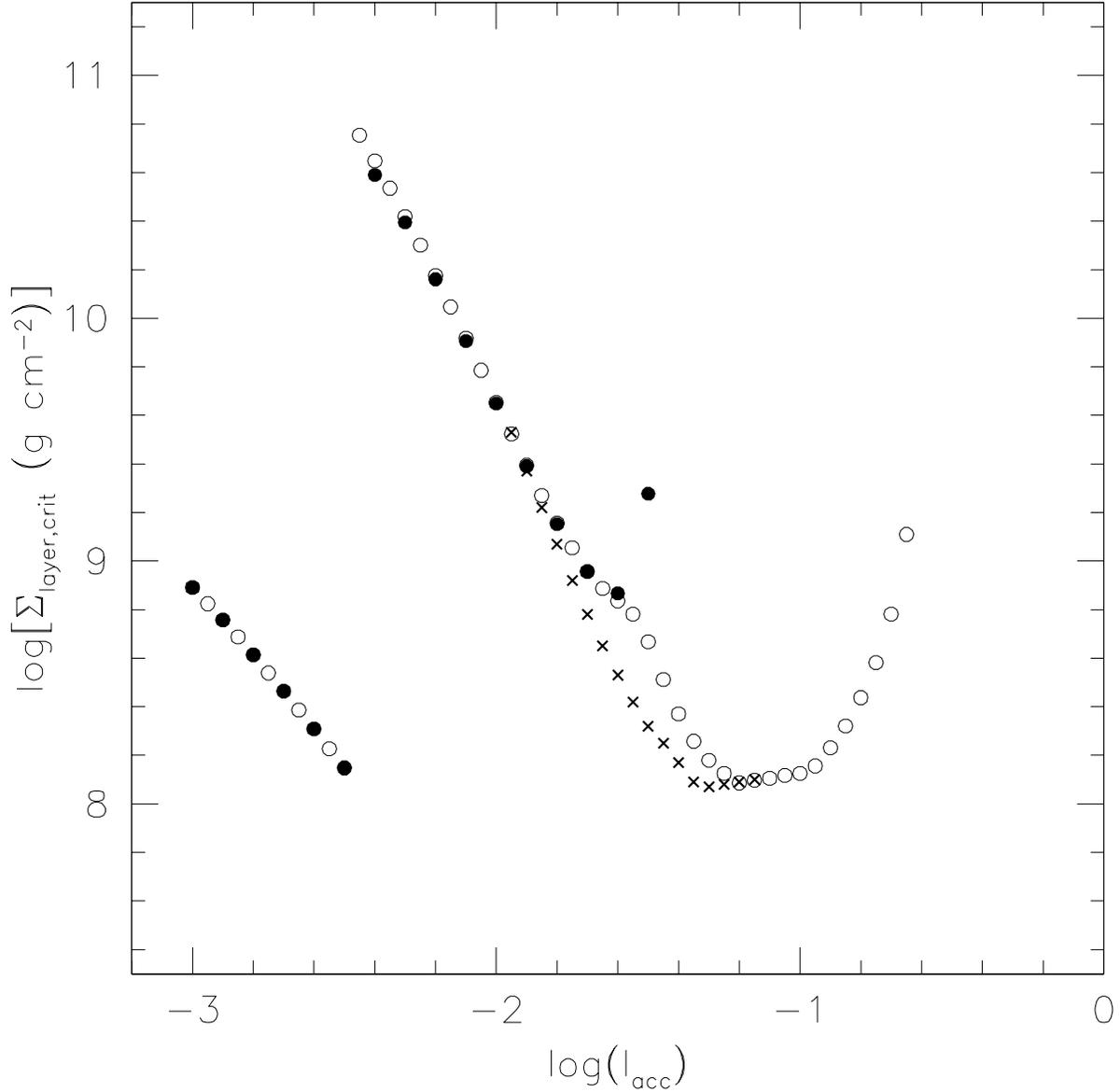}
\caption{Models for $\log R/R_s = 0.4$ and $M=1.4$M$_\odot$ with a
flux inner boundary condition.  It is assumed that a constant flux of
150 keV per accreted baryon flows out of the star into the accretion
layer.  The open circles correspond to the full calculation where all
variables are perturbed, and the filled circles to a calculation in
which the composition is not allowed to vary (thermal modes).  The
crosses trace the results corresponding to the instability criterion
of \citet{2000ApJ...544..453C}.}
\label{fig:fluxbc}
\end{figure}

We use the equilibria as computed with our code using our particular
prescriptions for opacity, equation of state, etc.
(\S\ref{sec:quant}) for all the calculations.  We thereby ensure a
fair comparison of the different methods.  We consider the
calculations done with our full linear perturbation analysis as the
``correct'' answer.  Indeed, since the two approximate criteria
described above are both restricted to thermal perturbations, we feel
that the ``correct'' answer is the set of {\it thermal} modes
calculated with our perturbation analysis (filled circles in
Figs. \ref{fig:gammacrit} and \ref{fig:fluxbc}).

Figure \ref{fig:gammacrit}(a) shows results corresponding to a
temperature boundary condition, with the crosses depicting the results
obtained with the Fushiki \& Lamb (1987a) criterion.  For the latter,
we applied the local criterion (\ref{eq:criterion}), (\ref{eq:flcrit})
at the temperature maximum in the accreted layer.  (By choosing this
point rather than the bottom of the layer, we feel that the method has
a better chance of approximating the true global problem.)  The
agreement between the approximate criterion and our full calculations
(thermal plus nuclear modes) is quite good.  The Fushiki \& Lamb
(1987a) criterion finds a smaller range of $l_{\rm acc}$ for the
helium bursts and overestimates the range of the hydrogen bursts.
Also, not surprisingly, it does not find the delayed mixed burst
regime.  The most surprising result is that the method, which is
overtly limited to thermal perturbations, finds modes well outside the
range of thermal modes.  This may be viewed either as a serious error
in the method or as an unexpected strength!

Figure \ref{fig:fluxbc} shows the corresponding comparison for the
Cumming \& Bildsten (2000) criterion.  Since these authors use a flux
inner boundary condition, we calculated the equilibria as well as our
stability results for this boundary condition, assuming an outward
flux of 150 keV per accreted baryon at the bottom of the accretion
layer.  In the stability analysis we assumed that the perturbed flux
is zero at the bottom of the layer (which corresponds to a non-zero
temperature perturbation).  Once again, we should view the thermal
modes (filled circles) as the ``correct'' answers.  The crosses show
the range of instability according to the criterion
(\ref{eq:criterion}), (\ref{eq:cbcrit}) applied at the bottom of the
accreted layer (where the temperature reaches a maximum).  Wherever
the Cumming \& Bildsten (2000) criterion indicates that there is an
instability, the agreement in the value of $\Sigma_{\rm layer, crit}$
is fairly good.  However, the method appears to miss finding an
instability over a fairly wide range of $l_{\rm acc}$.  Also,
naturally, the method does not reproduce the delayed mixed bursts
(which we have identified as an effect associated with nuclear modes).

One comment is in order.  In calculating the above results for the
Fushiki \& Lamb (1987a) and Cumming \& Bildsten (2000) criteria, we
followed the philosophy described in \S3, namely to label a system
unstable only if it has unstable configurations on the ``wall'' of the
S-curve for values of $\Sigma_{\rm layer}$ above the hydrogen and
helium peaks.  Thus, a situation like Fig. 2(a) in which the wall
solutions are all stable is considered stable, even though there are
unstable solutions on the trailing slope of the helium peak.  We do
not know if the other authors used this criterion or not, but we
believe it is the correct approach to the problem (for reasons
discussed in \S3).

In summary, the two approximate criteria discussed above are useful
for quick estimates, but are not appropriate for detailed quantitative
results.  For the latter, one needs to do a full linear stability
analysis, as presented in this paper, or carry out time-dependent
simulations of the kind reviewed in \S1.
 
The good news is that, in the regions where both our and previous work
have found instability, the results agree quantitatively quite well.
Specifically, the results presented by \citet{1987ApJ...319..902F}
agree in detail with the equilibrium curve presented in
Fig.~\ref{fig:onetenth_detail}(a).  For low accretion rates and cool
cores, \citet{1986PASJ...38...13H} and \citet{1987ApJ...315..198F}
found very long X-ray bursts with durations similar to those shown in
Fig.~\ref{fig:bfluences}.  We also find good qualitative agreement
with the more comprehensive results of \citet{1987ApJ...323L..55F} and
the different burning regimes that they have identified, and
quantitative agreement with the tabulated column densities and
temperatures in Table 2 of Cumming \& Bildsten (2000).  However, the
overarching nature of the calculations presented here make it
difficult to compare very precisely with earlier work which were
generally focused on understanding specific ignition criteria, rather
than presenting predictions of the gross properties of Type I bursts
over a wide range of accretion rates, temperatures and stellar radii.

\subsection{Comparison with Observations}
\label{sec:comp-with-observ}

\subsubsection{Stable Burning Limit}
\label{sec:stable-burning}

Van Paradijs et al. (1988) have collected together a body of burst
data obtained with the EXOSAT satellite.  They define a quantity
$\gamma$ to be the ratio of the observed persistent flux $F_p$ to the
net peak flux $F_{\rm re}$, not including the persistent emission, of
bursts with radius expansion.  The quantity $F_p$ is proportional to
$L_{\rm acc}$, while $F_{\rm re}$ is proportional to $L_{\rm
Edd}^{'}$, where $L_{\rm Edd}^{'}$ is the Eddington luminosity in the
frame of the neutron star surface and is smaller than our $L_{\rm
Edd}$ (defined at infinity) by a redshift factor.  Thus, $\gamma$ is
larger than $L_{\rm acc}/L_{\rm Edd}$ by a variable factor that may be
$\sim 1.5$.  Van Paradijs et al. (1988) find that bursts occur for
$\log\gamma \lesssim -0.5$; sources with $\gamma$ above this limit
apparently burn fuel stably without bursting.  The observational limit
they have obtained on $\gamma$ should be compared with our prediction
that bursts occur only below $\log(L_{\rm acc}/L_{\rm Edd}) \sim
{-0.7}$ to $-0.6$ (\S\ref{sec:together}).  The agreement is fairly
good.  In a previous paper \citep{Nara01typeibh}, we found that bursts
should occur up to $\log(l_{\rm acc})\sim -0.5$.  However, those
calculations were done with a less sophisticated inner boundary
condition (see \S\ref{sec:boundary-conditions}).

The brightest low-mass X-ray binaries generally do not show burst
activity.  Matsuba et al. (1995) saw Type I bursts from the bright
source GX 13+1 when $\gamma\sim 0.3$, Kuulkers \& van der Klis (2000)
saw a radius-expansion burst from the source GX 3+1 when the
persistent luminosity was $0.17L_{\rm Edd}$, and Kuulkers, van der
Klis \& van Paradijs (1995) and Smale (1998) observed bursts from Cyg
X--2 with $\gamma$ as large as 0.74.  Tournear et al. (2003) have
recently used the USA satellite to carry out long-duration
observations of a number of neutron star systems.  They observed
bursts from sources with a range of $l_{\rm acc}$ extending up to
$\sim 0.3$.  Finally, Cornelisse et al. (2003) find that bursts cease
at a luminosity of $5.5\times10^{37} ~{\rm erg\,s^{-1}}$, which
corresponds again to $0.3L_{\rm Edd}$ (for $M=1.4M_\odot$ and our
definition of the Eddington luminosity).  These observations are all
generally consistent with the van Paradijs et al. (1988) results.  The
stable limit according to our model is encouragingly close to the data,
which is noteworthy since our model is a first-principles calculation
with no free parameters.

\subsubsection{Burst $\alpha$}
\label{sec:burst-alpha}

As mentioned in \S\ref{sec:results-burst-alpha}, both the predicted
recurrence time and the fluence of bursts are subject to an
uncertainty over whether the entire reservoir of fuel is consumed
during the burst.  Only fully time-dependent calculations can predict
what fraction of the fuel is consumed in the burst, and even then
probably only with multi-dimensional simulations.  This is well beyond
the scope of our calculation.  However, partial burning affects the
recurrence time and the burst fluence in an identical fashion, and
therefore their ratio should be insensitive to this uncertainty.  The
parameter $\alpha$ defined in \S\ref{sec:results-burst-alpha} is such
a quantity.

\begin{figure}
\plotone{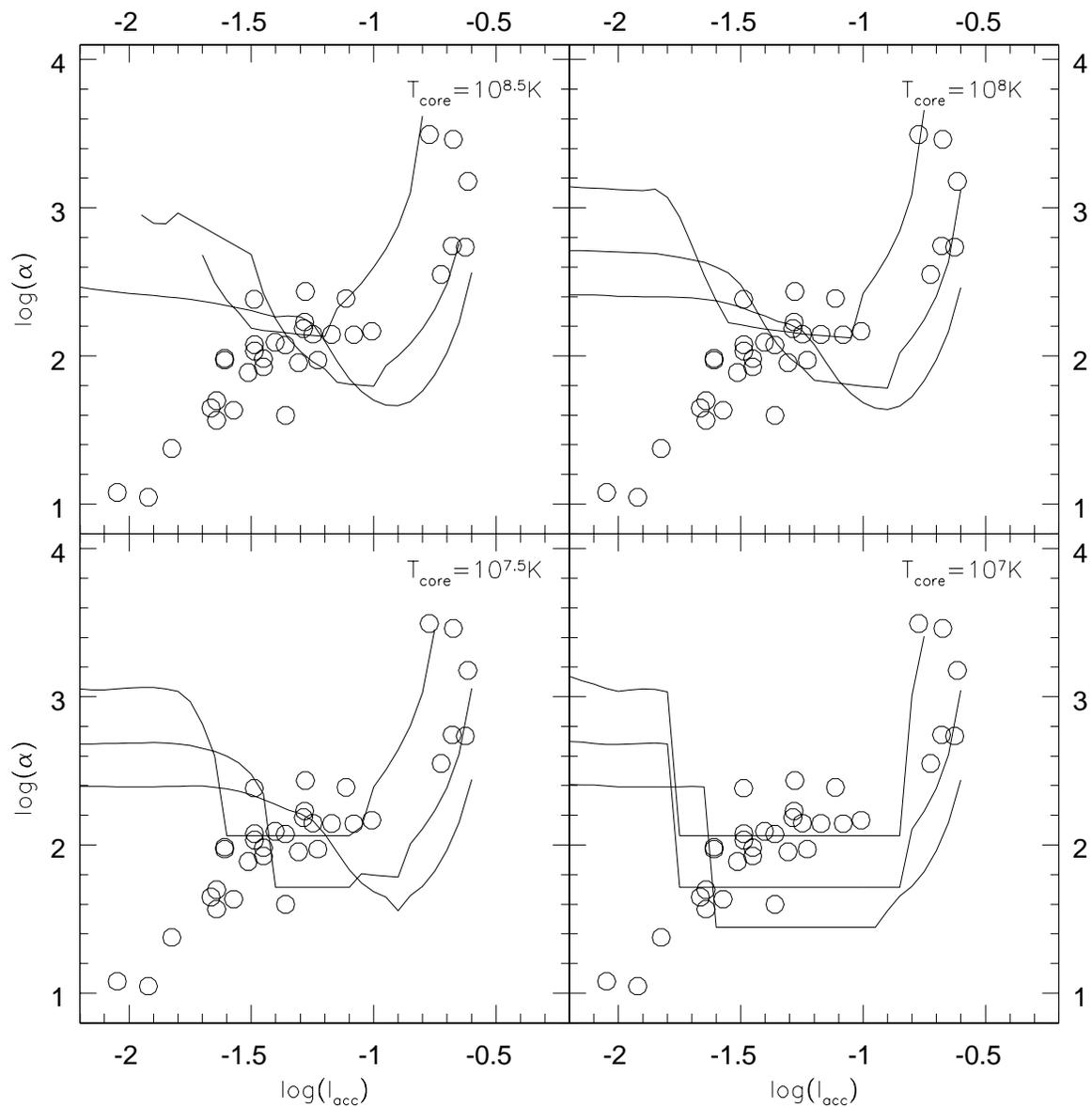}
\caption{A comparison of the model predictions for $\alpha$ with the
observational data presented in \citet{1988MNRAS.233..437V}.  The
circles show the data.  The lines in each panel correspond to the
model results shown in Fig. \ref{fig:alpha}, with the symbols replaced
by connected lines.  In each panel, the uppermost line corresponds to
a neutron star radius of 6.5 km, the middle line to 10.4 km, and the
lowest line to 16.4 km.}
\label{fig:alphadat}
\end{figure}

Van Paradijs et al. (1988) present results for $\alpha$ as a function
of $\gamma$.  Fig.~\ref{fig:alphadat} compares their data with our
predictions.  In plotting the data, we have assumed that the observed
$\gamma$ is equal to $l_{\rm acc}$.  As discussed in
\S\ref{sec:stable-burning}, the two may differ by a few tens of
percent (because of gravitational redshift), so the data may need to
be shifted to the left by this amount.  Allowing for this uncertainty,
we find that the data agree with the model predictions fairly well for
luminosities above $\log(l_{\rm acc})\sim -1.5$.  For accretion
luminosities in the range $\log(l_{\rm acc}) \sim -1.5$ to $-1$, the
data give $\alpha\sim 10^2$.  This value is roughly consistent with
our model predictions for smaller neutron star radii (6.5 km and 10.4
km); the largest radius we have tried (16.4 km) does not agree very
well with the data.

The most interesting feature of the data is that, for higher accretion
luminosities, $\alpha$ shoots up to values $\sim 10^3$.  Van Paradijs
et al. (1988) suggested that the increase is because much of the
nuclear energy in the accumulating fuel is released before the
configuration becomes unstable, i.e. between bursts during accretion.
The regime of increased $\alpha$ in the data very nicely coincides
with the regime of delayed mixed bursts in our models.  Indeed, in
this regime, much of the nuclear fuel does burn steadily, as suggested
by \cite{1988MNRAS.233..437V}, and it is only after a considerable
surface density has accreted that the burst is initiated.
Correspondingly, $\alpha$ goes up by a large factor.  According to our
models, in this regime, the critical $\Sigma_{\rm layer,crit}$ to
initiate a burst increases (Fig. \ref{fig:sigma}), as does the
recurrence time between bursts (Fig. \ref{fig:trec}), whereas the
burst fluence as measured by $t_{\rm H+He}$ is either unchanged or
goes down (Fig. \ref{fig:bfluences}).  All of these features are
consistent with the observations, and this is one of the successes of
the present work.

The sudden ``break'' from values of $\alpha\sim100$ to larger values
occurs at $\log(l_{\rm acc})\sim-1$ in the van Paradijs et al. (1988)
data, though the precise location is hard to determine.  A similar
break has been seen in recent data presented by \citet{cor03} who
quote a break luminosity of $2.1\times10^{37} ~{\rm erg\,s^{-1}}$
(they actually present data for burst rates and burst durations, but
it is straightforward to translate these to $\alpha$).  In our units,
their break occurs at $\log(l_{\rm acc})=-0.9$, which is close to
where we find a break in our models.  This quantitative agreement is
very encouraging.  \citet{cor03} interpret the break in terms of a
transition from unstable to stable hydrogen burning, even though such
a transition is expected to occur at very low luminosities, not at
$\log(l_{\rm acc}) \sim -1$, and it does not predict the particular
behavior seen in the data.  In our model, the break signals the switch
from the regime of prompt mixed bursts to that of delayed mixed
bursts.  Both the position of the break and the signatures we predict
are in encouraging agreement with the observations.

At low accretion luminosities $\sim0.01L_{\rm Edd}$,
\citet{1988MNRAS.233..437V} find very low values of $\alpha\sim10$.
We are unable to reproduce this trend and are not aware of any other
studies that succeed.  Our models predict an increase in $\alpha$ as
we move into the regime of helium bursts (where there is little or no
hydrogen to burn), which is just the opposite of what is seen in the
data.  It would be useful to probe this regime in more detail with
modern observations.

\begin{figure}
\plotone{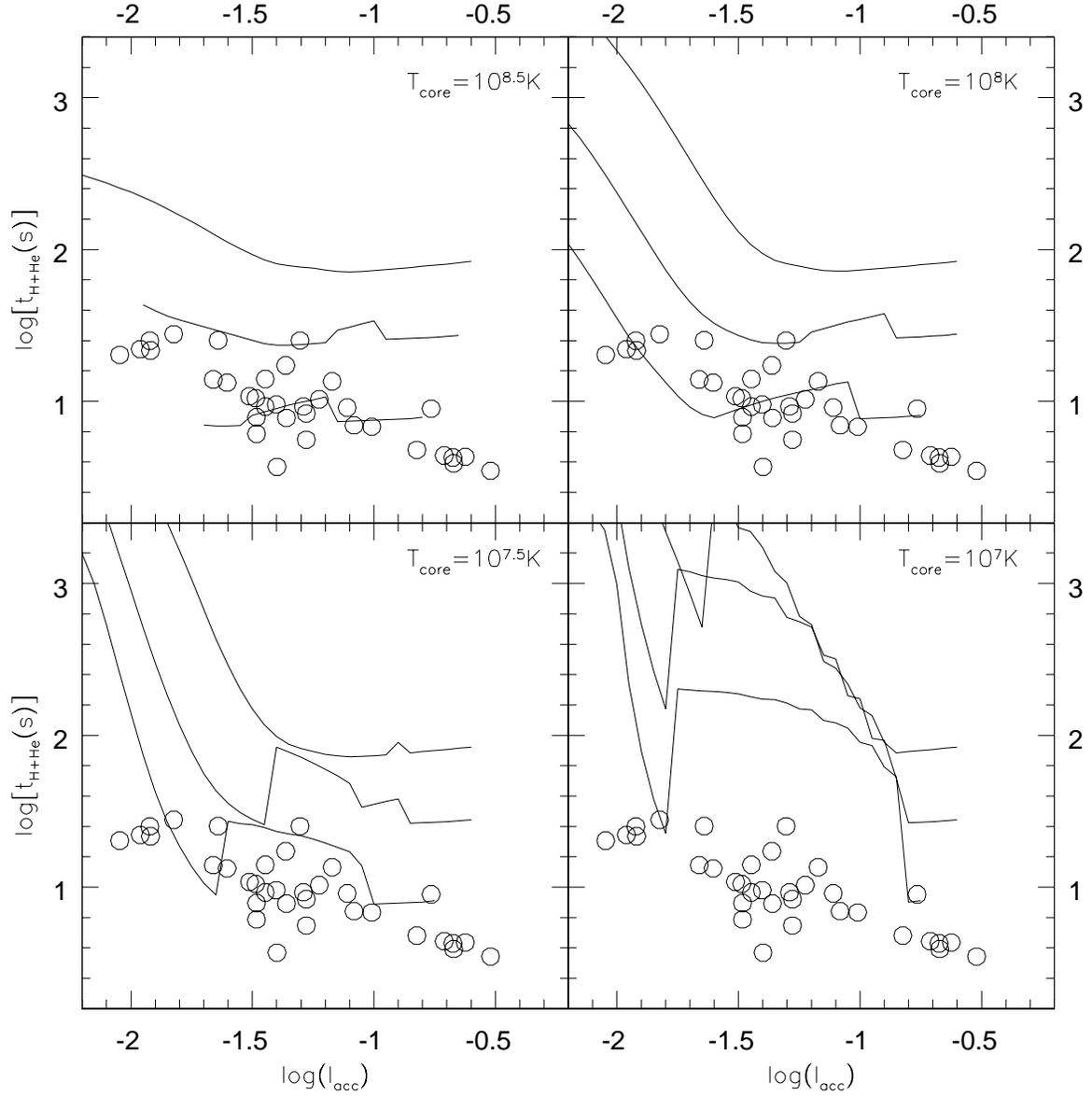}
\caption{A comparison of the model burst durations $t_{\rm H+He}$ with
the corresponding data presented in \citet{1988MNRAS.233..437V}.  The
model values are from Fig. \ref{fig:bfluences}, with the symbols
replaced by connected lines.}
\label{fig:tautotdat}
\end{figure}

\subsubsection{Recurrence Time}
\label{sec:recurrence}

The results on $t_{\rm rec}$ in Fig. \ref{fig:trec} show a variety of
patterns that could be tested against observations, but we are not
aware of appropriate data in the literature.  Continuous monitoring of
sources, e.g., the recent work of Tournear et al. (2003) and
\citet{cor03}, would be very useful in this regard.  The latter paper
finds that the peak burst rate in several sources is about 10 bursts
per day, corresponding to a recurrence time of about 2.5 hours.  (They
also see sources with a peak burst rate of 2.5 per day, but these seem
to belong to a separate class.)  In our model, the shortest recurrence
time is about 5 hours (Fig. 11), corresponding to a burst rate of
about 5 bursts per day; this particular rate is found for smaller
neutron star radii (6.5 km and 10.4 km) at the break point between the
prompt mixed bursts and delayed mixed bursts at $\log(l_{\rm
acc})\sim-1$.  The location of the peak is in good agreement with the
observations, but there is a factor of 2 discrepancy in the maximum
burst rate.  The latter may indicate that some of the approximations
we have made (e.g., the opacity or the inner boundary condition) are
still inadequate.  It might also indicate that only a fraction of the
fuel is burned during the burst.

The observations of \citet{cor03} indicate in a few sources that the
burst rate decreases proportional to the luminosity below the break
point (e.g., GX 354--0 in their Fig. 2); equivalently, their
recurrence times vary inversely as $t_{\rm rec}\propto l_{\rm
acc}^{-1}$.  Such a trend corresponds to a constant value of
$\Sigma_{\rm layer,crit}$, and arises naturally in our model (and in
other models, e.g., see Figs.~\ref{fig:gammacrit}, \ref{fig:fluxbc}).
Among the models we have calculated, it appears that smaller stellar
radii (6.5 km, 10.4 km) are more consistent with the observed trend
than larger radii (16.4 km).  Also, of the four core temperatures we
have tried, we find best agreement with the data for the hotter
temperatures $10^{7.5}, ~10^8, ~10^{8.5}$ K.  A core temperature of
$10^7$ K appears unlikely.

Above the break, \citet{cor03} find a decrease in the burst rate by a
factor 4.  This is consistent with the increase in $t_{\rm rec}$ seen
in the models.  Another interesting feature to check would be the
sudden increase of $t_{\rm rec}$ with decreasing $l_{\rm acc}$ in the
helium burst regime ($l_{\rm acc}<-1.5$).  This regime has regions of
very long recurrence times, which for all practical purposes may be
considered burst-stable since observations are unlikely to detect
these very rare bursts.  We thus predict a ``gap'' in the occurrence
of bursts for accretion luminosities in the vicinity of
$\sim10^{-2}L_{\rm Edd}$ \citep{Nara01typeibh}.  It is not that there
are no bursts here, but that bursts are very rare and correspondingly
very luminous (Fig. \ref{fig:bfluences}).  It is not clear that the
helium burst regime has been seen in the observations, or that there
is a gap.  Indeed, the van Paradijs et al. (1988) give anomalously
{\it low} values of $\alpha$ here, whereas helium bursts should have
large values of $\alpha$.

\subsubsection{Burst Fluence}
\label{sec:fluence}

Figure \ref{fig:tautotdat} compares observed burst durations $\tau$
from the \citet{1988MNRAS.233..437V} compilation with our model
predictions for the Eddington-scaled duration $t_{\rm H+He}$.  We see
that the predicted durations are generally larger than the observed
ones, though the models with a stellar radius of 6.5 km (the lowest
lines in the panels) come pretty close.  As in the case of the
recurrence time, this comparison again is compromised by the
possibility of partial burning of the fuel.  Modulo this caveat, we
conclude once again that somewhat hotter cores $T_{\rm core} \gtrsim
10^{7.5}$ K and smaller neutron star radii $R \lesssim 10$ km are
preferred.

One interesting feature in the models is that the burst duration drops
by about a factor of 1.5--2 above the break point at $\log(l_{\rm
acc})\sim-1$.  This is precisely the same point at which the burst
recurrence time increases by a large factor.  In the data presented by
\citet{cor03}, such a decrease in the burst durations is seen (it is
less obvious in the van Paradijs et al. data), but the amount of the
decrease appears to be larger than we predict.

%\begin{figure}
%\plotone{f20.eps}
%\caption{Similar to Fig. \ref{fig:tautotdat}, but with burst durations
%$t_{\rm He}$ calculated using only helium burning.  The model values
%of $t_{\rm He}$ are taken from Fig. \ref{fig:hefluences}, with symbols
%replaced by connected lines.}
%\label{fig:tauhedat}\end{figure}

%\begin{figure}
%\plotone{f21.eps}
%\caption{Plot of burst durations vs recurrence times, showing a
%comparison of the model predictions with the data presented in
%\citet{1988MNRAS.233..437V}.  In each panel, the uppermost line
%corresponds to a neutron star radius of 16.4 km, the middle line to a
%radius of 10.4 km, and the lowermost line to a radius of 6.5 km.  The
%model results assume that all the available fuel is burned.  The large
%arrows indicate the amount by which the model results would move if
%only 1\% of the fuel is burned in a burst. }
%\label{fig:tauubdat}
%\end{figure}

\subsubsection{Overstable Modes}
\label{sec:onverstable}

\citet{2001A&A...372..138R} have reported a new class of low-frequency
oscillations in burst sources that might be related to oscillatory
behavior of the accreted layer prior to a burst.  The oscillations
seem to be present for accretion luminosities of about a tenth
Eddington, and the frequency is about 0.01 Hz.

Because our analysis computes the complex frequencies of unstable
modes, it is natural to ask if we can explain the observed
oscillations.  The oscillations seem to be observed mostly in the
regime that we have identified as delayed mixed bursts, above the
break discussed in the previous subsections.  In this regime, we do
find complex values of $\gamma$ with the imaginary parts often larger
than the real parts.  However, our mode frequencies are generally
lower than the observed frequencies by a factor of 10 or more.  We do
not know if this is an indication that some of our input physics is
inadequate (\S6.1) or perhaps that our numerical technique for finding
eigenmodes is unable to converge on modes with very high frequencies
(high compared to the accretion frequency $\gamma_{\rm acc}$).  More
work is needed.  Alternatively, the oscillations may be related to
slow propagation of the burning front on the stellar surface (see
Bildsten 1995).

\subsection{Future Prospects}
\label{sec:future-prospects}

The richness of the properties of nuclear burning on neutron stars as
revealed by the {\it ab initio} linear stability analysis presented
here is remarkable.  Considering the encouraging agreement that we
find with observational data, the model deserves further exploration.
Since we have used a variety of prescriptions for calculating the
opacity, the equation of state, the nuclear burning, etc., it might be
useful to improve these aspects of the model.  The radiative free-free
opacity in the outer layers seems to have a large effect
(\S\ref{sec:sensitivity}), so this is one area that needs more work.
The treatment of the inner boundary condition is also a serious issue
that needs to be considered in greater detail, especially for cool
neutron star cores (see \S\ref{sec:boundary-conditions}).

One simplification in the present calculations is that we consider
only a limited number of reactions for H-burning and He-burning
(\S2.3.3).  At a sufficiently high temperature, breakout from the hot
CNO cycle is expected to occur and hydrogen will be burned via the
rapid proton capture (rp) process (Wallace \& Woosley 1982; Schatz et
al. 1999).  Figure 5 of Schatz et al. (1999) shows the temperature
limits above which the reaction rates for processes such as
$^{15}$O$(\alpha,\gamma)^{19}$Ne and $^{14}$O$(\alpha,p)^{17}$F exceed
the usual $\beta$-decay and proton capture rates for these nuclei.  We
have confirmed that all the burst-unstable configurations we have
calculated in the present paper lie to the left of the corresponding
lines in the Schatz et al. plot.  Hence, none of our burst results are
affected by the neglect of these additional reactions.  However, close
to the Eddington accretion rate, where our model predicts that the
systems should be burst-stable, some of our equilibria cross into the
zone where the breakout reactions, especially
$^{15}$O$(\alpha,\gamma)^{19}$Ne, become important.  It is thus
conceivable that some of the very high $\dot M$ configurations that we
have identified as stable may turn out to be unstable when breakout
reactions are included.  This topic needs further investigation.

In terms of further calculations, the most obvious next step is to
explore the effect of varying the composition of the accreting gas.
Variations in metallicity between burst sources in globular clusters
and those in the Galactic disk may possibly have observational
consequences.  As a more extreme example, ultracompact binaries with
degenerate helium secondaries should have very different properties
from the hydrogen-dominant systems we have considered here.
Obviously, without hydrogen, there will be no unstable hydrogen
burning, so the hydrogen burst region will be missing from the
results.  Instead, we imagine the helium burst regime will extend both
to lower and higher accretion rates than in the results presented in
this paper.  Because helium requires a higher temperature/density to
ignite, the bursts should occur more rarely than for
hydrogen-accreting neutron stars.

In \citet{Nara01typeibh}, we examined the burst properties of more
massive primaries (black hole candidates) and found that the threshold
for stable nuclear burning moves to a higher accretion rate.
\citet{1982PASJ...34..495H} argue that the properties of the nuclear
burning to lowest order are functions of the surface gravity and the
mass accretion rate; therefore, one can understand the mass dependence
of burning regimes by keeping the mass fixed and varying the radius
and accretion rate; specifically, increasing the mass of the primary
shifts a diagram like Fig.~\ref{fig:regime} down and to the right
which helps to account for the results of \citet{Nara01typeibh}.  Even
though this argument gives a qualitative understanding of the mass
dependence of the burning regimes, serious quantitative applications
requires the kind of detailed calculations we have presented here.

Recently, a class of very long X-ray bursts has been discovered
(Cornelisse et al. 2000; Wijnands 2001) which have been interpreted as
carbon-burning bursts (Taam \& Picklum 1978; Strohmayer 2000; Cumming
\& Bildsten 2001; Strohmayer \& Brown 2002).  In most cases, the
bursts cannot be due to helium burning of a very thick layer since
normal Type I bursts are seen at the same accretion rate, sometimes
just before the superburst occurs.  It would be interesting to include
carbon-burning physics in our model to see what kinds of bursts are
predicted.  It is worth noting, however, that not all long bursts
necessarily arise from carbon-burning.  As we see in
Fig.~\ref{fig:bfluences}, we predict very long helium and even
hydrogen bursts at low accretion luminosities under appropriate
conditions.  Some of the observed long bursts (e.g., Gotthelf \&
Kulkarni 1997) may correspond to these.

On the observational front, there are numerous tests that one could
envisage based on the results shown in
Figs. \ref{fig:regime}--\ref{fig:alpha}.  The different regimes of
bursting --- delayed mixed bursts, prompt mixed bursts, helium bursts,
hydrogen bursts --- reveal very distinct patterns in various
observables.  If these patterns are seen in the data, then one might
be able to constrain the neutron star radius and/or the core
temperature fairly well.  This would have important implications for
the neutron star equation of state and the nature of neutrino cooling
in the core.  We have made a beginning along these lines in this paper
and have argued that perhaps $T_{\rm core}$ is $\gtrsim 10^{7.5}$ K
and neutron star radii are $\lesssim 10$ km.  More work along these
lines is worthwhile.

If we are to extend this model, which has been developed for neutron
stars, to thermonuclear instabilities in accreting white dwarfs
(classical novae), it would be necessary to include Coulomb
corrections in the equation of state, a more accurate treatment of the
semi-degenerate regime, and more detailed radiative and conductive
opacities.  Understanding whether there is a regime of stable nuclear
burning at high accretion rates onto a white dwarf is a key ingredient
of any scenario in which Type Ia supernovae result from stable
accretion of mass onto a white dwarf \citep{2000ARA&A..38..191H}.
Because classical novae typically eject the material accreted along
with some of the substrate, a regime of stable nuclear burning is
required for accretion to cause the white dwarf to grow in mass and to
end up in a supernova.  Unfortunately, our current prescriptions are
too crude to treat this important burning regime accurately, but the
potential of the technique for white dwarfs is tantalizing.

Finally, the methods that we have described here have deliberately
avoided examination of the physics of the burst itself, during which
many complications arise, including other nuclear reaction channels,
convection and hydrodynamics.  An important question to examine is
whether our techniques could be extended in any simple way to study
the properties of the bursts.

\section{Summary}
\label{sec:summary}

We have presented a comprehensive treatment of the stability of
nuclear burning on the surface of neutron stars.  For the first time,
we have calculated the linearly unstable eigenmodes of an accretion
layer in quasi-steady state, making no ad hoc assumptions regarding
the criterion for instability.  The model reproduces the various
previously known regimes of nuclear burning on neutron stars, and
agrees with earlier results where there is overlap.  Additionally, we
have been able to probe in detail the behavior of accreting neutron
stars at high mass accretion rates, near the threshold of stable
nuclear burning.  Here, we find a hitherto unrecognized regime of
delayed mixed bursts, with very distinct properties compared to the
more standard prompt mixed bursts.

For accretion rates greater than one percent of the Eddington rate, we
find encouraging agreement between the model predictions and
observations of bursts.  The existence of the regime of delayed mixed
bursts provides a natural explanation for the observed dramatic
increase of the burst parameter $\alpha$ at high accretion
luminosities (Fig.~\ref{fig:alphadat} and Cornelisse et al. 2003).  In
addition, there is some indication from the preliminary comparisons
presented here that burst systems have hot cores $\gtrsim 10^{7.5}$ K,
consistent with cooling in the neutron star interior being dominated
by the modified URCA process or a similar low-efficiency cooling
mechanism.  Cool cores with $T_{\rm core}\sim10^7$ K, as might be
present if direct URCA cooling were to operate, are less likely.  We
also find a number of indications for small neutron star radii
$\lesssim 10$ km.  These results could be tightened with more careful
modeling, e.g., by improving some of the prescriptions we use for the
input physics (\S\ref{sec:future-prospects}), and with more extensive
and better quality data.

\acknowledgments We thank Deepto Chakrabarty, Duncan Galloway, Erik
Kuulkers, Feryal Ozel and Dimitrios Psaltis for useful discussions and
comments.  We are also grateful to the referee for a number of
suggestions on the prescriptions used in \S2.3 and for comments that
helped to improve the presentation of the paper.  R.N. was supported
in part by NASA grant NAG5-10780.  J.S.H. was supported by the
National Aeronautics and Space Administration through Chandra
Postdoctoral Fellowship Award Number PF0-10015 issued by the Chandra
X-ray Observatory Center, which is operated by the Smithsonian
Astrophysical Observatory for and on behalf of NASA under contract
NAS8-39073.

\appendix

\section{Equilibrium and Perturbations}
\label{sec:app}

Here we discuss in more detail the basic equations describing the
equilibrium and perturbations of the accreted layer.  The governing
equations are given in equations
(\ref{eq:fulleqs1})--(\ref{eq:fulleqs5}).  We use the notation defined
in Table~\ref{tab:sym} and in equations (14), (15), where quantities
like $\rho_0$, $T_0$ refer to the equilibrium, $\rho$, $T$ refer to
the corresponding quantities in the perturbed state, and $\rho_1$,
$T_1$ refer to the spatial component of the linear perturbations.  The
equilibrium quantities have no time dependence, while the
perturbations have a time dependence of the form $\exp(\gamma t)$.
Since $\gamma$ is in general complex, the quantities $\rho$, $T$,
$\rho_1$, $T_1$, etc. are complex, whereas $\rho_0$, $T_0$, etc. are
real.  Also, since $X_0+Y_0+Z_0=1$ and $X+Y+Z=1$, we have
$X_1+Y_1+Z_1=0$.

Let us begin with the radiative transfer equation (\ref{eq:fulleqs2}).
For the equilibrium, this equation gives
\begin{equation}
{\partial T_0 \over \partial \Sigma} =
{3\kappa_0 F_0 \over 16\sigma T_0^3},
\label{app1}
\end{equation}
where all quantities are real.  For the perturbations, we linearize
equation (\ref{eq:fulleqs2}) and consider first-order deviations.
This gives
\begin{equation}
{\partial T_1 \over \partial \Sigma} =
{3\kappa_0\over16\sigma T_0^3} F_1 + {3F_0\over16\sigma} \left[
{\partial\over\partial\rho}\left({\kappa\over T^3}\right)_0 \rho_1
+{\partial\over\partial T}\left({\kappa\over T^3}\right)_0 T_1
+{\partial\over\partial X}\left({\kappa\over T^3}\right)_0 X_1
+{\partial\over\partial Y}\left({\kappa\over T^3}\right)_0 Y_1
\right], \label{app2}
\end{equation}
where now the various quantities are in general complex.  For
instance, $\partial(\kappa/T^3)/\partial\rho$ refers to the derivative
of the complex quantity $\kappa/T^3$ with respect to complex $\rho$
(the derivative is well-defined since all the quantities are
analytic).  In practice, we calculate such derivatives numerically.
Since $\kappa$ in general depends on all three quantities $X$, $Y$,
$Z$, it is necessary to replace $Z$ by $1-X-Y$ or $Z_1$ by $-X_1-Y_1$
before computing the partial derivatives $\partial/\partial X$,
$\partial/\partial Y$.

Instead of writing an equation for the linear perturbation $T_1$, we
could equally well consider the equation for the total (complex)
perturbed temperature $T = T_0 + T_1 \exp(\gamma t) T_1$.  This is
nothing but the original equation
\begin{equation}
{\partial T\over\partial\Sigma} =
{3\kappa F\over 16\sigma T^3}, \label{app3}
\end{equation}
where $F$ is now the total complex perturbed flux, and $\kappa$ is the
opacity corresponding to the perturbed $\rho$, $T$, $X$, $Y$.  Within
the linear approximation, this equation, coupled with equation
(\ref{app1}), has the same content as equation (\ref{app2}) for the
linear perturbations (equation \ref{app2} multiplied by $\exp(\gamma
t)$ is just the difference of equations \ref{app3} and \ref{app1}).
Equation (\ref{app3}) has the advantage of being more compact than
equation (\ref{app2}).  It is also numerically more convenient, since
the compactness of the equation translates to relative simplicity of
the corresponding computer code.

Consider next the H-evolution equation (\ref{eq:fulleqs4}), which
involves a time derivative.  The equilibrium is described by
\begin{equation}
\dot\Sigma {\partial X_0\over\partial\Sigma} =
-{\epsilon_{\rm H,0}\over E_{\rm H}^*}, \label{app4}
\end{equation}
where all quantities are real.  The linear first order perturbation
$X_1$ satisfies
\begin{equation}
\gamma X_1 + \dot\Sigma {\partial X_1\over\partial\Sigma} =
-{1\over E_{\rm H}^*} \left[
\left( {\partial\epsilon_H\over\partial \rho} \right)_0 \rho_1
+\left( {\partial\epsilon_H\over\partial T} \right)_0 T_1
+\left( {\partial\epsilon_H\over\partial X} \right)_0 X_1
+\left( {\partial\epsilon_H\over\partial Y} \right)_0 Y_1
\right], \label{app5}
\end{equation}
where the term $\gamma X_1$ on the left comes from the time derivative
$\partial/\partial t$ operating on $\exp(\gamma t)X_1$.  Once again,
instead of considering the equation for the perturbation $X_1$, we may
write down the equation for the total perturbed quantity
$X=X_0+\exp(\gamma t)X_1$:
\begin{equation}
\gamma (X-X_0) + \dot\Sigma {\partial X\over\partial\Sigma} =
-{\epsilon_{\rm H}\over E_{\rm H}^*}, \label{app6}
\end{equation}
which is nearly identical in form to equation (\ref{app4}) for the
equilibrium, except that (i) it has an extra term proportional to
$\gamma$ because of the time dependence of the perturbations, (i)
$\epsilon_{\rm H}$ is evaluated at the perturbed $\rho$, $T$, $X$,
$Y$, and (iii) all quantities are complex.

Following the above examples, the other three equations are
straightforward.  The hydrostatic equilibrium equation
(\ref{eq:fulleqs1}) gives for the equilibrium
\begin{equation}
{\partial P_0 \over \partial \rho_0} {\partial \rho_0 \over \partial \Sigma}
+{\partial P_0 \over \partial T_0} {\partial T_0 \over \partial \Sigma}
+{\partial P_0 \over \partial X_0} {\partial X_0 \over \partial \Sigma}
+{\partial P_0 \over \partial Y_0} {\partial Y_0 \over \partial \Sigma}
=g, \label{app7}
\end{equation}
and for the perturbations
\begin{equation}
{\partial P \over \partial \rho} {\partial \rho \over \partial \Sigma}
+{\partial P \over \partial T} {\partial T \over \partial \Sigma}
+{\partial P \over \partial X} {\partial X \over \partial \Sigma}
+{\partial P \over \partial Y} {\partial Y \over \partial \Sigma}
=g. \label{app8}
\end{equation}

The energy conservation equation gives
\begin{equation}
{\partial F_0\over\partial\Sigma} =  -\dot\Sigma T_0 \left[
\left({\partial s\over\partial \rho}\right)_0 
{\partial\rho_0\over\partial\Sigma}
+\left({\partial s\over\partial T}\right)_0 
{\partial T_0\over\partial\Sigma}
+\left({\partial s\over\partial X}\right)_0 
{\partial X_0\over\partial\Sigma}
+\left({\partial s\over\partial Y}\right)_0 
{\partial Y_0\over\partial\Sigma}
\right] -\epsilon_{\rm H,0} -\epsilon_{\rm He,0}, \label{app9}
\end{equation}
\begin{eqnarray}
{\partial F\over\partial\Sigma} & = & -\gamma T \left[
\left({\partial s\over\partial \rho}\right)
(\rho - \rho_0)
+\left({\partial s\over\partial T}\right)
(T-T_0)
+\left({\partial s\over\partial X}\right)
(X-X_0)
+\left({\partial s\over\partial Y}\right)
(Y-Y_0)
\right] \nonumber \\
& & -\dot\Sigma T \left[
\left({\partial s\over\partial \rho}\right)
{\partial\rho\over\partial\Sigma}
+\left({\partial s\over\partial T}\right)
{\partial T\over\partial\Sigma}
+\left({\partial s\over\partial X}\right)
{\partial X\over\partial\Sigma}
+\left({\partial s\over\partial Y}\right)
{\partial Y\over\partial\Sigma}
\right] -\epsilon_{\rm H} -\epsilon_{\rm He}, \label{app10}
\end{eqnarray}

Finally, the He-evolution equation gives
\begin{equation}
\dot\Sigma {\partial Y_0\over\partial\Sigma} =
{\epsilon_{\rm H,0}\over E_{\rm H}^*}
-{\epsilon_{\rm He,0}\over E_{\rm He}^*}, \label{app11}
\end{equation}
\begin{equation}
\gamma (Y-Y_0) + \dot\Sigma {\partial Y\over\partial\Sigma} =
{\epsilon_{\rm H}\over E_{\rm H}^*} -{\epsilon_{\rm He}\over E_{\rm
He}^*}. \label{app12}
\end{equation}

Equations (\ref{app1}), (\ref{app4}), (\ref{app7}), (\ref{app9}) and
(\ref{app11}) are five differential equations for the equilibrium
quantities $F_0$, $\rho_0$, $T_0$, $X_0$, $Y_0$.  We assume a value
for the outgoing flux $F_{\rm out,0}$ at the surface of the layer,
solve for $\rho_{\rm out,0}$, $T_{\rm out,0}$, $X_{\rm out,0}$,
$Y_{\rm out,0}$ from the outer boundary conditions, and then integrate
the 5 differential equations down to the bottom of the layer and then
into the substrate to a depth equal to $\Sigma_{\rm diff}$ defined in
equation (\ref{ibc1}).  At this depth, we require the temperature
$T_0(\Sigma_{\rm diff})$ to be equal to the required core temperature
$T_{\rm core}$.  If it is not, we change the value of $F_{\rm out,0}$
and repeat until the inner boundary condition is satisfied.  We then
have a valid equilibrium solution.

For the perturbations, we work with equations (\ref{app3}),
(\ref{app6}), (\ref{app8}, (\ref{app10}), (\ref{app12}), which are
five differential equations for the total perturbed quantities $F$,
$\rho$, $T$, $X$, $Y$.  At the surface, we set the flux equal to
$F_{\rm out} = F_{\rm out,0}+F_{\rm out,1}$, where $F_{\rm out,1} \ll
F_{\rm out,0}$ in order to satisfy the assumption of a linear
perturbation of the equilibrium.  We solve for the other variables at
the surface, assume a value for the eigenvalue $\gamma$, and integrate
the equations down to a depth $\Sigma_{\rm max}$ in the substrate (see
eq.~\ref{ibc2}).  At this depth we require $T(\Sigma_{\rm
max})=T_0(\Sigma_{\rm max})$.  We vary $\gamma$ until this inner
boundary condition is satisfied, at which point we have a solution
for the complex eigenvalue $\gamma$, and also the shape of the
eigenfunction ($F_1$, $\rho_1$, etc.).  The search in $\gamma$ space
is tailored to find eigenvalues with positive real parts since only
such modes are unstable.

Let us define a turning point in the sequence of equilibria (S-curve)
to be a point at which the locus of equilibrium solutions satisfies
the condition $d\Sigma_{\rm layer}/dF_{\rm out,0} = 0$.  Since the
derivative is zero, it means that two equilibria with escaping surface
fluxes equal to $F_{\rm out,0}$ and $F_{\rm out} = F_{\rm
out,0}+F_{\rm out,1}$, where $F_{\rm out,1} \ll F_{\rm out,0}$, both
satisfy the differential equations as well as the boundary conditions
for the same value of $\Sigma_{\rm layer}$.  Now, the solution for
$F_{\rm out,0}$ satisfies the equilibrium equations (\ref{app1}),
(\ref{app4}), (\ref{app7}), (\ref{app9}), (\ref{app11}), while the
solution for $F_{\rm out}$ satisfies the perturbed equations
(\ref{app3}), (\ref{app6}), (\ref{app8}), (\ref{app10}),
(\ref{app12}).  The only difference between the two sets of equations
is the presence of various terms involving $\gamma$.  If the second
solution is also an equilibrium solution, then it implies that it must
satisfy the perturbation equations with $\gamma$ precisely equal to 0,
i.e., one of the modes of the system corresponds to 0 frequency.  This
result is not surprising.  Since the initial model is at a turning
point, there are neighboring equilibria with the same value of
$\Sigma_{\rm layer}$ but different values of $F_{\rm out,0}$, i.e.,
the system has a linear zero-frequency mode.

The above theorem, that a mode with $\gamma=0$ exists at turning
points of the S-curve, is well-known for one-zone models, e.g., see
the discussion of \citet{1983ApJ...264..282P} for an application to
bursts.  Our discussion shows that the same result is valid even when
one is considering the more complex model described here.  The reason
it works is that our equilibria are ultimately labeled by a single
parameter, the value of $F_{\rm out,0}$ at the surface.  This is all
that matters, and the fact that our solutions involve many variables
and are described by continuous functions is not relevant.

\bibliographystyle{apj} 
\bibliography{ns,mine,lmxb,typei,fus,math}

\begin{thebibliography}{45}
\expandafter\ifx\csname natexlab\endcsname\relax\def\natexlab#1{#1}\fi

\bibitem[Allen(2000)]{all00}
Allen, C. W. 2000, Allen's Astrophysical Quantities, fourth edition, 
ed. A. N. Cox, AIP Press (New York: Springer-Verlag)

\bibitem[{{Antia}(1993)}]{1993ApJS...84..101A}
{Antia}, H.~M. 1993, \apjs, 84, 101

\bibitem[Bildsten(1995)]{bil95}
Bildsten, L. 1995, \apj, 438, 852

\bibitem[{Bildsten(1998)}]{Bild98}
Bildsten, L. 1998, in The Many Faces of Neutron Stars, ed. A.~Alpar,
  L.~Buccheri, \& J.~van Paradij (Dordrecht: Kluwer), 419, astro-ph/9709094

\bibitem[{{Bildsten}(2000)}]{2000arxt.confE..65B}
{Bildsten}, L. 2000, in Rossi2000: Astrophysics with the Rossi X-ray Timing
  Explorer. March 22-24, 2000 at NASA's Goddard Space Flight Center, Greenbelt,
  MD USA, p.E65

\bibitem[{{Bildsten} \& {Cumming}(1998)}]{1998ApJ...506..842B}
{Bildsten}, L. \& {Cumming}, A. 1998, \apj, 506, 842

\bibitem[Brown(2000)]{bro00}
Brown, E. F. 2000, \apj, 531, 988

\bibitem[{{Brown} \& {Bildsten}(1998)}]{1998ApJ...496..915B}
{Brown}, E.~F. \& {Bildsten}, L. 1998, \apj, 496, 915

\bibitem[{{Brown} {et~al.}(1998){Brown}, {Bildsten}, \&
  {Rutledge}}]{1998ApJ...504L..95B}
{Brown}, E.~F., {Bildsten}, L., \& {Rutledge}, R.~E. 1998, \apjl, 504, L95

\bibitem[{Clayton(1983)}]{Clay83}
Clayton, D.~D. 1983, Principle of Stellar Evolution and Nucleosynthesis
  (Chicago: The University of Chicago Press)

\bibitem[Cornelisse et al.(2000)]{cor00}
  Cornelisse, R., Heise, J., Kuulkers, E., Verbunt, F., \& in't Zand,
  J. J. M. 2000, \aa, 357, L21  

\bibitem[Cornelisse et al.(2003)]{cor03}
  Cornelisse, R., in't Zand, J. J. M., Verbunt, F., Kuulkers, E.,
  Heise, J., et al. 2003, \aa, in press (astro-ph/0304500)  

\bibitem[{{Cumming} \& {Bildsten}(2000)}]{2000ApJ...544..453C}
{Cumming}, A. \& {Bildsten}, L. 2000, \apj, 544, 453

\bibitem[{{Dewitt} {et~al.}(1973){Dewitt}, {Graboske}, \&
  {Cooper}}]{1973ApJ...181..439D}
{Dewitt}, H.~E., {Graboske}, H.~C., \& {Cooper}, M.~S. 1973, \apj, 181, 439

\bibitem[{{Epstein} {et~al.}(1983){Epstein}, {Gudmundsson}, \&
  {Pethick}}]{1983MNRAS.204..471E}
{Epstein}, R.~I., {Gudmundsson}, E.~H., \& {Pethick}, C.~J. 1983, \mnras, 204,
  471

\bibitem[{{Ergma} \& {Tutukov}(1980)}]{1980A&A....84..123E}
{Ergma}, E.~V. \& {Tutukov}, A.~V. 1980, \aap, 84, 123

\bibitem[{{Fujimoto} {et~al.}(1987{\natexlab{a}}){Fujimoto}, {Hanawa}, {Iben},
  \& {Richardson}}]{1987ApJ...315..198F}
{Fujimoto}, M.~Y., {Hanawa}, T., {Iben}, I.~J., \& {Richardson}, M.~B.
  1987{\natexlab{a}}, \apj, 315, 198

\bibitem[{{Fujimoto} {et~al.}(1981){Fujimoto}, {Hanawa}, \&
  {Miyaji}}]{1981ApJ...247..267F}
{Fujimoto}, M.~Y., {Hanawa}, T., \& {Miyaji}, S. 1981, \apj, 247, 267

\bibitem[{{Fujimoto} {et~al.}(1987{\natexlab{b}}){Fujimoto}, {Sztajno},
  {Lewin}, \& {van Paradijs}}]{1987ApJ...319..902F}
{Fujimoto}, M.~Y., {Sztajno}, M., {Lewin}, W.~H.~G., \& {van Paradijs}, J.
  1987{\natexlab{b}}, \apj, 319, 902

\bibitem[{{Fushiki} \& {Lamb}(1987{\natexlab{a}})}]{1987ApJ...323L..55F}
{Fushiki}, I. \& {Lamb}, D.~Q. 1987{\natexlab{a}}, \apjl, 323, L55

\bibitem[{{Fushiki} \& {Lamb}(1987{\natexlab{b}})}]{1987ApJ...317..368F}
---. 1987{\natexlab{b}}, \apj, 317, 368

\bibitem[Gotthelf \& Kulkarni(1997)]{got97}
  Gotthelf, E. V., \& Kulkarni, S. R. 1997, \apj, 490, L161

\bibitem[{{Grindlay} {et~al.}(1976){Grindlay}, {Gursky}, {Schnopper},
  {Parsignault}, {Heise}, {Brinkman}, \& {Schrijver}}]{1976ApJ...205L.127G}
{Grindlay}, J., {Gursky}, H., {Schnopper}, H., {Parsignault}, D.~R., {Heise},
  J., {Brinkman}, A.~C., \& {Schrijver}, J. 1976, \apjl, 205, L127

\bibitem[{{Hanawa} \& {Fujimoto}(1982)}]{1982PASJ...34..495H}
{Hanawa}, T. \& {Fujimoto}, M.~Y. 1982, \pasj, 34, 495

\bibitem[{{Hanawa} \& {Fujimoto}(1986)}]{1986PASJ...38...13H}
---. 1986, \pasj, 38, 13

\bibitem[{{Hansen} \& {van Horn}(1975)}]{1975ApJ...195..735H}
{Hansen}, C.~J. \& {van Horn}, H.~M. 1975, \apj, 195, 735

\bibitem[{Hernquist \& Applegate(1984)}]{Hern84b}
Hernquist, L. \& Applegate, J.~H. 1984, ApJ, 287, 244

\bibitem[{{Hillebrandt} \& {Niemeyer}(2000)}]{2000ARA&A..38..191H}
{Hillebrandt}, W. \& {Niemeyer}, J.~C. 2000, \araa, 38, 191

\bibitem[{{Iben}(1975)}]{1975ApJ...196..525I}
{Iben}, I. 1975, \apj, 196, 525

\bibitem[{{Itoh} {et~al.}(2003){Itoh}, {Tomizawa}, {Wanajo}, \&
  {Nozawa}}]{2003ApJ...586.1436I}
{Itoh}, N., {Tomizawa}, N., {Wanajo}, S., \& {Nozawa}, S. 2003, \apj, 586, 1436

\bibitem[{{Joss}(1977)}]{1977Natur.270..310J}
{Joss}, P.~C. 1977, \nat, 270, 310

\bibitem[{{Joss}(1978)}]{1978ApJ...225L.123J}
---. 1978, \apjl, 225, L123

\bibitem[{{Joss} \& {Li}(1980)}]{1980ApJ...238..287J}
{Joss}, P.~C. \& {Li}, F.~K. 1980, \apj, 238, 287

\bibitem[Kuulkers \& van der Klis(2000)]{KvK00}
  Kuulkers, E., \& van der Klis, M. 2000, \aa, 356, L45

\bibitem[Kuulkers, van der Klis \& van Paradijs (2000)]{Ketal00}
  Kuulkers, E., van der Klis, M., \& van Paradijs, J. 1995, \apj, 450, 748

\bibitem[{{Lewin} {et~al.}(1993){Lewin}, {van Paradijs}, \&
  {Taam}}]{1993SSRv...62..223L}
{Lewin}, W.~H.~G., {van Paradijs}, J., \& {Taam}, R.~E. 1993, Space Science
  Reviews, 62, 223+

\bibitem[Matsuba et al.(1995)]{Metal95}
  Matsuba, E., Dotani, T., Misuda, K., Asai, K., Lewin, W.~H.~G.,
  van Parakijs, J., \& van der Klis, M. 1995, \pasj, 47, 575

\bibitem[{Mathews \& Dietrich(1984)}]{1984ApJ...287..969M}
Mathews, G.~J. \& Dietrich, F.~S. 1984, ApJ, 287, 969

\bibitem[{M\"uller(1956)}]{Mull56}
M\"uller, D.~E. 1956, Mathematical Tables and Aids to Computation, 10, 208

\bibitem[Muno et al.(2001)]{mun01}
  Muno, M. P., Chakrabarty, D., Galloway, D. K., \& Savov, P. 2001,
  \apj, 553, L157

\bibitem[{Narayan \& Heyl(2002)}]{Nara01typeibh}
Narayan, R. \& Heyl, J.~S. 2002, \apjl, 574, 139

\bibitem[{{Paczynski}(1983{\natexlab{a}})}]{1983ApJ...264..282P}
{Paczynski}, B. 1983{\natexlab{a}}, \apj, 264, 282

\bibitem[{{Paczynski}(1983{\natexlab{b}})}]{1983ApJ...267..315P}
---. 1983{\natexlab{b}}, \apj, 267, 315

\bibitem[{{Potekhin}(1999)}]{1999A&A...351..787P}
{Potekhin}, A.~Y. 1999, \aap, 351, 787

\bibitem[{{Regev} \& {Livio}(1984)}]{1984A&A...134..123R}
{Regev}, O. \& {Livio}, M. 1984, \aap, 134, 123

\bibitem[{{Revnivtsev} {et~al.}(2001){Revnivtsev}, {Churazov}, {Gilfanov}, \&
  {Sunyaev}}]{2001A&A...372..138R}
{Revnivtsev}, M., {Churazov}, E., {Gilfanov}, M., \& {Sunyaev}, R. 2001, \aap,
  372, 138

\bibitem[{{Schatz} {et~al.}(1999){Schatz}, {Bildsten}, {Cumming}, \&
  {Wiescher}}]{1999ApJ...524.1014S}
{Schatz}, H., {Bildsten}, L., {Cumming}, A., \& {Wiescher}, M. 1999, \apj, 524,
  1014

\bibitem[{Shapiro \& Teukolsky(1983)}]{Shap83}
Shapiro, S.~L. \& Teukolsky, S.~A. 1983, Black Holes, White Dwarfs, and Neutron
  Stars (New York: Wiley-Interscience)

\bibitem[Smale(1998)]{S98}
  Smale, A.~P. 1998, \apj, 498, L141

\bibitem[Strohmayer et al.(1996)]{Stro96}
  Strohmayer, T., Zhang, W., Swank, J. H., Smale, A., Titarchuk, L.,
  Day, C., \& Lee, U. 1996, \apj, 469, L9

\bibitem[Strohmayer(2001a)]{Stro01a}
  Strohmayer, T. 2001, Adv. Space Res., 28, 511

\bibitem[Strohmayer(2001b)]{stro01b}
  Strohmayer, T. 2000, AAS HEAD Meeting, 32, 24.10

\bibitem[Strohmayer \& Brown(2002)]{stro02}
  Strohmayer, T., \& Brown, E. F. 2002, \apj, 566, 1045

\bibitem[{Strohmayer {et~al.}(1998)Strohmayer, Swank, \& Zhang}]{Stro98}
Strohmayer, T.~E., Swank, J.~H., \& Zhang, W. 1998, Nucl. Phys. B (Proc.
  Suppl.), 69, 129, astro-ph/9801219

\bibitem[{{Taam}(1982)}]{1982ApJ...258..761T}
{Taam}, R.~E. 1982, \apj, 258, 761

\bibitem[{{Taam} \& {Picklum}(1978)}]{1978ApJ...224..210T}
{Taam}, R.~E. \& {Picklum}, R.~E. 1978, \apj, 224, 210

\bibitem[{{Taam} \& {Picklum}(1979)}]{1979ApJ...233..327T}
---. 1979, \apj, 233, 327

\bibitem[{{Taam} {et~al.}(1996){Taam}, {Woosley}, \&
  {Lamb}}]{1996ApJ...459..271T}
{Taam}, R.~E., {Woosley}, S.~E., \& {Lamb}, D.~Q. 1996, \apj, 459, 271

\bibitem[{{Taam} {et~al.}(1993){Taam}, {Woosley}, {Weaver}, \&
  {Lamb}}]{1993ApJ...413..324T}
{Taam}, R.~E., {Woosley}, S.~E., {Weaver}, T.~A., \& {Lamb}, D.~Q. 1993, \apj,
  413, 324

\bibitem[Tournear et al.(2003)]{Tetal03}
  Tournear, D., et al. 2003, preprint

\bibitem[{{van Paradijs} {et~al.}(1988){van Paradijs}, {Penninx}, \&
  {Lewin}}]{1988MNRAS.233..437V}
{van Paradijs}, J., {Penninx}, W., \& {Lewin}, W.~H.~G. 1988, \mnras, 233, 437

\bibitem[van Straaten et al.(2001)]{vanstr01}
  van Straaten, S., van der Klis, M., Kuulkers, E., \& Mendez, M.
  2001, \apj, 551, 907

\bibitem[Wijnands, R.(2001)]{wij01}
  Wijnands, R. 2001, \apj, 554, L59

\bibitem[Wallace \& Woosley(1982)]{wal82}
  Wallace, R. K., \& Woosley, S. E. 1982, \apj, 258, 696

\bibitem[{{Woosley} \& {Taam}(1976)}]{1976Natur.263..101W}
{Woosley}, S.~E. \& {Taam}, R.~E. 1976, \nat, 263, 101

\bibitem[Zingale et al.(2001)]{zin01}
  Zingale, M., Timmes, F. X., Fryxell, B., Lamb, D. Q., Olson, K.,
  et al. 2001, ApJS, 133, 195

\end{thebibliography}

\end{document}